\documentclass[12pt]{article}
\usepackage{color,upref,amsmath,amssymb,amscd}
\usepackage{amsthm}
\usepackage[dvipdfmx]{hyperref}
\usepackage{graphicx}
\usepackage{color}

\def\hybrid{\topmargin 0pt      \oddsidemargin 0pt
        \headheight 0pt \headsep 0pt
       \voffset-1cm
        \textwidth 6.25in       
       \textheight 9.5in       
        \marginparwidth 0.0in
        \parskip 5pt plus 1pt   \jot = 1.5ex}
\catcode`\@=11
\def\marginnote#1{}

\newcount\hour
\newcount\minute
\newtoks\amorpm
\hour=\time\divide\hour by60
\minute=\time{\multiply\hour by60 \global\advance\minute by-\hour}
\edef\standardtime{{\ifnum\hour<12 \global\amorpm={am}%
        \else\global\amorpm={pm}\advance\hour by-12 \fi
        \ifnum\hour=0 \hour=12 \fi
        \number\hour:\ifnum\minute<10 0\fi\number\minute\the\amorpm}}
\edef\militarytime{\number\hour:\ifnum\minute<10 0\fi\number\minute}

\def\draftlabel#1{{\@bsphack\if@filesw {\let\thepage\relax
   \xdef\@gtempa{\write\@auxout{\string
      \newlabel{#1}{{\@currentlabel}{\thepage}}}}}\@gtempa
   \if@nobreak \ifvmode\nobreak\fi\fi\fi\@esphack}
        \gdef\@eqnlabel{#1}}
\def\@eqnlabel{}
\def\@vacuum{}
\def\draftmarginnote#1{\marginpar{\raggedright\scriptsize\tt#1}}

\def\draftlabel#1{{\@bsphack\if@filesw {\let\thepage\relax
   \xdef\@gtempa{\write\@auxout{\string
      \newlabel{#1}{{\@currentlabel}{\thepage}}}}}\@gtempa
   \if@nobreak \ifvmode\nobreak\fi\fi\fi\@esphack}
        \gdef\@eqnlabel{#1}}
\def\@eqnlabel{}
\def\@vacuum{}
\def\draftmarginnote#1{\marginpar{\raggedright\scriptsize\tt#1}}

\def\draft{\oddsidemargin -.5truein
        \def\@oddfoot{\sl preliminary draft \hfil
        \rm\thepage\hfil\sl\today\quad\militarytime}
        \let\@evenfoot\@oddfoot \overfullrule 3pt
        \let\label=\draftlabel
        \let\marginnote=\draftmarginnote
   \def\@eqnnum{(\theequation)\rlap{\kern\marginparsep\tt\@eqnlabel}%
\global\let\@eqnlabel\@vacuum}  }

\def\numberbysection{\@addtoreset{equation}{section}
        \def\theequation{\thesection.\arabic{equation}}}

\def\underline#1{\relax\ifmmode\@@underline#1\else
        $\@@underline{\hbox{#1}}$\relax\fi}

\def\titlepage{\@restonecolfalse\if@twocolumn\@restonecoltrue\onecolumn
     \else \newpage \fi \thispagestyle{empty}\c@page\z@
        \def\thefootnote{\fnsymbol{footnote}} }

\def\endtitlepage{\if@restonecol\twocolumn \else  \fi
        \def\thefootnote{\arabic{footnote}}
        \setcounter{footnote}{0}}  
\relax

\numberbysection
\hybrid

\newfont{\Bbbb}{msbm7 scaled 1\@ptsize00}

\newcommand{\ZZ}{\mathbb Z}
\newcommand{\z}{\raise-1pt\hbox{$\mbox{\Bbbb Z}$}}
\newcommand{\RR}{\mathbb R}
\newcommand{\RRR}{\raise-1pt\hbox{$\mbox{\Bbbb R}$}}
\newcommand{\CC}{\mathbb C}
\newcommand{\CCC}{\raise-1pt\hbox{$\mbox{\Bbbb C}$}}
\newcommand{\SSSS}{\mathbb S}
\newcommand{\SSS}{\raise-1pt\hbox{$\mbox{\Bbbb S}$}}

\def\beq{\begin{equation}}
\def\eeq{\end{equation}}
\def\p{\partial}

\def\rbr{\right >}

\def\lvac{\left <0\right |}
\def\rvac{\left |0\right >}

\def\lvacn{\left <n\right |}
\def\rvacn{\left |n\right >}

\newcommand{\boldPsi}{{\boldsymbol\Psi}}
\newcommand{\boldpsi}{{\boldsymbol\psi}}
\newcommand{\boldPsistar}{{\boldsymbol\Psi}^{*}}
\newcommand{\boldpsistar}{{\boldsymbol\psi}^{*}}
\newcommand{\boldphi}{{\boldsymbol\phi}}
\newcommand{\boldPhi}{{\boldsymbol\Phi}}

\newcommand{\boldphihat}{{\boldsymbol {\hat \phi}}}
\newcommand{\boldPhihat}{{\boldsymbol {\hat \Phi}}}

\begin{document}

\begin{titlepage}

\title{Revisiting B\"acklund-Darboux transformations
for KP and BKP integrable hierarchies}

\author{A. Zabrodin\thanks{National Research University 
Higher School of Economics,
20 Myasnitskaya Ulitsa, Moscow 101000, Russia,
and NRC ``Kurchatov institute'', Moscow, Russia;
e-mail: zabrodin@itep.ru}}

\date{}
\maketitle


\begin{abstract}

We consider B\"acklund-Darboux transformations for 
integrable hierarchies of nonlinear equations such as KP,
BKP and their close relatives referred to as
modified KP and Schwarzian KP. We work in the 
framework of the bilinear formalism based on the 
bilinear equations for the tau-function. This approach allows one
to extend the theory to fully difference (or discrete) versions
of the integrable equations and their hierarchies in a 
natural way. 
We also
show how to construct the B\"acklund-Darboux transformations
in the operator approach developed by the Kyoto school, in which
the tau-functions are represented as vacuum expectation values
of certain operators made of free fermionic 
fields (charged for KP and neutral for BKP).

\end{abstract}

\end{titlepage}

\vspace{5mm}

%

\tableofcontents

\vspace{5mm}

\section{Introduction}

B\"acklund-Darboux (BD) transformations acting in the space of 
exact solutions to integrable nonlinear partial 
differential or difference 
equations is an old and well studied subject. Generally speaking,
such a transformation sends any solution of
one or another integrable equation to another solution of the same
or maybe a different equation. As soon as integrable equations 
are known to form infinite hierarchies of compatible equations,
one may extend the notion of a BD transformation to the whole
hierarchy. General solutions to such hierarchies as 
Kadomtsev-Petviashvili (KP), its B-version (BKP), 2D Toda lattice
and other are provided by {\it tau-functions} $\tau$, which are
special 
functions of infinitely many independent variables (``times'')
parametrizing commuting flows of the hierarchies. 
All other objects related to the integrable equations and
their hierarchies such as Baker-Akhiezer functions, potentials
and other types of dependent variables can be expressed through
the tau-function by acting to it by certain differential or
difference operators. 
In this paper, we consider the BD transformations of common 
solutions 
to the whole hierarchies: KP, modified KP (mKP), BKP
as well as their difference/discrete analogs;
and it is natural to do this on the level of tau-functions.

The existing literature on the subject is enormous, and there
is no chance to cite the most part of it here. The standard sources
are books \cite{book,RS02}, which illuminate the subject
from different points of view, and where many
related references can be found. Among research 
papers, we mention here
\cite{Nimmo97}--\cite{HSS84b} and also
\cite{BK98,BK99,KS02,Schief03,WTS97} for the reason that they 
had a great impact on the contents of the present paper.

This paper is an attempt to systematize the existing knowledge
on the subject on the basis of the bilinear approach to the 
integrable hierarchies developed by Hirota and 
researches of the Kyoto school (see \cite{Hirota-book},
\cite{DJKM83,JM83,DJKM82} and also the book
\cite{HarnadBalogh}), where the main hero is the 
tau-function $\tau ({\bf t})$, with ${\bf t}=
\{t_1, t_2, t_3, \ldots \}$ being an infinite set of
independent variables (``times'') which parametrize
different commuting flows. 
The tau-function can be regarded as a
universal dependent variable of the hierarchies. 
To much extent, the present paper 
should be regarded
as a review; however, we do not exclude that some formulas 
and results, or their interpretation given here, 
were never mentioned in the literature in this form, 
so some of them are probably new.

The main motivation to write this paper was the need to 
achieve a better and a more detailed understanding of some
elements of the theory which are related to BD transformations;
such a requirement arose after writing the paper \cite{Z25},
where the BD transformations related to the classical mKP hierarchy 
appeared naturally in applications to spectral problems for
integrable quantum spin chains. These quantum models are
usually solved by means of one or another version of 
the celebrated Bethe ansatz, see, e.g.
\cite{Faddeev,Gaudin,book1, Slavnov}.
The above mentioned applications of the BD transformations
go back to papers \cite{KLWZ97}--\cite{TZZ15}, where it was 
realized that the nested Bethe ansatz method has a purely
classical reincarnation as a chain of iterated BD transformations
of trigonometric or rational solutions to 
the modified KP (for generalized spin chains with rational or
trigonometric quantum $R$-matrices) 
or KP (for models of the Gaudin type) hierarchy that 
allow one to ``undress'' the initial problem to a trivial one,
hence obtaining a solution of the former. 
The nested Bethe equations 
themselves appear in this approach as equations of motion
for poles of trigonometric or rational solutions to the 
hierarchies in discrete time. Eigenvalues of a special 
generating function of commuting operators (integrals of
motion) in the quantum model appear as tau-functions 
of the classical hierarchies, with different eigenvalues in
the spectrum corresponding to different solutions.
We hope that the approach developed in this paper
might help to clarify some yet uncovered secrets 
of Bethe ansatz, including a hypothetical extension of
the approach initiated in \cite{AKLTZ13} to
quantum models with elliptic $R$-matrices, which remains 
problematic so far.

The theory of BD transformations is closely related to 
the problem of integrable discretization of nonlinear
partial differential equations such as KP, modified KP (mKP),
BKP. A powerful general method for such discretization was
suggested in \cite{DJM82} on the level of tau-functions.
The key ingredient of this method is Miwa's change
of variables from the continuous times 
${\bf t}$ to a set of discrete variables
$s_i\in \ZZ$ first suggested in \cite{Miwa82}:
\beq\label{miwa}
t_k =\frac{1}{k}\sum_i s_i  a_i^{-k}, \quad s_i\in \ZZ ,
\eeq
where $a_i \in \CC$ are parameters playng the role of inverse
lattice spacings for the lattice variables $s_i$.

Remarkably, the discrete variable $n\in \ZZ$ 
that numbers different
steps of a chain of iterated BD transformations enters 
the equations connecting neighboring members of the chain
in the same way as the discrete variables that are introduced
via Miwa's substitution (\ref{miwa}). 
Moreover, if we consider two or more
such chains (each having its own discrete variable), then
it is possible to write bilinear equations for the tau-function
as function of these variables (and other variables of the 
hierarchy under consideration), and, again, the discrete
variables corresponding to the chains enter the equations 
in a very similar way as those coming from the Miwa's 
substitution. In particular, the equations obtained in 
this way that contain
only shifts of the discrete variables numbering levels
of the BD chains have precisely the same form as 
equations of the multi-component KP hierarchy 
introduced in \cite{DJKM81} and further studied in 
\cite{KL93}--\cite{KZ24}.

As it was discovered by the Kyoto school \cite{DJKM83,JM83},
the integrable hierarchies have a natural realization as
quantum field theories of chiral 2D free fermions. In this
approach, the tau-functions enter the game as vacuum expectation 
values of certain operators built from the Fermi-operators.
The BD transformations have their natural interpretation
in this language, too. Moreover, in this setup they become
especially visible: such a transformation is basically an
insertion (between the left and right vacua in the 
expectation value) of an operator which is a 
linear combination (or even an integral with some density
function) of the original
Fermi operator fields. In this paper, we have devoted a special 
section (section 4) to re-interpretation of the content
of the previous sections 2 and 3 within the fermionic
approach.

Section 2 is devoted to the KP hierarchy and its close 
relatives like modified KP (mKP) and Schwarzian KP (SKP), and
to the BD transformations for them.
The difference between these equations consists in choosing
the dependent variables. Our starting point is the 
celebrated generating bilinear integral 
equation for the KP tau-function
(equation (\ref{g1}) below). 

For the mKP equation and its discrete version 
(section \ref{section:mKP}), the dependent variable is the
wave function $\phi ({\bf t})$ (a common 
solution to the auxiliary linear problems for the KP hierarchy). 
In general, it can be represented as an integral, over the
complex plane of the spectral parameter $z\in \CC$, of the
product of the Baker-Akhiezer (BA) function $\psi ({\bf t}; z)$ 
and some ``density function'' $\rho (z)$ having a compact support:
\beq\label{int1}
\phi ({\bf t})=\int_{\CCC}\psi ({\bf t}; z)\rho (z)d^2 z.
\eeq
In the most interesting applications the density function is actually 
a distribution with the support concentrated at 1D lines or
isolated points.
Being rewritten in terms of the tau-functions, this equation 
simultaneously defines the ``forward'' BD transformation
${\bf B}: \tau ({\bf t})\to \hat \tau ({\bf t})$;
to define the ``backward'' one, 
${\bf B}^*: \tau ({\bf t})\to \check \tau ({\bf t})$,
we should integrate 
the dual BA function $\psi^*({\bf t}, z)$ instead of 
$\psi ({\bf t}; z)$ in the similar way.
The BD transformations send any solution to the hierarchy
to another solution and can be iterated.
Iterations of such transformations (the BD chains) are considered 
in section \ref{section:BacklundKP}. 
The forward transformations are unified in a half-infinite
chain (from $n=0$ to $\infty$). Combining them with the
backward ones, one can extend this chain to the inverse
direction (from $-\infty$ to $0$) in such a way that any
neighboring tau-functions of the whole infinite chain
are connected by the discrete mKP equation. 

For the SKP equation and its discrete version 
(section \ref{section:SKP}), 
the dependent variable $\Phi ({\bf t})$ 
is defined as a double integral of the so-called Baker-Akhiezer (BA)
kernel $\Psi ({\bf t}; z, w)$ depending on two spectral parameters
$z,w\in \CC$ with two density functions $\rho , \rho^*$:
\beq\label{int2}
\Phi ({\bf t})=\int_{\CCC}\! \int_{\CCC}\rho (z)
\Psi ({\bf t}; z, w)\rho^* (w) \, d^2 z d^2 w.
\eeq
The fully discretized version of the SKP equation for the
$\Phi ({\bf t})$ has an especially simple and nice form
(equation ({\ref{s5}) in section \ref{section:SKP}). 
It has a remarkable geometrical
meaning clarified in the works \cite{KS02,Schief03}. 
Being rewritten in terms of tau-functions, equation (\ref{int2}) 
simultaneously defines the BD transformation that is sometimes
called the ``binary'' BD transformation \cite{OS93,O93}. In fact it can be 
represented as a composition of a forward and a backward ones.
The connection between BA kernels and BD transformations was
discussed in \cite{AvM92,LW97,ANP98}.

Applying the forward BD transformations $n$ times, starting
from an initial KP tau-function $\tau (0, {\bf t})$, it is
possible to express the tau-function $\tau (n, {\bf t})$ obtained
at the $n$th step in the form of an $n$-fold integral over
$\CC^n$ (\ref{b4}) which,
however, can be represented as determinant of an $n\times n$
matrix whose matrix elements are single integrals.
The most interesting case for the known applications is
the one when we start from the trivial solution
$\tau (0, {\bf t})=1$, and then ``dress'' it by the 
sequential BD transformations. In this case, the series
in inverse powers of the spectral parameter $z$ for the 
BA function is truncated at the $n$th term:
\beq\label{int3}
\psi ({\bf t}; z)=e^{\xi ({\bf t}, z)}\Bigl (
1+ \frac{\xi_1({\bf t}}{z}+ \ldots
\, + \frac{\xi_n({\bf t}}{z^n}\Bigr ), \quad 
\xi ({\bf t}, z)=\sum_{k\geq 1}t_k z^k.
\eeq
This truncation turns out to be very important for the 
above mentioned applications
to models of random matrices and to quantum spin chains.

In section \ref{section:BKP}, a similar approach is developed 
applying to the KP equation of type B (BKP) introduced
in \cite{DJKM82}. Again, the starting
point is the bilinear integral equation for the tau-function,
which is in fact equivalent to the 4-term bilinear 
functional relation obtained as one of its corollaries.
This relation allows one
to introduce, in a natural way, the fully discrete
version of the BKP equation and also its modified and Schwarzian
analogs which in the BKP case 
turn out to be almost the same. The BD transformations are
introduced in the same way as in the KP case, i.e., through
the representation of general solutions to the auxiliary
linear problem as 
integrals of the BA function with density functions. However,
in the BKP case there exist only transformations which 
are similar to the ones of the 
``forward'' type, so the chain 
of such transformations is half-infinite, from 
$n=0$ to $\infty$. Nevertheless, this chain can be extended 
symmetrically to the negative integer $n$ from $0$ to $-\infty$
by setting $\tau (-n)=\tau (n)$.

Section 4 is devoted to the operator approach to the BD
transformations. The fermionic technique developed by the 
Kyoto school \cite{JM83} allows one to represent the tau-functions
as vacuum expectation values of certain operators constructed
from free fermions $\boldpsi (z), \boldpsistar (z)$ and
$\boldphi (z)$ (charged for KP and neutral for BKP). 
This realization makes the BD transformations especially visible:
each such transformation is obtained by inserting the operators
of the form 
\beq\label{int10}
\boldPsi = \int_{\CCC}\boldpsi (z)\rho (z) d^2 z, \quad
\boldPsistar = \int_{\CCC}\boldpsistar (z)\rho^* (z) d^2 z
\eeq
for the KP case\footnote{The first paper that deals with the fermionic
approach to BD transformations for the KP hierarchy is \cite{WTLS98}.} 
and 
\beq\label{int11}
\boldPhi = \int_{\CCC}\boldphi (z)\rho (z) d^2 z
\eeq
for the BKP case. 

There are also two appendices. Appendix A contains some 
information about Pfaffians which are extensively used in 
construction of solutions to the BKP hierarchy. In Appendix B
the algebra of neutral fermions is realized as a subalgebra
of the charged ones.

\section{Equations and hierarchies of the KP type}

This section is devoted to the KP hierarchy and its close 
relatives such as mKP and SKP, as well
as to the fully discretized versions of these equations. 
Our approach is based on bilinear equations for the tau-function.
The general references are \cite{DJKM83,JM83,KS02,Schief03}.

\subsection{The KP hierarchy and its discretization}
\label{section:KP}

The generating bilinear equation for the KP tau-function
has the form
\beq\label{g1}
\oint_{C_{\infty}}e^{\xi ({\bf t}-{\bf t}', z)}
\tau ({\bf t}-[z^{-1}])\tau ({\bf t}'+[z^{-1}])\, dz =0,
\eeq
where 
\beq\label{xi}
\xi ({\bf t}, z) =\sum_{k\geq 1} t_k z^{k},
\eeq
\beq\label{shift}
{\bf t}\pm [z^{-1}]=\Bigl \{ t_1\pm z^{-1}, \,
t_2\pm \frac{1}{2}z^{-2}, \, t_3\pm \frac{1}{3}z^{-3}\Bigr \},
\eeq
and $C_{\infty}$ is a big circle around $\infty$.
This equation is valid for all ${\bf t}$ and ${\bf t}'$.

The bilinear equations of the Hirota-Miwa type follow from
(\ref{g1}) if one sets ${\bf t}-{\bf t}'$ in a special way.
For example, let us put ${\bf t}-{\bf t}'=[a^{-1}]+[b^{-1}]+
[c^{-1}] - [d^{-1}]$, where the complex numbers $a,b,c,d$ 
are assumed to belong to some neighborhood of infinity.
Then we have
$$
e^{\xi ({\bf t}-{\bf t}', z)}=\frac{abc \, (d-z)}{d(a-z)
(b-z)(c-z)}
$$
and the integral in (\ref{g1}) can be found by means of the 
residue calculus. There are simple poles at the points $a,b,c$,
and the residue at infinity is equal to 0. A simple calculation
yields a three-term bilinear equation for the tau-function,
which after a shift of the times acquires the form
\beq\label{g2}
\begin{array}{l}
(a-d)(b-c)\tau \Bigl ({\bf t}+[b^{-1}]+[c^{-1}]\Bigr )
\tau \Bigl ({\bf t}+[a^{-1}]+[d^{-1}]\Bigr )
\\ \\
\phantom{aaaaaaaaa}
+(b-d)(c-a)\tau \Bigl ({\bf t}+[a^{-1}]+[c^{-1}]\Bigr )
\tau \Bigl ({\bf t}+[b^{-1}]+[d^{-1}]\Bigr )
\\ \\
\phantom{aaaaaaaaaaaaaaaaaa}
+(c-d)(a-b)\tau \Bigl ({\bf t}+[a^{-1}]+[b^{-1}]\Bigr )
\tau \Bigl ({\bf t}+[c^{-1}]+[d^{-1}]\Bigr )=0.
\end{array}
\eeq
This equation is well-known in the literature. Sometimes 
it is called the Fay identity\footnote{This is because of the fact
that for the algebraic-geometrical (quasiperiodic) 
solutions in terms of Riemann theta
functions it just becomes the Fay identity for the 
theta-functions associated with smooth algebraic curves.}.
Letting $d$ to $\infty$, we obtain a simpler version of this equation:
\beq\label{g3}
\begin{array}{l}
(b-c)\tau \Bigl ({\bf t}+[b^{-1}]+[c^{-1}]\Bigr )
\tau \Bigl ({\bf t}+[a^{-1}]\Bigr )
\\ \\
\phantom{aaaaaaaaa}
+(c-a)\tau \Bigl ({\bf t}+[a^{-1}]+[c^{-1}]\Bigr )
\tau \Bigl ({\bf t}+[b^{-1}]\Bigr )
\\ \\
\phantom{aaaaaaaaaaaaaaaaaa}
+(a-b)\tau \Bigl ({\bf t}+[a^{-1}]+[b^{-1}]\Bigr )
\tau \Bigl ({\bf t}+[c^{-1}]\Bigr )=0.
\end{array}
\eeq
Note that the simplest solution is $\tau ({\bf t})=1$.

These equations have two different interpretations. From one
point of view, they are regarded as generating equations for the 
KP hierarchy. Expanding them in inverse powers of $a,b,c,d$ (as
these parameters tend to infinity), one is able to obtain all
differential bilinear equations of the hierarchy. 
Another 
point of view is regarding (\ref{g2}) or
(\ref{g3}) as an integrable discretization of the KP hierarchy.
To this end, it is convenient to introduce three discrete variables
$s_1, s_2, s_3 \in \ZZ$ 
associated with three parameters $a_1, a_2, a_3$
that are assumed to be distinct. These parameters play
the role of the inverse lattice spacings for
the discrete variables. For any function $f({\bf t})$ we 
introduce the function
\beq\label{g4}
f(s_1, s_2, s_3)=f \Bigl ({\bf t}+s_1[a_1^{-1}]+s_2[a_2^{-1}]
+s_3[a_3^{-1}]\Bigr )
\eeq
of the discrete variables. All the continuous variables
${\bf t}$ are regarded here as constants and serve as parameters. 
The following short-hand notation is convenient:
\beq\label{g5}
\begin{array}{c}
f_1=f(s_1+1, s_2, s_3), \quad 
f_2=f(s_1, s_2+1, s_3), \quad 
f_3=f(s_1, s_2, s_3+1), 
\\ \\
f_{12}=f(s_1\! +\! 1, s_2\! +\! 1, s_3), \quad 
f_{13}=f(s_1\! +\! 1, s_2, s_3\! +\! 1), \quad 
f_{23}=f(s_1, s_2\! +\! 1, s_3\! +\! 1).
\end{array}
\eeq
The similar notation will be used in the cases when there are
more parameters $a_i$ and 
discrete variables introduced in the same way as in (\ref{g4}).
Sometimes, when it is important 
to indicate the dependence on ${\bf t}$ explicitly,
we will also
write $f_1({\bf t})$, $f_2({\bf t})$, etc. 
In this notation, equation (\ref{g3}) acquires the compact form
\beq\label{g6}
a_{12}\tau_{12}\tau_{3} +a_{23}\tau_{23}\tau_{1}
+a_{31}\tau_{31}\tau_{2}=0, \qquad a_{ij}=a_i-a_j
\eeq
and is usually regarded as 
an integrable discretization of the KP equation. 
Equation (\ref{g2}) can be written as
\beq\label{g2a}
a_{14}a_{23}\tau_{14}\tau_{23} +a_{24}a_{31}\tau_{24}\tau_{31}
+a_{34}a_{12}\tau_{34}\tau_{12}=0.
\eeq
Let us prove that equations (\ref{g6}) and (\ref{g2a}) are equivalent.
Indeed, putting $a_4=\infty$ in (\ref{g2a}), we obtain (\ref{g6}).
The inverse statement, that (\ref{g6}) implies (\ref{g2a}),
can be proved by the following argument.
Consider the equations of the form (\ref{g6}) for each triple
of indices from the set $\{1,2,3,4\}$:
$$
\begin{array}{l}
a_{12}\tau_{12}\tau_{3} +a_{23}\tau_{23}\tau_{1}
+a_{31}\tau_{31}\tau_{2}=0, 
\\ \\
a_{12}\tau_{12}\tau_{4} +a_{24}\tau_{24}\tau_{1}
+a_{41}\tau_{41}\tau_{2}=0,
\\ \\
a_{13}\tau_{13}\tau_{4} +a_{34}\tau_{34}\tau_{1}
+a_{41}\tau_{41}\tau_{3}=0,
\\ \\
a_{23}\tau_{23}\tau_{4} +a_{34}\tau_{34}\tau_{2}
+a_{42}\tau_{42}\tau_{3}=0.
\end{array}
$$
This is a homogeneous 
linear system for $\tau_1, \tau_2, \tau_3, \tau_4$, and the 
condition of existence of non-zero solutions is
$$
\left |
\begin{array}{cccc}
0 & a_{34} \tau_{34} & a_{42} \tau_{42} & a_{23} \tau_{23}
\\ &&& \\
a_{43} \tau_{43} &0 & a_{14} \tau_{14} & a_{31} \tau_{31}
\\ &&& \\
a_{24} \tau_{24} & a_{41} \tau_{41} & 0  & a_{12} \tau_{12}
\\ &&& \\
a_{32} \tau_{32} & a_{13} \tau_{13} &  a_{1} \tau_{21} & 0 
\end{array} 
\right | =0.
$$
Note that this $4\times 4$ matrix is skew-symmetric. As is well known,
the determinant of a 
skew-symmetric matrix of an even 
size is square of its Pfaffian, hence we have:
$$
\Bigl (a_{14}a_{23}\tau_{14}\tau_{23} +a_{24}a_{31}\tau_{24}\tau_{31}
+a_{34}a_{12}\tau_{34}\tau_{12}\Bigr )^2=0.
$$ 
This is the equation (\ref{g2a}). 

At this point, two remarks are in order.

\noindent
{\bf Remark 1.} 
First, the values of the coefficients in (\ref{g6}) (or (\ref{g2a})
are actually not that very important because the simple transformation
$\tau \to \tau '$, where
$$
\tau ' (s_1, s_2, s_3)=\tau  (s_1, s_2, s_3)\prod_{i<j}^3
\Bigl (a_{ij}/b_{ij}\Bigr )^{s_is_j}
$$
allows one to change the coefficients $a_{ij}$ in (\ref{g6})
to any other $b_{ij}$. (As we shall see later, of particular 
importance is the
case when the coefficients are equal to $\pm 1$.) However, as far as
analytic properties of the solutions as functions of 
the $s_i$'s are concerned, the particular 
values of the coefficients may be important. 

\noindent
{\bf Remark 2.} In fact an even simpler version of equation 
(\ref{g6}) is equivalent to (\ref{g2a}). Namely, if we tend
$a_3\to \infty$ in (\ref{g6}), then we obtain:
\beq\label{t1}
a_{12}\, \frac{\tau_{12}\tau}{\tau_1 \tau_2}=a_{12}
+\p_{t_1}\log \frac{\tau_2}{\tau_1}.
\eeq
It is easy to see that this equation is equivalent to 
(\ref{g2a}). Indeed, write it for the pairs of indices 
$(12), (13), (23)$ and sum the equations obtained in this way.
As a result, we obtain (\ref{g2a}).

The higher 
equations of the discrete KP hierarchy can be also considered.
To obtain them, for any integer $m\geq 2$ 
fix $m+1$ distinct points $a_i$, $m-1$ 
distinct points $b_i$ and set
$$
{\bf t}-{\bf t}'=\sum_{i=1}^{m+1}[a_i^{-1}]-
\sum_{i=1}^{m-1}[b_i^{-1}], 
\quad 
\mbox{then} \quad
e^{\xi ({\bf t}-{\bf t}', z)}=C\,
\frac{\prod\limits_{i=1}^{m-1}(b_i-z)}{\prod\limits_{i=1}^{m+1}(a_i-z)},
$$
where $C$ is an irrelevant constant. The residue calculus converts
(\ref{g1}) to the following equation:
\beq\label{g2b}
\begin{array}{l}
\displaystyle{
\sum_{k=1}^{m+1}(-1)^{k-1}
\Delta_m (a_1, \ldots , \hat a_k, \ldots , a_{m+1})
\prod_{i=1}^{m-1}(b_i-a_k)}
\\ \\
\phantom{aaaaaaaaaaaaaaaaaa}
\displaystyle{\times \,
\tau \Bigl ({\bf t}+\sum_{j=1, \neq k}^{m+1}[a_j^{-1}]\Bigr )
\tau \Bigl ({\bf t}+[a^{-1}_k]+ 
\sum_{j=1}^{m-1}[b_j^{-1}]\Bigr )=0}
\end{array}
\eeq
(equation ({\ref{g2}) is its particular case for $m=2$).
In (\ref{g2b}),
\beq\label{vandermonde}
\Delta_m (a_1, \ldots , a_m)=\det_{m\times m}(a_i^{j-1})=
\prod_{i>j}^m (a_i-a_j)
\eeq
is the Vandermonde determinant, and hat above $a_k$ means that 
$a_k$ is omitted. Putting $b_i=\infty$ for all $i=1, \ldots , m-1$,
we obtain from (\ref{g2b}) the higher equations of the discrete
KP hierarchy which first appeared in \cite{OHTI93}.
In our short-hand notation
they can be written in the form
\beq\label{g7}
\left |
\begin{array}{cccccc}
1 & a_1 & a_1^2 & \ldots & a_1^{n-2} & \tau_1 \tau_{\hat 1}
\\ 
1 & a_2 & a_2^2 & \ldots & a_2^{n-2} & \tau_2 \tau_{\hat 2}
\\ 
\vdots & \vdots & \vdots & \ddots & \vdots & \vdots 
\\ 
1 & a_n & a_n^2 & \ldots & a_n^{n-2} & \tau_n \tau_{\hat n}
\end{array}
\right |=0,
\eeq
where $\tau_{\hat k}=\tau_{1\ldots \hat k \ldots n}$ (the index
$k$ is omitted). 
One can show that these higher equations
for any $n>3$ are algebraic corollaries of the equations
of the form
(\ref{g6}) obtained from (\ref{g7}) at $n=3$ assuming that 
they are valid for
all triples of the indices from 1 to $n$. This means that 
equation (\ref{g6}) is already enough for introducing the whole
hierarchy. 

The fact that equation (\ref{g2a}), or, equation 
(\ref{g6}),
is equivalent to the whole hierarchy (i.e., to the generating equation
(\ref{g1})), was proved in \cite{TT95,Shigyo13}. The idea of the
direct proof given in \cite{Shigyo13} is as follows. 
The first step is to prove that the 
initial integral equation (\ref{g1}) is equivalent to the 
general relation (\ref{g2b}).
This can be done using properties of symmetric functions. Next,
one can show that the particular case (\ref{g2a}) of (\ref{g2b})
implies a determinant representation for 
$\displaystyle{
\tau \Bigl ({\bf t}+\sum_{i=1}^m ([a_i^{-1}]-[b_i^{-1}])\Bigr )}$
in terms of $\tau \Bigl ({\bf t}+[a_i^{-1}]-[b_j^{-1}]\Bigr )$
(see (\ref{s1a})).
At last, certain determinant identities for these determinants
(the Pl\"ucker relations) imply (\ref{g2b}).

As usual for non-linear integrable equations, equation (\ref{g6})
can be obtained as compatibility condition for an overdetermined
system of linear problems for a ``wave function'' 
$\phi =\phi (s_1, s_2, s_3)$. These problems
can be written in the form
\beq\label{lin1}
a_{\alpha}\phi_{\beta}-a_{\beta}\phi_{\alpha}=(a_{\alpha}-a_{\beta})
\frac{\tau \tau_{\alpha \beta}}{\tau_{\alpha}
\tau_{\beta}}\phi_{\alpha \beta},
\qquad \alpha , \beta =1,2,3.
\eeq
Indeed, consider, for  example, the equations for $(\alpha , \beta )=
(1,3)$ and $(\alpha , \beta )=(2,3)$
written in the form
$$
\begin{array}{l}
\displaystyle{
a_1\phi_{\bar 1}=
a_3\phi_{\bar 3}+ (a_1-a_3)\frac{\tau 
\tau_{\bar 1 \bar 3}}{\tau_{\bar 1} \tau_{\bar 3}}}\, \phi ,
\\ \\
\displaystyle{
a_2\phi_{\bar 2}=a_3\phi_{\bar 3} +(a_2-a_3)
\frac{\tau \tau_{\bar 2 \bar 3}}{\tau_{\bar 2} \tau_{\bar 3}}}\, \phi ,
\end{array}
$$
where
$
\phi_{\bar 1}=\phi (s_1-1, s_2, s_3)$, 
$\phi_{\bar 2}=\phi (s_1, s_2-1, s_3)$ etc. (comparing to (\ref{lin1}),
we have shifted the arguments as $s_i\to s_i-1$).
These equations allow one to represent the function 
$\phi_{\bar 1 \bar 2}$ as a linear combination of
$\phi$, $\phi_{\bar 3}$ and $\phi_{\bar 3 \bar 3}$ in two different ways.
Compatibility of the linear problems means that the results must coincide.  
Equating to each other the expressions obtained in the 
two ways, we see that
the terms proportional to 
$\phi$ and $\phi_{\bar 3 \bar 3}$ cancel identically while the terms
proportional to $\phi_{\bar 3}$ give a non-trivial relation
(provided that $\phi_{\bar 3}$ is not identically zero):
$$
(a_1-a_3)\frac{\tau_{\bar 1\bar 3 \bar3}
\tau_{\bar 3}}{\tau_{\bar 1\bar 3}\tau_{\bar 3\bar 3}} +
(a_2-a_3)\frac{\tau_{\bar 1\bar 2\bar 3}
\tau_{\bar 1}}{\tau_{\bar 1\bar 2}\tau_{\bar 1\bar 3}}=
(a_1-a_3)\frac{\tau_{\bar 1\bar 2\bar 3}
\tau_{\bar 2}}{\tau_{\bar 1\bar 2}\tau_{\bar 2\bar 3}}+
(a_2-a_3)\frac{\tau_{\bar 2\bar 3\bar 3}
\tau_{3}}{\tau_{\bar 2\bar 3}\tau_{\bar 3\bar 3}},
$$
which means that
$$
\frac{(a_3-a_2)\tau_{\bar 1}\tau_{\bar 2 \bar 3}+(a_1-a_3)
\tau_{\bar 2}\tau_{\bar 1 \bar 3}}{\tau_{\bar 1 \bar 2}\tau_{\bar 3}}
$$
is a periodic function of
$s_3$ with period $1$ and an arbitrary function of 
$s_1, s_2$.
Since we do not assume any special
periodicity properties of the 
solutions, we should conclude that this function
does not depend on $s_3$. Therefore, shifting the 
arguments as $s_i \to s_i+1$, we come to the relation
$$
(a_2 -a_3)\tau_1 \tau_{23} 
+(a_3-a_1)\tau_2\tau_{13}=
h(s_1, s_2) \tau_3\tau_{12},
$$
where $h$ may be an arbitrary function of $s_1, s_2$.
The compatibility with the third linear problem implies that
$h$ must be a constant equal to 
$a_2-a_1$, hence equation (\ref{g6}) 
follows. 

A comment is in order.
Generally speaking, compatibility of linear problems
follows from existence of a continuous family of 
common solutions. In our case
the coefficient functions in the difference operators
are such that the compatibility is equivalent to existence 
of {\it at least one} 
non-trivial solution.

The continuum limit of equation (\ref{g6}) is most easily
taken in three steps. At the first step we tend $a_1 \to \infty$
keeping the other two parameters fixed. This gives:
\beq\label{con1}
\p_{t_1}\log \frac{\tau_2}{\tau_3} =
(a_2-a_3)\Bigl (1-\frac{\tau \tau_{23}}{\tau_2 \tau_3}\Bigr ).
\eeq
Next we tend $a_2\to \infty$ in (\ref{con1}) 
keeping $a_3$ fixed. The first non-vanishing order yields:
\beq\label{con2}
\Bigl (\p_{t_2}-2a_3 \p_{t_1}\Bigr )\log \frac{\tau}{\tau_3}=
\p_{t_1}^2 \log (\tau \tau_3)+\Bigl (\p_{t_1}\log 
\frac{\tau}{\tau_3}\Bigr )^2.
\eeq
At last, we tend $a_3\to \infty$. 
The first non-vanishing order yields, after a relatively long
calculation:
\beq\label{con3}
4\p_{t_1}\p_{t_3}\log \tau =3\p_{t_2}^2 \log \tau +
6(\p_{t_1}^2 \log \tau )^2 +\p_{t_1}^4 \log \tau .
\eeq
This equation is bilinear in the tau-function:
\beq\label{con4}
4\tau_{xt}\tau -4\tau_x \tau_t=
3\tau_{yy}\tau -3\tau_y^2 +3\tau_{xx}^2 +
\tau_{xxxx}\tau -4\tau_{xxx}\tau_x,
\eeq
where we have denoted
$x=t_1$, $y=t_2$, $t=t_3$. In terms of the function
$
u=\p_x^2 \log \tau
$
the equation (more precisely, its second $x$-derivative) becomes
the standard KP equation:
\beq\label{con5}
3u_{yy}=\Bigl ( 4u_t -12uu_x -u_{xxx}\Bigr )_x.
\eeq

An important role in the theory is played by the 
functions
\beq\label{mkp1}
\begin{array}{l}
\displaystyle{
\psi ({\bf t}; z)=e^{\xi ({\bf t}, z)}\,
\frac{\tau ({\bf t}-[z^{-1}])}{\tau ({\bf t})},}
\\ \\
\displaystyle{
\psi^* ({\bf t}; z)=e^{-\xi ({\bf t}, z)}\,
\frac{\tau ({\bf t}+[z^{-1}])}{\tau ({\bf t})},}
\end{array}
\eeq
where $z\in \CC$ is a spectral parameter.
We call $\psi$ (respectively, $\psi^*$)
the Baker-Akhiezer (BA) function\footnote{This terminology 
comes from the theory of Riemann surfaces and 
is standard in the algebro-geometric approach to the construction
of periodic solutions; in this case $\psi$ becomes a special
function on an algebraic Riemann surface.}
(respectively,
the adjoint Baker-Akhiezer function). For brevity, 
we will call both $\psi$ and $\psi^*$ the BA functions.
In general, these functions, as functions of $z$,
are assumed to be holomorphic in a sufficiently small
neighborhood ${\sf U}_{\infty}$ of $\infty$ 
(i.e., for $|z|>R$ with some sufficiently
large $R$) except maybe the very point $z=\infty$, 
where an essential singularity is allowed.
In this neighborhood, they are given by equations (\ref{mkp1}).
It is known that the BA functions satisfy
an infinite system of linear differential equations with 
$z$-independent coefficients (the compatibility of these equations
is just equivalent to the equations of the KP hierarchy).
For example, the simplest nontrivial equations are
\beq\label{mkp2}
\begin{array}{l}
\phantom{-}\p_{t_2}\psi = \p_{t_1}^2 \psi +2 u \psi ,
\\ \\
-\p_{t_2}\psi^* = \p_{t_1}^2 \psi^* +2 u \psi^*, 
\qquad
u=\p_{t_1}^2 \log \tau .
\end{array}
\eeq

It is worth noting that for some important classes
of solutions the BA functions
can be analytically continued outside ${\sf U}_{\infty}$
as analytic functions on $\CC \setminus {\sf S}$, where
${\sf S}\subset \CC$ is a set of points of nonzero 
codimension (which is $2$ for a number of isolated points or
$1$ for lines). The well known examples are soliton 
and rational solutions. To avoid unnecessary analytical 
complications in this work, we will have in mind 
the classes of solutions for which such analytic
continuation is possible.

For completeness,
let us say some words about the general case, when
the BA functions can not be analytically continued from
${\sf U}_{\infty}$ to bigger
domains of the complex plane. In this case 
$\p_{\bar z}\psi ({\bf t}, z)$ and $\p_{\bar z}\psi^* ({\bf t}, z)$
may be nonzero in some compact 2D domain ${\sf D}\subset \CC$,
and the BA functions (which are now functions of both $z$ and 
$\bar z$ for $z\in {\sf D})$ are characterized by certain integral
or integro-diffrential equations which were suggested in
\cite{ZM85} and are known in the literature as non-local
$\bar \p$-problem. They are:
\beq\label{dbar}
\begin{array}{l}
\displaystyle{
\p_{\bar z}\psi ({\bf t}, z)=\int_{\CCC}
K(z, \zeta )\psi ({\bf t}, \zeta )\, d^2 \zeta,
}
\\ \\
\displaystyle{
\p_{\bar z}\psi^* ({\bf t}, z)=-\int_{\CCC}
\psi^* ({\bf t}, \zeta )
K(\zeta , z)\, d^2 \zeta.
}
\end{array}
\eeq
The kernel $K(z, \zeta )$ is assumed to have a compact support
with respect to the both arguments, i.e., $K(z, \zeta )=0$ if
$z$ or $\zeta$ belong to ${\sf U}_{\infty}$. This condition allows
one to require the holomorphic in $z$ asymptotics as $z\to \infty$.
Under certain conditions imposed on the
kernel these integro-differential equations are known to have 
nontrivial solutions, and the linear space of solutions is in general
one-dimensional.

The BA functions are particular and very special 
solutions to this linear system. More general solutions,
which we call wave functions and denote as
$\phi ({\bf t})$, $\phi^*({\bf t})$, 
can be obtained by integrating 
the BA functions with respect to the
spectral parameter $z$ with arbitrary functions or distributions
$\rho (z)$, $\rho^*(z)$ of $z, \bar z$ (we call them density functions):
\beq\label{mkp2a}
\begin{array}{l}
\displaystyle{
\phi ({\bf t})=\int_{\CCC}\psi ({\bf t}; z)\rho (z)d^2 z,}
\\ \\
\displaystyle{
\phi^* ({\bf t})=\int_{\CCC}\psi^* ({\bf t}; z)\rho^* (z)d^2 z,}
\end{array}
\eeq
where $d^2z \equiv dx dy$ is the standard measure 
in the complex plane.
(We assume that the density functions $\rho (z)$,
$\rho^*(z)$ are such that the integrals converge; for example,
we may require that they have a compact support.)

More precisely, we have in mind the following possibilities
for the density functions:
\begin{itemize}
\item[--]
Its support is a compact domain ${\sf D}\in \CC$:
$$
\rho (z)=\mu (z)\Theta_{\sf D}(z),
$$
where $\mu (z)$ is an integrable bounded function of $z, \bar z$
and $\Theta_{\sf D}(z)$ is the characteristic function
of the domain ${\sf D}$: 
$\Theta_{\sf D}(z)=1$ if $z\in {\sf D}$ and $0$ otherwise.
In this case we encounter the non-local 
$\bar \p$-problem, as is explained above. 
We do not consider this case in this
paper and hope to address it elsewhere.
\item[--]
The support of $\rho(z)$ is a contour $\Gamma$ 
in $\CC$, and $\rho (z)$ is a distribution represented 
in terms of the delta-function $\delta_{\Gamma}(z)$
with the support on $\Gamma$
as follows\footnote{The delta-function $\delta_{\Gamma}(z)$
is defined by the rule  
$\displaystyle{
\int_{\CCC}f(z)\delta_{\Gamma}(z)d^2 z=\int_{\Gamma}f(z)|dz|}
$
for any integrable function $f$.}:
$$
\rho (z)=\nu (z)\delta_{\Gamma}(z),
$$
where $\nu (z)$ is some integrable function.
\item[--]
The support of $\rho(z)$ is a finite number of points $p_i\in \CC$,
and $\rho (z)$ is a distribution represented as a linear combination
of 2D delta-functions:
$$
\rho (z)=\sum_i c_i \delta^{(2)}(z-p_i).
$$
\end{itemize}
Various combinations of these possibilities are also admissible.
The same is assumed for $\rho^* (z)$.

A few words about reductions are in order. As is known, the KP
hierarchy admits a lot of interesting stationary reductions. 
In particular, the $N$-Gelfand-Dikii (or $N$-KdV) hierarchy
is obtained if one requires that the tau-function does not depend 
on the times $t_{kN}$, $k\geq 2$. (The case $N=2$ is the 
celebrated KdV hierarchy.) For the reduced hierarchy, the 
choice of possible density functions becomes very restrictive.
For example, for the $N$-KdV hierarchy the admissible 
density functions are distributions of the form
\beq\label{omega}
\rho (z)=\sum_{j=0}^{N-1} \alpha_j \delta^{(2)}(z-\omega^j p),
\quad \omega =e^{2\pi i/N},
\eeq
with arbitrary $p\in \CC$. They correspond to adding a soliton
to the initial solution.

\subsection{The mKP hierarchy and its discretization}
\label{section:mKP}

The tau-function of the mKP hierarchy, apart from the times 
${\bf t}$, depends on a discrete variable $n\in \ZZ$:
$\tau (n, {\bf t})$. The generating bilinear equation has the form
\beq\label{m1}
\oint_{C_{\infty}}z^k e^{\xi ({\bf t}-{\bf t}', z)}
\tau (n+k, {\bf t}-[z^{-1}])
\tau (n, {\bf t}'+[z^{-1}])dz =0, \quad k\geq 0.
\eeq
In particular, setting $k=0$ we see that $\tau (n, {\bf t})$ 
as a function of ${\bf t}$ is a tau-function of
the KP hierarchy for any fixed $n$.
Note that the mKP tau-function is defined up to a multiplier
$c_n$ that depend only on $n$: equation (\ref{m1})
is invariant under the transformations $\tau (n, {\bf t})
\to c_n \tau (n, {\bf t})$.
 
Setting $\displaystyle{
{\bf t}-{\bf t}'=\sum_{\alpha =1}^{k+2}[a_{\alpha}^{-1}]}$, we
obtain from (\ref{m1}) by residue calculus:
\beq\label{m2}
\sum_{\alpha =1}^{k+2}(-1)^{\alpha}\Delta_{k+1}(a_1, \ldots ,
\hat a_{\alpha}, \ldots , a_{k+2})
\tau \Bigl (n+k, {\bf t}+\sum_{\beta \neq \alpha}^{k+2}
[a_{\beta}^{-1}]\Bigr )\tau \Bigl (n, {\bf t}+[a_{\alpha}^{-1}]\Bigr )=0,
\eeq
where $\Delta_m$ is the Vandermonde determinant 
(\ref{vandermonde})
and the hat means that $a_{\alpha}$ 
is omitted. In particular, at $k=1$ (\ref{m2}) gives:
\beq\label{m3}
\begin{array}{l}
a_1(a_3-a_2) \tau \Bigl (n+1, {\bf t}+[a_2^{-1}]+[a_3^{-1}]\Bigr )
\tau \Bigl (n, {\bf t}+[a_1^{-1}]\Bigr )
\\ \\
\phantom{aaaaaaaa}
+a_2(a_1-a_3)\tau \Bigl (n+1, {\bf t}+[a_1^{-1}]+[a_3^{-1}]\Bigr )
\tau \Bigl (n, {\bf t}+[a_2^{-1}]\Bigr )
\\ \\
\phantom{aaaaaaaaaaaaaaaaaaa}
+a_3(a_2-a_1)\tau \Bigl (n+1, {\bf t}+[a_1^{-1}]+[a_2^{-1}]\Bigr )
\tau \Bigl (n, {\bf t}+[a_3^{-1}]\Bigr )=0.
\end{array}
\eeq
Letting $a_3\to \infty$, we obtain a simpler equation:
\beq\label{m4}
\begin{array}{l}
a_1\tau \Bigl (n+1, {\bf t}+[a_2^{-1}]\Bigr )
\tau \Bigl (n, {\bf t}+[a_1^{-1}]\Bigr )
\\ \\
\phantom{aaaaaaaa}
-a_2\tau \Bigl (n+1, {\bf t}+[a_1^{-1}]\Bigr )
\tau \Bigl (n, {\bf t}+[a_2^{-1}]\Bigr )
\\ \\
\phantom{aaaaaaaaaaaaaaaaaaa}
=(a_1-a_2)\tau \Bigl (n+1, {\bf t}+[a_1^{-1}]+[a_2^{-1}]\Bigr )
\tau \Bigl (n, {\bf t}\Bigr ),
\end{array}
\eeq
which in the short-hand notation introduced in the previous
subsection acquires the form
\beq\label{m5}
a_1 \tau_2(n+1)\tau_1(n)-a_2\tau_1(n+1)\tau_2(n)=(a_1-a_2)
\tau_{12}(n+1)\tau (n).
\eeq
Let us show that equations (\ref{m3}) and (\ref{m4}) are in fact
equivalent, i.e. that (\ref{m4}) (or (\ref{m5})) implies (\ref{m3}).
Indeed, let $(\alpha \beta \gamma )$ be any 
cyclic permutation of $(123)$, then writing equations (\ref{m5}) 
for $\alpha \beta$, multiplying both sides by 
$a_{\gamma}\tau_{\gamma}(n)$ and summing the results 
over the three cyclic permutations, we obtain 
$$
\sum_{(\alpha \beta \gamma )}a_{\gamma}(a_{\alpha}-a_{\beta})
\tau_{\alpha \beta}(n+1)\tau_{\gamma}(n)=0
$$
which is (\ref{m3}). We also note that it was proven 
in \cite{Shigyo13}) that 
the three-term equation (\ref{m3}) (and, therefore, (\ref{m4}))
is equivalent to the whole hierarchy.

The transition from KP to the modified KP essentially
consists in passing from the tau-function to the BA functions
(\ref{mkp1}),
or, more generally, to the wave functions
as dependent variables of the hierarchy. Technically this can
be done in the following way. From the definition (\ref{mkp1})
together with the 3-term bilinear equation (\ref{g3}) one can 
derive, after some simple algebra, the following relations:
\beq\label{mkp3}
\begin{array}{l}
\displaystyle{
a\psi ({\bf t} -[a^{-1}];z)-b\psi ({\bf t}-[b^{-1}];z)=
(a-b)\frac{\tau ({\bf t}-[a^{-1}]-
[b^{-1}]) \, \tau ({\bf t})}{\tau ({\bf t}-[a^{-1}])
\tau ({\bf t}-[b^{-1}])}\, \psi ({\bf t}; z)},
\\ \\
\displaystyle{
a\psi^* ({\bf t}+[a^{-1}];z)-b\psi^* ({\bf t}+[b^{-1}];z)=
(a-b)\frac{\tau ({\bf t}+[a^{-1}]+
[b^{-1}]) \, \tau ({\bf t})}{\tau ({\bf t}+[a^{-1}])
\tau ({\bf t}+[b^{-1}])}\, \psi^* ({\bf t}; z)}.
\end{array}
\eeq
In our short-hand notation they can be written in the form
\beq\label{mkp4}
\begin{array}{l}
\displaystyle{
a_1 \psi_2 -a_2 \psi_1 =a_{12}\, \frac{\tau \tau_{12}}{\tau_1 \tau_2}
\, \psi_{12}},
\\ \\
\displaystyle{
a_1 \psi^*_{1} -a_2 \psi^*_{2} =
a_{12}\, \frac{\tau_{12} \tau_{}}{\tau_{1} \tau_{2}}
\, \psi^*_{}},
\end{array}
\eeq
where $a_{12}=a_1-a_2$.
Integrating the both sides of relations (\ref{mkp4}) with
some functions $\rho (z)$,
$\rho^*(z)$, we obtain the same relations for the wave functions:
 \beq\label{mkp4a}
\begin{array}{l}
\displaystyle{
a_1 \phi_2 -a_2 \phi_1 =a_{12}\, \frac{\tau \tau_{12}}{\tau_1 \tau_2}
\, \phi_{12}},
\\ \\
\displaystyle{
a_1 \phi^*_{1} -a_2 \phi^*_{2} =
a_{12}\, \frac{\tau_{12} \tau_{}}{\tau_{1} \tau_{2}}
\, \phi^*_{}}.
\end{array}
\eeq
Note that the first equation coincides with the linear problem
(\ref{lin1}) which was discussed above in Section \ref{section:KP}.
Now we see that the linear problem is solved in terms of the 
BA function (\ref{mkp1}).

From (\ref{mkp4a}) and (\ref{g6}) it immediately follows that
the following equations hold:
\beq\label{mkp5}
\frac{a_1 \phi_2 -a_2 \phi_1}{\phi_{12}}+
\frac{a_2 \phi_3 -a_3 \phi_2}{\phi_{23}}+
\frac{a_3 \phi_1 -a_1 \phi_3}{\phi_{13}}=0,
\eeq
\beq\label{mkp5a}
\frac{a_2 \phi^{*}_{12} -a_3 \phi^{*}_{13}}{\phi^{*}_{1}}+
\frac{a_3 \phi^{*}_{23} -a_1 \phi^{*}_{12}}{\phi^{*}_{2}}+
\frac{a_1 \phi^{*}_{13} -a_2 \phi^{*}_{23}}{\phi^{*}_{3}}=0.
\eeq
They are the discrete modified KP equations. 
Note that $\phi^{-1}({\bf t})$ satisfies the same 
equation (\ref{mkp5a}) as $\phi^*({\bf t})$. 
Moreover, the two equations exchange the forms under the 
change of variables $s_i \to -s_i$.
Therefore, in what follows
we can deal with equation (\ref{mkp5}) only.
It is easy to see that the equivalent forms of equations
(\ref{mkp5}), (\ref{mkp5a}) are
\beq\label{mkp5b}
\left | 
\begin{array}{ccc}
1 & a_1 \phi_{23} & \phi_1 \phi_{23}
\\ && \\
1 & a_2 \phi_{31} & \phi_2 \phi_{31}
\\ && \\
1 & a_3 \phi_{12} & \phi_3 \phi_{12}
\end{array}
\right | =0,
\qquad
\left | 
\begin{array}{ccc}
1 & a_1 \phi_{1}^{*} & \phi^*_1 \phi^*_{23}
\\ && \\
1 & a_2 \phi_{2}^* & \phi^*_2 \phi^*_{31}
\\ && \\
1 & a_3 \phi_{3}^* & \phi^*_3 \phi^*_{12}
\end{array}
\right | =0.
\eeq
The simple change of the dependent variable $\phi \to \phi '$, where
$$
\phi '(s_1, s_2 , s_3)=\phi (s_1, s_2 , s_3)
\prod_{i=1}^3 a_{i}^{-s_i}
$$
(and similarly for 
$\phi^{*}$) makes the coefficients in these equations equal to 1.

As before, the continuum limit of equation (\ref{mkp5a}) is most easily
taken in steps. At the first step we tend $a_1 \to \infty$
keeping the other two parameters fixed. This gives:
\beq\label{mkp6}
\p_{t_1}\log \frac{\phi_3}{\phi_2} =a_2 \Bigl (\frac{\phi}{\phi_2}
-\frac{\phi_3}{\phi_{23}}\Bigr )-
a_3 \Bigl (\frac{\phi}{\phi_3}
-\frac{\phi_2}{\phi_{23}}\Bigr ).
\eeq
The further limit $a_2\to \infty$  yields:
\beq\label{mkp7}
2a_3 \p_{t_1}\Bigl (\frac{\phi}{\phi_3}\Bigr )=
\p_{t_1}\log \Bigl (\frac{\phi_3}{\phi}\Bigr )
\p_{t_1}\log \Bigl (\phi_3 \phi\Bigr )-
\p_{t_1}^2\log \Bigl (\phi_3 \phi\Bigr )-
\p_{t_2}\Bigl (\frac{\phi_3}{\phi}\Bigr ).
\eeq
At last, the limit $a_3\to \infty$ yields the 
mKP equation for $v=\log \phi$ \cite{JM83}:
\beq\label{mkp8}
v_{xxxx}-4v_{xt} +3v_{yy}+6v_{xx}v_{y}-6v_x^2 v_{xx}=0,
\eeq
where $x=t_1, y=t_2, t=t_3$.

It follows from (\ref{mkp4a}) that the discrete mKP equations
(\ref{mkp5}), (\ref{mkp5a}) can be solved in terms of the 
mKP tau-function. Indeed, set
\beq\label{mkp9}
\phi ({\bf t})=\frac{\hat \tau ({\bf t})}{\tau ({\bf t})},
\eeq
where $\hat \tau ({\bf t})$ is some other tau-function (in a moment
we will see that it is a KP tau-function, too). Plugging this into
the first equation in (\ref{mkp4a}), we obtain the relation
\beq\label{mkp10}
a_1\hat \tau_2 \tau_1-a_2\hat \tau_1 \tau_2 =
a_{12}\hat \tau_{12} \tau
\eeq
As before, let $\{\alpha \beta \gamma \}$ be any 
cyclic permutation of $\{123\}$, then writing equations (\ref{mkp10}) 
for $\alpha \beta$, multiplying both sides by 
$\hat \tau_{\gamma}$ and summing the results 
over the three cyclic permutations, we obtain 
\beq\label{mkp11}
a_{12}\hat \tau_{12}\hat \tau_3 
+a_{23}\hat \tau_{23}\hat \tau_1
+a_{31}\hat \tau_{13}\hat \tau_2=0,
\eeq
i.e. the equation for a tau-function of the KP hierarchy.
From (\ref{mkp2a}) and (\ref{mkp9}) it is clear that
\beq\label{mkp9b}
\hat \tau ({\bf t})=\int_{\CCC}e^{\xi ({\bf t}, z)}
\rho (z)\tau ({\bf t}-[z^{-1}])d^2z.
\eeq
We can say that the transition from $\tau$ to $\hat \tau$ is a
BD transformation sending a solution of the KP hierarchy
to another solution. 

In a similar way, setting
\beq\label{mkp9a}
\phi^* ({\bf t})=\frac{\check \tau ({\bf t})}{\tau ({\bf t})},
\eeq
we obtain from the second equation in (\ref{mkp4a}):
\beq\label{mkp12}
a_1\tau_2 \check \tau_1 -a_2\tau_1 \check \tau_2  =
a_{12}\tau_{12} \check \tau ,
\eeq
from which it follows, in the same way as before, 
that $\check \tau ({\bf t})$ is a KP
tau-function. 
From (\ref{mkp2a}) and (\ref{mkp9a}) it is clear that
\beq\label{mkp9c}
\check \tau ({\bf t})=\int_{\CCC}e^{-\xi ({\bf t}, z)}
\rho^* (z)\tau ({\bf t}+[z^{-1}])d^2z.
\eeq

Now we see that relations (\ref{mkp10}) and (\ref{mkp12}) 
can be identified with the 3-term bilinear equations for the
tau-function $\tau (n, {\bf t})$ 
of the mKP hierarchy (\ref{m5}). Namely, fix some $n$ and 
identify the tau-functions as follows:
$$
\tau ({\bf t})=\tau (n, {\bf t}), \quad
\hat \tau ({\bf t})=\tau (n+1, {\bf t}), \quad
\check \tau ({\bf t})=\tau (n-1, {\bf t}), 
$$
then (\ref{mkp10}) and (\ref{mkp12}) acquire the form
(\ref{m5}). Therefore, we can say that the mKP hierarchy describes
a chain of BD transformations of the KP hierarchy, and the
discrete variable $n$ of the former hierarchy 
numbers subsequent steps of the chain.
More details on this are given below 
in section \ref{section:BacklundKP}.

Let us mention the following two determinant formulas:
\beq\label{mkp13}
\tau \Bigl (n, {\bf t}-\sum_{\alpha =1}^k [a_{\alpha}^{-1}]\Bigr )
= \frac{\det\limits_{k\times k}\Bigl (a_i^{j-1}
\tau (n\! +\! j\! -\! 1, {\bf t}-
[a_i^{-1}]\Bigr )}{\det\limits_{k\times k}
(a_i^{j-1}) \prod\limits_{\alpha =1}^{k-1}\tau (n+\alpha , {\bf t})},
\eeq
\beq\label{mkp13a}
\tau \Bigl (n, {\bf t}+\sum_{\alpha =1}^k [a_{\alpha}^{-1}]\Bigr )
= \frac{\det\limits_{k\times k}\Bigl (a_i^{j-1}
\tau (n\! -\! j\! +\! 1, {\bf t}+
[a_i^{-1}]\Bigr )}{\det\limits_{k\times k}
(a_i^{j-1}) \prod\limits_{\alpha =1}^{k-1}\tau (n-\alpha , {\bf t})}.
\eeq
The determinant in the denominators
is the Vandermonde determinant (\ref{vandermonde}).
These formulas 
can be proved by induction using (\ref{m5}) (as the basis
of the induction) and (\ref{m2}).

\subsection{An infinite chain of BD trans\-for\-ma\-ti\-ons
as the mKP hierarchy}
\label{section:BacklundKP}

In the previous subsection, we have seen that the tau-functions
of the KP hierarchy
$\check \tau ({\bf t})$, $\tau ({\bf t})$, $\hat \tau ({\bf t})$
defined by (\ref{mkp9}), (\ref{mkp9a})
are subsequent members of a chain of BD transformations,
which forms a set of solutions to the mKP hierarchy.
The functional parameters that define a particular chain
is a sequence of functions $\{ \rho_1(z), \rho_2(z), \rho_3(z),
\ldots \}$, which are assumed to be functions or distributions
of $z, \bar z$ with a compact support.
More precisely, fix some ``initial'' solution to the KP hierarchy
with the tau-function $\tau ({\bf t})=\tau (0, {\bf t})$ and set,
according to (\ref{mkp9}):
$$
\tau (1, {\bf t})=\int_{\CCC}e^{\xi ({\bf t}, z)}\rho_1(z)
\tau (0, {\bf t}-[z^{-1}]) \, d^2 z.
$$
As was proved in the previous subsection, $\tau (1, {\bf t})$ 
is a solution of the KP hierarchy.
The corresponding BA function is
\beq\label{b1}
\psi (1, {\bf t}; z)=ze^{\xi ({\bf t}, z)}
\frac{\tau (1, {\bf t}-[z^{-1}])}{\tau (1, {\bf t})}.
\eeq
Integrating it with a function $\rho_2(z)$, we get
$$
\phi (2, {\bf t})=\int_{\CCC}\rho_2(z)\psi (1, {\bf t}; z)d^2z
$$
which is used for introducing the tau-function $\tau (2, {\bf t})$
via
$$
\phi (2, {\bf t})=\frac{\tau (2, {\bf t})}{\tau (1, {\bf t})},
$$
hence
\beq\label{b2}
\tau (2, {\bf t})=\int_{\CCC}e^{\xi ({\bf t}, z)}z\rho_2(z)
\tau (1, {\bf t}-[z^{-1}]) \, d^2 z.
\eeq
Given an infinite sequence of functions (or distributions)
$\{ \rho_1(z), \rho_2(z), \rho_3(z),
\ldots \}$, we can 
repeat this procedure $n$ times and obtain a sequence of
KP tau-functions $\tau (n, {\bf t})$ with the corresponding BA
functions
$$
\psi (n, {\bf t}; z)=z^ne^{\xi ({\bf t}, z)}
\frac{\tau (n, {\bf t}-[z^{-1}])}{\tau (n, {\bf t})}.
$$
The recurrence relation is
\beq\label{b3}
\tau (n+1, {\bf t})=\int_{\CCC}e^{\xi ({\bf t}, z)}z\rho_{n+1}(z)
\tau (n, {\bf t}-[z^{-1}]) \, d^2 z,
\eeq
which can be resolved as
\beq\label{b4}
\tau (n, {\bf t})=\int_{\CCC} \! \ldots \! \int_{\CCC}
\Bigl (\prod_{i=1}^n d^2 z_i e^{\xi ({\bf t}, z_i)} \rho_i (z_i)\Bigr )
\Delta_{n}(z_1, \ldots , z_n)\tau (0, {\bf t}-\sum_{\alpha =1}^n
[z_{\alpha}^{-1}]\Bigr ).
\eeq
Using (\ref{mkp13}), this can be represented in the determinant form
\beq\label{b5}
\tau (n, {\bf t})=\left (\prod_{\alpha =1}^{n-1}\tau (\alpha , {\bf t})
\right )^{-1}\det_{n\times n}\left (
\int_{\CCC}d^2 z e^{\xi ({\bf t}, z)} \rho_i(z)z^{j-1}
\tau \Bigl (j\! -\! 1, {\bf t}-[z^{-1}]\Bigr )\right ),
\eeq
or
\beq\label{b6}
\frac{\tau (n, {\bf t})}{\tau (0, {\bf t})}=
\det_{n\times n} \left (\int_{\CCC}
\rho_i(z)\psi (j\! -\! 1, {\bf t};z)d^2 z \right ).
\eeq

Let us mention some important particular cases of 
the tau-function given by (\ref{b4}). 
If $\tau (0, {\bf t})=1$, this formula
yields:
\beq\label{b7}
\tau (n, {\bf t})=
\det_{n\times n} \left (\int_{\CCC}
e^{\xi ({\bf t}, z)}
\rho_i(z)z^{j-1}d^2 z \right ).
\eeq
If $\tau (0, {\bf t})=1$ and $\rho_m(z)=z^{m-1}\mu (z)$, where
$\mu (z)$ is a measure with the support on the real axis, i.e.,
$\mu (x+iy)=\delta (y)e^{V(x)}$, then 
\beq\label{b8}
\tau (n, {\bf t})=\frac{1}{n!}\int_{\RRR}\! \ldots \! \int_{\RRR}
\Delta_n^2(x_1, \ldots , x_n)\prod_{i=1}^n e^{\xi ({\bf t}, x_i)
+V(x_i)}dx_i
\eeq
is the partition function of the Hermitean matrix model 
represented as the $n$-fold integral over eigenvalues $x_i$ 
after integration over ``angle variables''. Equation (\ref{b7})
gives its well known representation as determinant of the moments
matrix:
\beq\label{b7a}
\tau (n, {\bf t})=
\det_{n\times n} \left (\int_{\RRR}
e^{\xi ({\bf t}, x)+V(x)}
x^{i+j-2}dx \right ).
\eeq
If $\tau (0, {\bf t})=1$ and $\rho_m(z)=z^{-m+1}\nu (z)$, where
$\nu (z)=\delta (|z|-1)$ is the measure with the support on the unit
circle, then
\beq\label{b9}
\tau (n, {\bf t})=\frac{1}{n!}\oint_{|z_1|=1}\! \ldots \! \oint_{|z_n|=1}
\Bigl |\Delta_n(z_1, \ldots , z_n)\Bigr |^2
\prod_{i=1}^n e^{\xi ({\bf t}, z_i)}|dz_i|
\eeq
is the partition function of the unitary matrix model.
Other important examples are provided by distributions
with support at two points. For example, set
\beq\label{rho-soliton}
\rho_m(z)=\delta^{(2)}(z-p_m)+\alpha_m \delta^{(2)}(z-q_m),
\eeq
where $p_1, p_2, p_3 ,\ldots$, $q_1, q_2, q_3, \ldots$ are
some distinct points in the complex plane and $\alpha_m$ are
arbitrary parameters. In this case we have 
from (\ref{b7}) (if $\tau (0, {\bf t})=1$):
\beq\label{b10}
\tau (n, {\bf t})=\det_{n\times n}
\Bigl (p_i^{j-1}e^{\xi ({\bf t}, p_i)}+\alpha_i 
q_i^{j-1}e^{\xi ({\bf t}, q_i)}\Bigr ).
\eeq
This is the $n$-soliton solution of the KP hierarchy.
The mKP hierarchy connects them for different values of $n$.
If the support of a density distribution consists of more than
two points, we get in the same way more general soliton-like
solutions. A limiting procedure when these points merge in
a special way leading to the support of each $\rho_j$ 
concentrated just at a point, but with
higher derivatives of the delta-function, is also possible.
In this case equation (\ref{b7}) gives rational solutions,
for which the tau-function is a polynomial in the time variables
multiplied by an exponential function.
Applying the transformations with density distributions
of the form (\ref{rho-soliton}) to 
a non-trivial $\tau (0, {\bf t})$, one obtains, using 
(\ref{b4}), solutions which are sometimes called ``solitons
on a nontrivial background'' (such as, for example, solitons on the
background of algebraic-geometrical quasiperiodic solutions
constructed from algebraic curves of finite genus).

So far we have considered the BD transformations
in the positive direction (we call them forward transformations), 
i.e., $\tau (n)\to \tau (n+1)$
with $n\geq 0$,
starting from some ``initial'' tau-function $\tau(0)$.
In fact this chain can be extended to the negative direction 
as well, i.e. it is possible to define BD transformations
of another type (the backward BD transformations)
that send $\tau(n)\to \tau (n-1)$ for $n\leq 0$,
starting again from $\tau (0)$. In fact, the two 
half-infinite chains are glued together
in a single double-infinite 
chain of KP tau-functions $\tau (n, {\bf t})$ 
with $n\in \ZZ$ such that any two 
of them are connected by equations of the mKP
hierarchy.

This can be done using the adjoint wave function $\phi^*({\bf t})$.
Fix a set of functional parameters 
$\{ \rho_1^*(z), \rho_2^*(z), \rho_3^*(z),
\ldots \}$ (functions or distributions
of $z, \bar z$ with a compact support) and set
according to (\ref{mkp9a}):
\beq\label{b11}
\tau (-m-1, {\bf t})=\int_{\CCC}e^{-\xi ({\bf t}, z)}z^m
\rho_{m+1}^*(z)
\tau (-m, {\bf t}+[z^{-1}]) \, d^2 z, \quad m\geq 0.
\eeq
This recurrence relation 
defines a sequence of KP tau-functions starting from the initial
$\tau (0, {\bf t})=\tau ({\bf t})$. Then equation (\ref{mkp12})
states that the sequence $\tau (n, {\bf t})$ for all $n\in \ZZ$
is a solution of the mKP hierarchy (\ref{m5}).
This solution is defined by fixing the two sets of functional
parameters $\{ \rho_1(z), \rho_2(z), \rho_3(z),
\ldots \}$, $\{ \rho_1^*(z), \rho_2^*(z), \rho_3^*(z),
\ldots \}$.
The recurrence relation (\ref{b11}) can be resolved as
\beq\label{b4a}
\tau (-m, {\bf t})=\int_{\CCC} \! \ldots \! \int_{\CCC}
\Bigl (\prod_{i=1}^m d^2 z_i e^{-\xi ({\bf t}, z_i)} \rho_i^* (z_i)\Bigr )
\Delta_{m}(z_1, \ldots , z_n)\tau (0, {\bf t}+\sum_{\alpha =1}^m
[z_{\alpha}^{-1}]\Bigr ).
\eeq
Using (\ref{mkp13a}), the right-hand side can 
be represented in the determinant form
\beq\label{b5a}
\tau (-m, {\bf t})=\left (\prod_{\alpha =1}^{m-1}
\tau (-\alpha , {\bf t})
\right )^{-1}\det_{m\times m}\left (
\int_{\CCC}d^2 z e^{-\xi ({\bf t}, z)} \rho_i^* (z)z^{j-1}
\tau \Bigl (-j\! +\! 1, {\bf t}+[z^{-1}]\Bigr )\right ),
\eeq
or
\beq\label{b6a}
\frac{\tau (-m, {\bf t})}{\tau (0, {\bf t})}=
\det_{m\times m} \left (\int_{\CCC}
\rho_i^*(z)\psi^* (-j\! +\! 1, {\bf t};z)d^2 z \right ),
\eeq
where
$$
\psi^*(-m, {\bf t}; z)=
z^me^{-\xi ({\bf t}, z)}
\frac{\tau (-m, {\bf t}+[z^{-1}])}{\tau (-m, {\bf t})}, \quad m\geq 0.
$$
Formulas similar to (\ref{b7})--(\ref{b10}) can be easily obtained.
The chain of BD transformations is illustrated by Fig. 
\ref{figure:chain}.

\begin{figure}[t]
\centering{\includegraphics[scale=0.8]{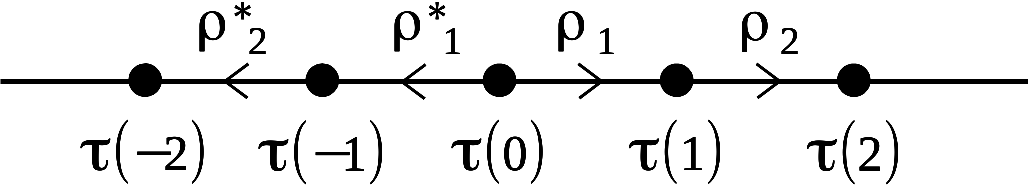}}
\vspace{1cm}
\caption{The chain of BD transformations.}
\label{figure:chain}
\end{figure}

Let us consider a chain of the 
forward B\"acklund transformations defined
by a sequence of the density functions 
$\{\rho_1(z), \rho_2(z), \ldots \}$. A natural question is whether 
the backward BD transformations can be understood 
as transformations inverse
to the forward ones. More precisely, let ${\bf B}$ be the forward
transformation, i.e.,
$$
{\bf B}\tau (n, {\bf t})=\tau (n+1, {\bf t})=
\int e^{\xi ({\bf t}, z)}z^n \rho_{n+1}(z)\tau (n, {\bf t}-[z^{-1}])
d^2 z.
$$
The question is what density $\rho^*(z)$ one should assign to 
a backward transformation ${\bf B}^*$ in order to come back to
$\tau (n, {\bf t})$, applying it to $\tau (n+1, {\bf t})$:
${\bf B}^*{\bf B}\tau (n, {\bf t})=\tau (n, {\bf t})$. The answer
is rather simple and explicit in the case when supports ${\sf D}_n$
of the functions $\rho_n(z)$ do not intersect and
$\tau (0, {\bf t})=1$. In general we have:
\beq\label{b13}
{\bf B}^*{\bf B}\tau (n, {\bf t})=\int \! \int
e^{\xi ({\bf t}, z)-\xi ({\bf t}, w)}\frac{z^n w^{n+2}}{w-z}
\, \rho_{n+1}(z)\rho^{*}(w) \tau (n, {\bf t}\! +\! 
[w^{-1}]\! - \![z^{-1}])
d^2 z d^2 w.
\eeq
By assumption, the function $\rho_{n+1}(z)$ has a compact support
${\sf D}_{n+1}\subset \CC$ (which can be a compact domain in $\CC$ or
a set of points) such that ${\sf D}_{n+1}\cap {\sf D}_{k}=\emptyset$
for all $k=1, \ldots , n$. Let us find the function $\rho^*$ in the form
$$
w^2 \rho^*(w)=\p_{\bar w}f(w),
$$
where the function $f$ has a support ${\sf D}\supset {\sf D}_{n+1}$
and is such that $f(w)=0$ if $w\in \p {\sf D}$. We also assume that
${\sf D}\cap {\sf D}_{k}=\emptyset$ for all $k=1, \ldots , n$. 
With these assumptions, equation (\ref{b13}) acquires the form
\beq\label{b14}
{\bf B}^*{\bf B}\tau (n, {\bf t})=\int_{{\sf D}_{n+1}}
d^2 z e^{\xi ({\bf t}, z)}z^n \rho_{n+1}(z)
\int_{{\sf D}}d^2 w e^{-\xi ({\bf t}, w)}
\frac{w^{n}}{w-z} \p_{\bar w}f(w)\tau (n, {\bf t}\! +\! 
[w^{-1}]\! - \![z^{-1}]).
\eeq
The integral in $w$ can be transformed by taking ``by parts''
and using the formula
$$
\p_{\bar w}\Bigl (\frac{1}{w-z}\Bigr )=\pi \delta^{(2)}(w-z).
$$
In this way we get:
\beq\label{b15}
\begin{array}{l}
\displaystyle{
\int_{{\sf D}}d^2 w e^{-\xi ({\bf t}, w)}
\frac{w^{n}}{w-z} \p_{\bar w}f(w)\tau (n, {\bf t}\! +\! 
[w^{-1}]\! - \![z^{-1}])}
\\ \\
\phantom{aaaaaaaaa}\displaystyle{
=\frac{1}{2i}\oint_{\p {\sf D}}dw 
e^{-\xi ({\bf t}, w)}
\frac{w^{n}}{w-z} f(w)\tau (n, {\bf t}\! +\! 
[w^{-1}]\! - \![z^{-1}])}
\\ \\
\phantom{aaaaaaaaaaaaaaaa}\displaystyle{
-\int_{{\sf D}}d^2 w e^{-\xi ({\bf t}, w)}
\frac{w^{n}}{w-z} f(w)\p_{\bar w}\tau (n, {\bf t}\! +\! 
[w^{-1}]\! - \![z^{-1}])}
\\ \\
\phantom{aaaaaaaaaaaaaaaaaaaaaaaaaaaaaaa}\displaystyle{
-\pi e^{-\xi ({\bf t}, z)} z^n f(z) \tau (n, {\bf t}).}
\end{array}
\eeq
The first integral in the right-hand side vanishes because
$f(w)=0$ on $\p {\sf D}$. In fact, under our assumptions 
the second
integral also vanishes. To see this,
we use the explicit formula
(\ref{b4}) in which we put $\tau (0, {\bf t})=1$:
\beq\label{b4b}
\tau (n, {\bf t})=\int_{\CCC} \! \ldots \! \int_{\CCC}
\Delta_{n}(z_1, \ldots , z_n)
\prod_{i=1}^n e^{\xi ({\bf t}, z_i)} \rho_i (z_i)d^2 z_i.
\eeq
After some transformations, we find:
$$
\p_{\bar w}\tau (n, {\bf t}\! +\! 
[w^{-1}]\! - \![z^{-1}])=\pi (z-w)e^{\xi ({\bf t}, w)}z^{-1}w^n
\sum_{j=1}^n (-1)^{n-j}\rho_j(w)A_j({\bf t}, z),
$$
where
$$
A_j({\bf t}, z)=\int_{\CCC} \! \ldots \! \int_{\CCC}
\Delta_{n-1}(z_1, \ldots , \hat z_j, \ldots ,z_n)
\prod_{i=1, \neq j}^n e^{\xi ({\bf t}, z_i)} 
\Bigl (1-\frac{z_i}{z}\Bigr )\rho_i (z_i)d^2 z_i.
$$
Plugging this into (\ref{b15}), we get:
$$
\begin{array}{l}
\displaystyle{
\int_{{\sf D}}d^2 w e^{-\xi ({\bf t}, w)}
\frac{w^{n}}{w-z} f(w)\p_{\bar w}\tau (n, {\bf t}\! +\! 
[w^{-1}]\! - \![z^{-1}])}
\\ \\
\phantom{aaaaaaaaa}\displaystyle{
=\pi z^{-1}\sum_{j=1}^n(-1)^{n-j} A_j({\bf t}, z)
\int_{{\sf D}}w^{2n}f(w)\rho_j(w)d^2w=0}
\end{array}
$$
because the supports of the functions $\rho_j$ and $f$ do not
intersect by the assumption. Finally, we have from (\ref{b14}):
\beq\label{b16}
{\bf B}^*{\bf B}\tau (n, {\bf t})=c_n \tau (n, {\bf t}),
\eeq
where
$$
c_n=-\pi \int_{{\sf D}_{n+1}} z^{2n}\rho_{n+1}(z)f(z)d^2 z
$$
is an irrelevant ${\bf t}$-independent common multiplier.

\subsection{Several chains of BD trans\-for\-ma\-ti\-ons
as a discrete integrable hierarchy}
\label{section:severalBacklundKP}

Let us now consider several infinite chains of BD
transformations. We begin with two chains in the forward direction
that are defined by fixing two sequences of functions
$\{\rho_1^{(1)}(z), \rho_2^{(1)}(z), \ldots \}$ and
$\{\rho_1^{(2)}(z), \rho_2^{(2)}(z), \ldots \}$. Starting from
an initial KP tau-function $\tau ({\bf t})$, one can apply 
the first transformation $n_1$ times and then the second one
$n_2$ times. We denote the resulting tau-function as 
$\tau (n_1, n_2, {\bf t})$. Let us denote the first and
the second transformations by ${\bf B}_1$ and
${\bf B}_2$:
\beq\label{B1B2}
\begin{array}{l}
\displaystyle{
{\bf B}_1\tau (n_1, n_2, {\bf t})=
\tau (n_1+1, n_2, {\bf t})=
\int_{\CCC}e^{\xi ({\bf t}, z)}z^{n_1+n_2}\rho^{(1)}_{n_1+1}(z)
\tau (n_1, n_2, {\bf t}-[z^{-1}])d^2 z,}
\\ \\
\displaystyle{
{\bf B}_2\tau (n_1, n_2, {\bf t})=
\tau (n_1, n_2+1, {\bf t})=
\int_{\CCC}e^{\xi ({\bf t}, z)}z^{n_1+n_2}\rho^{(2)}_{n_2+1}(z)
\tau (n_1, n_2, {\bf t}-[z^{-1}])d^2 z.}
\end{array}
\eeq
There are two ways to obtain $\tau (n_1+1, n_2+1, {\bf t})$
from $\tau (n_1, n_2, {\bf t})$ (first apply ${\bf B}_1$, then
${\bf B}_2$ or vice versa):
$$
\tau (n_1, n_2, {\bf t})\stackrel{{\bf B}_1}{\longrightarrow}
\tau (n_1+1, n_2, {\bf t})
\stackrel{{\bf B}_2}{\longrightarrow}
\tau (n_1+1, n_2+1, {\bf t}),
$$
$$
\tau (n_1, n_2, {\bf t})\stackrel{{\bf B}_2}{\longrightarrow}
\tau (n_1, n_2+1, {\bf t})
\stackrel{{\bf B}_1}{\longrightarrow}
\tau (n_1+1, n_2+1, {\bf t}).
$$
Writing the two composite transformations explicitly, one can
easily see that the results differ by a sign, i.e., the
anti-commutation relation
\beq\label{b18}
{\bf B}_1 {\bf B}_2=-{\bf B}_2 {\bf B}_1
\eeq
holds true. Therefore, given $\tau (n_1, n_2, {\bf t})$, 
the function $\tau (n_1+1, n_2+1, {\bf t})$ is defined
only up to a sign\footnote{However, since the BA functions
are ratios of two tau-functions, the two transformations
${\bf B}_1$, ${\bf B}_2$ commute when they act 
to the BA functions.}. 

\noindent
{\bf Remark.} The anti-commutation relation (\ref{b18})
becomes obvious in the realization of the BD transformations
via free fermions (section \ref{section:BDchargedfermions} below).

In order to fix this uncertainty and make 
$\tau (n_1, n_2, {\bf t})$ to be a well-defined single-valued
function on the $\ZZ_{\geq 0} \times \ZZ_{\geq 0}$ lattice, we 
proceed in the following way. Consider a set of $N$ 
chains of the forward BD transformations 
defined by the operators ${\bf B}_{\alpha}$
with sequences of functions
$\{\rho_1^{(\alpha )}(z), \rho_2^{(\alpha )}(z), \ldots \}$,
$\alpha =1, \ldots , N$. 
Let ${\bf n}=\{n_1, \ldots , n_N\}$ 
(where all $n_{\alpha}\geq 0$) be a point 
of the lattice $(\ZZ_{\geq 0})^{N}$, and 
${\bf e}_{\alpha}$ be the vector whose $\alpha$-th component
is 1 and all other components are equal to 0. Then we 
write, by definition:
$$
{\bf B}_{\alpha}\tau ({\bf n}, {\bf t})=
\tau ({\bf n}+{\bf e}_{\alpha}, {\bf t}).
$$
Let us fix the linear order in the set of the transformations
${\bf B}_{\alpha}$
according to the natural order in the set $\{1, \ldots , N\}$.
Then, given $\tau ({\bf n}, {\bf t})$, we can uniquely define 
$\tau ({\bf n}+{\bf e}_{\alpha}+{\bf e}_{\beta}, {\bf t})$
for $\alpha \neq \beta$ as follows:
\beq\label{b19}
{\bf B}_{\alpha}{\bf B}_{\beta}\tau ({\bf n}, {\bf t})=
\epsilon_{\alpha \beta}
\tau ({\bf n}+{\bf e}_{\alpha}+{\bf e}_{\beta}, {\bf t})
\eeq
where
$\epsilon_{\alpha \beta}=1$ if $\alpha \leq \beta$ and
$\epsilon_{\alpha \beta}=-1$ if $\alpha > \beta$.
This prescription allows one 
to unambiguously extend the tau-function to all points 
of the lattice, 
starting from a given tau-function 
at the origin ${\bf n}={\bf 0}$.

Let us now derive equations which connect the tau-functions
at different points of the $N$-dimensional lattice, as functions
of ${\bf n}$. First, we have the equation
\beq\label{b20}
a_1\Bigl (\tau ({\bf n}+{\bf e}_{\alpha})
\tau_1 ({\bf n}+{\bf e}_{\beta})-
\tau ({\bf n}+{\bf e}_{\beta})
\tau_1 ({\bf n}+{\bf e}_{\alpha})\Bigr )=
\epsilon_{\alpha \beta}
\tau_1 ({\bf n}+{\bf e}_{\alpha}+{\bf e}_{\beta})
\tau ({\bf n}),
\eeq
where we use the short-hand notation introduced in
section \ref{section:KP}. It can be obtained 
using the integral formulas of the type (\ref{B1B2}) and 
bilinear equation (\ref{g3}) for the KP tau-function.
In order to write this equation in
a more compact form, we can use even more short-hand notation
like $\tau ({\bf n}+{\bf e}_{\alpha})=\tau^{(\alpha )}$,
$\tau ({\bf n}+{\bf e}_{\alpha}+{\bf e}_{\beta})=
\tau^{(\alpha \beta )}$, etc. Then equation (\ref{b20})
acquires the form
\beq\label{b20a}
a_1\Bigl (\tau^{(\alpha )}\tau_1^{(\beta )}-
\tau^{(\beta )}\tau_1^{(\alpha )}\Bigr )=
\epsilon_{\alpha \beta}\tau_{1}^{(\alpha \beta )}\tau .
\eeq
Letting $a_1 \to \infty$, we obtain in the limit:
\beq\label{b21}
\epsilon_{\alpha \beta}
\p_{t_1}\log \frac{\tau^{(\beta )}}{\tau^{(\alpha )}}=
\frac{\tau \, \tau^{(\alpha \beta )}}{\tau^{(\alpha )}
\tau^{(\beta )}}.
\eeq

Using the short-hand notation from section \ref{section:KP}
and the ones introduced in the present section, we can write equation (\ref{m5})
in the form
\beq\label{b22}
a_1 \tau_{\bar 2}\tau_{\bar 1}^{(\alpha )}-
a_2 \tau_{\bar 1}\tau_{\bar 2}^{(\alpha )}=
a_{12}\tau_{\bar 1 \bar 2}\tau^{(\alpha )}.
\eeq
Put $a_2=z$, then the BA function can be represented as
$
\psi =z^n e^{\xi ({\bf t}, z)} \tau_{\bar 2}/\tau 
$,
and equation (\ref{b22}) can be read as a linear equation for
the BA function:
\beq\label{b23}
a_1 \psi_{\bar 1}=-\psi^{(\alpha )}+a_1
\frac{\tau_{\bar 1}^{(\alpha )}
\tau}{\tau^{(\alpha )}\tau_{\bar 1}}\psi .
\eeq
Subtracting it from the similar equation for 
another BD flow ${\bf B}_{\beta}$ 
with $\beta \neq \alpha$, we get:
$$
\psi^{(\beta )}=\psi^{(\alpha )}-a_1
\frac{\tau (\tau^{(\alpha )}_{\bar 1}\tau^{(\beta )}-
\tau^{(\beta )}_{\bar 1}\tau^{(\alpha )})}{\tau_{\bar 1}\, 
\tau^{(\alpha )}\, \tau^{(\beta )}}\, \psi .
$$
The right-hand side can be simplified with the help of
equation (\ref{b20a}). Then this relation becomes the following
linear equation for the BA function:
\beq\label{b24}
\psi^{(\alpha )}-\psi^{(\beta )}=
\epsilon_{\alpha \beta}\, \frac{\tau 
\tau^{(\alpha \beta )}}{\tau^{(\alpha )}\tau^{(\beta )}}\, \psi .
\eeq
Let us mention a slightly different form
of this equation which may be useful for applications.
Put $n_1=n$, $n_2=m$, then (\ref{b24}) can be written
as
\beq\label{b24a}
\psi (m+1,n)=\left (e^{\p_n}-
\frac{\tau(m,n)\tau (m+1, n+1)}{\tau (m,n+1)\tau (m+1,n)}\right )
\psi (m,n),
\eeq
which means that $\psi (m+1,n)$ is obtained from $\psi (m,n)$
by the action of the first order difference operator 
acting to it with respect to the variable $n$.

There also exists an equation containing only shifts of the 
``B\"acklund-Darboux variables'' ${\bf n}$. 
Let $(\alpha \beta \gamma )$ be any cyclic permutation of
indices $(123)$. 
We need
the integral formulas of the type (\ref{B1B2}) for 
$\tau^{(\alpha )}$, $\tau^{(\beta )}$, $\tau^{(\gamma )}$ and
the formulas
$$
\tau^{(\alpha \beta )}=\epsilon_{\alpha \beta}
\int_{\CCC} \! \int_{\CCC}
e^{\xi ({\bf t}, z_{\beta})+\xi ({\bf t}, z_{\alpha})} 
(z_{\beta}-z_{\alpha})
\rho^{(\beta )}(z_{\beta})\rho^{(\alpha )}(z_{\alpha})\tau
\Bigl ({\bf t}-[z_{\alpha}^{-1}]-[z_{\beta}^{-1}]\Bigr )d^2z_{\alpha}
d^2z_{\beta}
$$
that follow from them. Taking into account the bilinear
equation (\ref{g3}) for the KP tau-function, it is not difficult
to see that the equation
\beq\label{b25}
\epsilon_{\alpha \beta}\tau^{(\alpha \beta )}\tau^{(\gamma )}+
\epsilon_{ \beta \gamma}\tau^{( \beta \gamma )}\tau^{(\alpha )}+
\epsilon_{\gamma \alpha}\tau^{(\gamma \alpha )}\tau^{(\beta )}=0
\eeq
holds true. In the more explicit notation it reads
\beq\label{b25a}
\begin{array}{c}
\epsilon_{\alpha \beta}\tau ({\bf n}+{\bf e}_{\alpha}+
{\bf e}_{\beta})\tau ({\bf n}+{\bf e}_{\gamma})
+\epsilon_{\beta \gamma}\tau ({\bf n}+{\bf e}_{\beta}+
{\bf e}_{\gamma})\tau ({\bf n}+{\bf e}_{\alpha})
\\ \\
+\, \epsilon_{\gamma \alpha}\tau ({\bf n}+{\bf e}_{\gamma}+
{\bf e}_{\alpha})\tau ({\bf n}+{\bf e}_{\beta})=0.
\end{array}
\eeq
In particular, we have:
\beq\label{b25b}
\tau ^{(12)} \tau^{(3)}-\tau ^{(13)} \tau^{(2)}+
\tau ^{(23)} \tau^{(1)}=0.
\eeq

Let us note that equation (\ref{b25}) is the compatibility
condition for the linear problems of the form (\ref{b24}).
With the help of the same method that was used in section
\ref{section:KP} to prove that equations of the form  
(\ref{g3}) imply
equations (\ref{g2}), one can show that equations of the form
(\ref{b25}) imply the following equations:
\beq\label{b26}
\epsilon_{\alpha \delta}\epsilon_{\beta \gamma}
\tau^{(\alpha \delta )}\tau^{(\beta \gamma )}+
\epsilon_{\beta \delta}\epsilon_{\gamma \alpha}
\tau^{(\beta \delta )}\tau^{(\gamma \alpha )}+
\epsilon_{\gamma \delta}\epsilon_{\alpha \beta}
\tau^{(\gamma \delta )}\tau^{(\alpha \beta )}=0.
\eeq
This is nothing else than the standard discrete Hirota equation for the 
KP case.

Finally, we note that equation (\ref{b25a}) is the 
equation of the multi-component mKP hierarchy \cite{Z19} 
restricted 
to the sector of discrete variables. To see this, we recall
that the tau-function 
$\tau ({\bf n}, {\bf t})$ of the $N$-component mKP hierarchy
depends on $N$ discrete variables ${\bf n}=\{n_1, \ldots , n_N\}$
and on $N$ infinite sets of continuous variables
$${\bf t}=\{{\bf t}_1, \ldots , {\bf t}_N\}, 
\qquad
{\bf t}_{\alpha}=\{t_{\alpha ,1}, t_{\alpha ,2}, \ldots \},
\qquad \alpha =1, \ldots , N.
$$
The integral bilinear equation for the tau-function
has the form \cite{Z19}
\beq\label{b27}
\begin{array}{l}
\displaystyle{
\sum_{\mu =1}^N \epsilon_{\alpha \mu} ({\bf n})
\epsilon_{\beta \mu} ({\bf n}')
\oint_{C_{\infty}}
z^{n_{\mu}-
n'_{\mu}+\delta_{\alpha \mu}+\delta_{\beta \mu}-2}}
e^{\xi ({\bf t}_{\mu}-{\bf t}_{\mu}',z)}
\\ \\
\phantom{aaaaaaaaaaa}\displaystyle{\times \,
\tau \Bigl ({\bf n}+{\bf e}_{\alpha}-{\bf e}_{\mu},
{\bf t}-[z^{-1}]_{\mu}\Bigr )
\tau \Bigl ({\bf n}'+{\bf e}_{\mu}-{\bf e}_{\beta},
{\bf t}'+[z^{-1}]_{\mu}\Bigr )\, dz=0,}
\end{array}
\eeq
where the notation ${\bf t}\pm [z^{-1}]_{\mu}$ means that
the times from the set ${\bf t}_{\mu}$ are shifted as in
(\ref{shift}) and all other times are untouched.
The sign factors are defined as follows:
\beq\label{eps}
\epsilon_{\alpha \mu}({\bf n})=\left \{
\begin{array}{ll}
\;\; (-1)^{n_{\alpha +1}+
\ldots +n_{\mu}} &\quad \mbox{if $\alpha <\mu$}
\\
\quad 1 &\quad \mbox{if $\alpha =\mu$}
\\
-(-1)^{n_{\mu +1}+\ldots +n_{\alpha}} &
\quad \mbox{if $\alpha >\mu$}
\end{array}\right.
\eeq
Equation (\ref{b27}) is valid for all ${\bf t}, {\bf t}'$ 
and ${\bf n}, {\bf n}'$ such that $n_{\alpha}\geq n_{\alpha}'$
for all $\alpha$. Putting ${\bf t}={\bf t}'$ and
${\bf n}-{\bf n}'={\bf e}_{\gamma}$, where $\gamma \neq \alpha ,
\beta$ (and also $\alpha \neq \beta$), we see that the sum 
in left-hand side of (\ref{b27}) has only three non-zero terms
(for $\mu =\alpha , \beta , \gamma$), and the integrals are
calculated by taking residues at infinity. Then it is not
difficult to see that the resulting 3-term bilinear equation is 
equivalent to (\ref{b25a}).

\subsection{The discrete Schwarzian KP hierarchy}
\label{section:SKP}

The Schwarzian form of the KP hierarchy is obtained if
one chooses the Baker-Akhiezer kernel (BA kernel) 
as the dependent variable. 
The BA kernel is the function
\beq\label{s1}
\Psi ({\bf t}; z,w)=e^{\xi ({\bf t}, z)-\xi ({\bf t}, w)}\,
\frac{\tau ({\bf t}+[w^{-1}]-[z^{-1}])}{(z-w) \tau ({\bf t})}.
\eeq
Moreover, for any integer $n\geq 2$ 
one can define ``higher'' (multi-point) BA kernels via
\beq\label{s1c}
\begin{array}{l}
\displaystyle{
\Psi_n ({\bf t}; z_1, \ldots , z_n , w_1, \ldots , w_n)=
\exp \Bigl (\sum\limits_{i=1}^n (\xi ({\bf t}, z_i)-
\xi ({\bf t}, w_i))\Bigr )\,}
\\ \\
\displaystyle{
\phantom{aaaaaaaaaaaaaaaaaaaaaaaaa}\times 
\frac{\tau \Bigl ({\bf t}+
\sum\limits_{\alpha =1}^n ([w_{\alpha}^{-1}]-
[z_{\alpha}^{-1}])\Bigr )}{\tau ({\bf t})}
\det\limits_{n\times n}
\Bigl (\frac{1}{z_i-w_j}\Bigr ),}
\end{array}
\eeq
where determinant of the Cauchy matrix in the right-hand side
is
\beq\label{s1b}
\det\limits_{n\times n}
\Bigl (\frac{1}{z_i-w_j}\Bigr )=\frac{\prod\limits_{l<m}
(z_l-z_m)(w_m-w_l)}{\prod\limits_{l,m}(z_l-w_m)}.
\eeq
We note the following determinant formula:
\beq\label{s1a}
\Psi_n ({\bf t}; z_1, \ldots , z_n , w_1, \ldots , w_n)=
\det\limits_{n\times n} 
\Bigl (\Psi ({\bf t}; z_i, w_j)\Bigr ).
\eeq
The proof can be found in \cite{Shigyo13}.

Using the definition and the bilinear equation (\ref{g2}), one
can prove, after some simple calculations, that the following identity
holds:
\beq\label{s2}
\begin{array}{c}
\displaystyle{
\Psi ({\bf t}+[a^{-1}]; z,w)-\Psi ({\bf t}+[b^{-1}]; z,w)}
\\ \\
\displaystyle{
=\,
(a^{-1}-b^{-1}) 
\frac{\tau ({\bf t}+[a^{-1}]+
[b^{-1}]) \, \tau ({\bf t})}{\tau ({\bf t}+[a^{-1}])
\tau ({\bf t}+[b^{-1}])}\,
\psi ({\bf t}+[a^{-1}]+[b^{-1}];z)\psi^*({\bf t};w)}.
\end{array}
\eeq
In particular, letting $b\to \infty$, we obtain 
a simpler identity:
\beq\label{s3}
\Psi ({\bf t}+[a^{-1}]; z,w)-\Psi ({\bf t}; z,w)=
a^{-1}\psi ({\bf t}+[a^{-1}];z)\psi^*({\bf t};w),
\eeq
which we will use in what follows. In the short-hand notation
it can be written as
\beq\label{s3a}
\Psi_i-\Psi=a_i^{-1}\psi_i(z)\psi^*(w),
\eeq
where $i=1,2,3$, etc. 

\paragraph{Remark 3.}
The connection between BA kernels and BD transformations
is well known in the literature. In particular, it was discussed in 
\cite{AvM92,LW97,ANP98}, where a plethora of relations as
(\ref{s2}) can be found.

In the same way as we have defined the wave functions, one can 
obtain from the BA kernel a more general object, integrating it
in $z$ and $w$ with the functions $\rho (z)$,
$\rho^*(w)$:
\beq\label{s4}
\Phi ({\bf t})=\int_{\CCC}\int_{\CCC}
\rho (z) \Psi ({\bf t}; z, w)\rho^*(w) \, d^2z \, d^2w.
\eeq
Integrating both sides of (\ref{s3a}) in $z$ and $w$, 
we obtain the following 
identity for $\Phi$:
\beq\label{s3b}
\Phi_i-\Phi=a_i^{-1}\phi_i \phi^*.
\eeq
Writing it for $i=1,2,3$ and shifting the arguments appropriately,
we obtain the relations
$$
\begin{array}{l}
\Phi_{12}-\Phi_1=a_2^{-1}\phi_{12}\phi_1^*,
\\ \\
\Phi_{23}-\Phi_2=a_3^{-1}\phi_{23}\phi_2^*,
\\ \\
\Phi_{13}-\Phi_3=a_1^{-1}\phi_{13}\phi_3^*
\end{array}
$$
and 
$$
\begin{array}{l}
\Phi_{13}-\Phi_1=a_3^{-1}\phi_{13}\phi_1^*,
\\ \\
\Phi_{12}-\Phi_2=a_1^{-1}\phi_{12}\phi_2^*,
\\ \\
\Phi_{23}-\Phi_3=a_2^{-1}\phi_{23}\phi_3^*.
\end{array}
$$
Combining them, we can exclude the wave functions and obtain
the equation which contains only $\Phi$-functions:
\beq\label{s5}
\frac{(\Phi_1 -\Phi_{12})
(\Phi_2 -\Phi_{23})(\Phi_3 -\Phi_{13})}{(\Phi_1 -\Phi_{13})
(\Phi_2 -\Phi_{12})(\Phi_3 -\Phi_{23})}=1.
\eeq
It is easy to check that it can be rewritten in the form
\beq\label{s5a}
\left |
\begin{array}{ccc}
1 & \Phi_{1}+\Phi_{23} & \Phi_1 \Phi_{23}
\\ && \\
1 & \Phi_{2}+\Phi_{31} & \Phi_2 \Phi_{31}
\\ && \\
1 & \Phi_{3}+\Phi_{12} & \Phi_3 \Phi_{12}
\end{array} \right |=0.
\eeq
Equation (\ref{s5}) (or (\ref{s5a})) 
is the discrete Schwarzian KP equation. Its interesting geometric
meaning was clarified in the seminal papers
\cite{KS02,Schief03}. 

The form (\ref{s5}) makes it almost obvious that 
this equation is invariant under frac\-ti\-o\-nal-li\-ne\-ar
(M\"obious) transformations
\beq\label{s6}
\Phi \to \frac{a\Phi +b}{c\Phi +d}, \qquad ad-bc \neq 0.
\eeq
It is enough to check this for the three transformations
$\Phi \to a\Phi$, $\Phi \to \Phi +c$ and $\Phi \to \Phi^{-1}$,
because the general transformation from the M\"obious group
can be represented as a composition of these three. 

Note also that equation (\ref{s5a}) does not contain the
``lattice spacing'' parameters $a_i$ explicitly. That is why this
equation is actually invariant with respect to any permutations
of the discrete variables $s_1, s_2, s_3$.

Combining the equations of the form (\ref{s5}) written for 
any triple of distinct indices from 
the set $\{1,2, \ldots , n\}$, one arrives at the equations
\beq\label{s7}
\prod_{i=1}^n \frac{\Phi_i - 
\Phi_{i, i+1}}{\Phi_i - \Phi_{i, i-1}}=1, \qquad n\geq 3,
\eeq
where the indices in the right-hand side are understood to be taken 
modulo
$n$. These equations can be regarded as the higher equations 
of the discrete Schwarzian KP hierarchy. Since all of them follow
from (\ref{s5}), the latter is not only the integrable discretization
of the Schwarzian KP equation, but also of the whole hierarchy.

The continuum limit of equation (\ref{s5}) is most easily
taken in three steps. At the first step we tend $a_1 \to \infty$
keeping the other two parameters fixed. This gives:
\beq\label{s8}
(\Phi -\Phi_2)(\Phi_2 -\Phi_{23})\p_{t_1}\Phi_3 =
(\Phi -\Phi_3)(\Phi_3 -\Phi_{23})\p_{t_1}\Phi_2.
\eeq
At the second step we tend $a_2 \to \infty$ and expand equation
(\ref{s8}) in inverse powers of $a_2$. The first non-vanishing
term arises at $a_2^{-2}$. After some calculation this limit
yields:
\beq\label{s9}
2\p_{t_1}\log \frac{\Delta_3 \Phi}{\p_{t_1}\Phi}=
\Delta_3 \Bigl ( \frac{\p_{t_2}\Phi +
\p_{t_1}^2 \Phi}{\p_{t_1}\Phi}\Bigr ), 
\eeq
where the difference operator $\Delta_3$ is defined as
$$
\Delta_3 f ({\bf t})=f({\bf t}+[a_3^{-1}])-f({\bf t}).
$$
Finally, one should tend $a_3\to \infty$ in (\ref{s9}).
After a relatively long calculation, the first non-vanishing term
yields the equation
\beq\label{s10}
4\Bigl (\frac{\Phi_t}{\Phi_x}\Bigr )_x=
\Bigl (S[\Phi ]\Bigr )_x +3 \frac{\Phi_x^2 \Phi_{yy}-
\Phi_y^2 \Phi_{xx}}{\Phi_x^3},
\eeq
where 
\beq\label{s10a}
S[\Phi]=\frac{\Phi_{xxx}}{\Phi_x}-\frac{3}{2}\,
\Bigl (\frac{\Phi_{xx}}{\Phi_x}\Bigr )^2
\eeq
is the Schwarzian 
derivative,
and we have denoted
$x=t_1$, $y=t_2$, $t=t_3$. This is the continuous 
Schwarzian KP equation in the which first appeared 
in \cite{Weiss83} in the context of Painlev\'e analysis
as the 
singularity manifold equation associated with the KP equation.
Since it was obtained from equation (\ref{s5}), it is invariant
under the fractional-linear transformations (\ref{s6}) by
construction. This fact can be also checked directly from
equation (\ref{s10}). Indeed, the Schwarzian derivative is known
to be invariant, and invariance of the other terms can be easily
checked for the transformations 
$\Phi \to a\Phi$, $\Phi \to \Phi +c$ and $\Phi \to \Phi^{-1}$.

Representing the $\Phi$-function defined by (\ref{s4}) as
\beq\label{s11}
\Phi ({\bf t})=\frac{\tilde \tau ({\bf t})}{\tau ({\bf t})},
\eeq
one can show that the transition $\tau \to \tilde \tau$ is a
BD transformation, i.e. $\tilde \tau ({\bf t})$ is another
solution of the KP hierarchy. This can be proved 
in the following way. From (\ref{s2}) it follows that
$$
\Phi_1-\Phi_2=(a_1^{-1}-a_2^{-1})\frac{\tau_{12}\tau}{\tau_1\tau_2}\,
\phi_{12}\phi^*.
$$
Plugging here
(\ref{mkp9}), (\ref{mkp9a}) and (\ref{s11}), we obtain the relation
$$
\tilde \tau_\alpha \tau_\beta -\tilde \tau_\beta 
\tau_\alpha =(a_\alpha^{-1}-a_\beta^{-1})
\hat \tau_{\alpha \beta}\check \tau ,
$$
where $\alpha , \beta$ may take values $1,2,3$. The three 
equations in the matrix form can be written as the following
linear system for $\tau_1 , \tau_2 , \tau_3$:
$$
\left ( \begin{array}{ccc}
0 & -\tilde \tau_3 & \tilde \tau_2
\\ 
\tilde \tau_3 & 0 & -\tilde \tau_1
\\ 
-\tilde \tau_2 & \tilde \tau_1 & 0
\end{array}
\right )
\left ( \begin{array}{c}
\tau_1 \\ \tau_2 \\ \tau_3 
\end{array}
\right )
=
\check \tau
\left ( \begin{array}{c}
(a_2^{-1}-a_3^{-1})\hat \tau_{23}
\\ 
 (a_3^{-1}-a_1^{-1})\hat \tau_{13}
 \\ 
 (a_1^{-1}-a_2^{-1})\hat \tau_{12}
\end{array}
\right ).
$$
The $3\times 3$ skew-symmetric matrix in the left-hand 
side is degenerate, and the eigenvector with zero eigenvalue
is
$(\tilde \tau_1, \tilde \tau_2 , \tilde \tau_3)$.
Therefore, for existence of 
nonzero solutions this vector must be orthogonal to the 
vector in the right-hand side,
i.e., the relation
$$
a_1a_{23} \hat \tau_{23}\tilde \tau_1 +
a_2a_{31} \hat \tau_{31}\tilde \tau_2
+ a_3a_{12} \hat \tau_{12}\tilde \tau_3=0
$$
holds. If $a_3 =\infty$, it reads
$$
a_1\hat \tau_{2}\tilde \tau_1 
-a_2 \hat \tau_{1}\tilde \tau_2
=a_{12} \hat \tau_{12}\tilde \tau .
$$
Let $(\alpha \beta \gamma )$ be any 
cyclic permutation of $(123)$, then writing these 
equations 
for $\alpha \beta$ with the overall shift
$n_{\gamma}\to n_{\gamma}+1$, multiplying both sides by 
$\tilde \tau_{\alpha \beta}$ and summing the results 
over the three cyclic permutations, we obtain
the 3-term bilinear equation for $\tilde \tau$.

The BD transformation $\tau \to \tilde \tau$ is known in the
literature as the binary BD transformation \cite{OS93,O93}.
From (\ref{s1}) and (\ref{s4}) one can represent the tau-function
$\tilde \tau ({\bf t})$ explicitly as a double integral:
\beq\label{s12}
\tilde \tau ({\bf t})=\int_{\CCC} \!  \int_{\CCC}
e^{\xi ({\bf t}, z)-\xi ({\bf t}, w)}\, 
\frac{\rho (z)\rho^*(w)}{z-w} \, \tau \Bigl (
{\bf t}+[w^{-1}]-[z^{-1}]\Bigr ) d^2 z d^2 w.
\eeq
It easy to see that any binary BD transformation can be represented as
a composition of one forward and one backward BD 
transformations ${\bf B}$,
${\bf B}^*$ introduced in 
section \ref{section:BacklundKP}.

\section{Equations of the BKP type}
\label{section:BKP}

This section is devoted to the BKP hierarchy, its discrete
version and BD transformations. The general references are
\cite{DJKM82}, see also \cite{DJKM82a}--\cite{Z21}.

\subsection{The BKP hierarchy and its discretization}

The set of independent variables of the BKP hierachy is the
infinite set ${\bf t}=\{t_1, t_3, t_5, \ldots \}$, where the times
carry only odd indices\footnote{In this section, 
we denote the set of times by the same
bold letter ${\bf t}$ as for the KP hierarchy and hope that 
this will not cause any misunderstanding since the KP hierarchy
will not be discussed in this section.}.

Let $\tau ({\bf t})=\tau (t_1, t_3, t_5, \ldots )$ be the tau-function
of the BKP hierarchy.
The generating bilinear equation for the BKP tau-function
has the form
\beq\label{bb1}
\oint_{C_{\infty}}\Bigl (e^{\xi_{\rm o} ({\bf t}-{\bf t}', z)}
-1\Bigr )
\tau ({\bf t}-2[z^{-1}]_{\rm o})\tau ({\bf t}'+2[z^{-1}]_{\rm o})\, 
\frac{dz}{z} =0,
\eeq
where 
\beq\label{xib}
\xi_{\rm o} ({\bf t}, z) =\sum_{k\geq 1, \, {\rm odd}} t_k z^{k},
\eeq
\beq\label{shiftb}
\begin{array}{l}
{\bf t}\pm 2[z^{-1}]_{\rm o}
=\Bigl \{ t_1\pm 2z^{-1}, \,
t_3\pm \frac{2}{3}z^{-3}, t_5\pm \frac{2}{5}z^{-5}\Bigr \},
\end{array}
\eeq
and $C_{\infty}$ is a big circle around $\infty$.
This equation is valid for all ${\bf t}$ and ${\bf t}'$.
Note
that in the BKP case $[-a^{-1}]_{\rm o}=-[a^{-1}]_{\rm o}$.

Choosing ${\bf t}-{\bf t}'$ in a special way, one can 
derive from (\ref{bb1}) various bilinear 
equations of the Hirota-Miwa type. For any integer $m\geq 3$
and $m$ distinct points $a_i$ 
set
$$
{\bf t}-{\bf t}'=2\sum_{i=1}^m [a_i^{-1}], \quad
\mbox{then} \quad
e^{\xi_{\rm o} ({\bf t}-{\bf t}', z)}=\prod_{i=1}^m 
\frac{a_i+z}{a_i-z}.
$$
With this substitution, 
the residue calculus converts equation (\ref{bb1}) into
the following form:
\beq\label{bb2}
\begin{array}{l}
\displaystyle{
\sum_{k=1}^m \prod_{i=1, \neq k}^m \frac{a_i+a_k}{a_i-a_k}
\, \tau \Bigl ({\bf t}+2\!\! 
\sum_{l=1, \, \neq k}[a_l^{-1}]_{\rm o}\Bigr )
\tau \Bigl ({\bf t}+2[a_k^{-1}]_{\rm o}\Bigr )}
\\ \\
\phantom{aaaaaaaaaaaaaaaaaaaaaaaaaa}\displaystyle{
=\frac{1}{2}(1-(-1)^m)\tau ({\bf t})
\tau \Bigl ({\bf t}+2\sum_{l=1}[a_l^{-1}]_{\rm o}\Bigr ).}
\end{array}
\eeq
In particular, at $m=3$ we have the equation
\beq\label{bb3}
\begin{array}{l}
\tilde a_{12}\tau \Bigl ({\bf t}+2[a_1^{-1}]_{\rm o}
+2[a_2^{-1}]_{\rm o}\Bigr )
\tau \Bigl ({\bf t}+2[a_3^{-1}]_{\rm o}\Bigr ) 
\\ \\
\phantom{aaaaaaa}
+\, \tilde a_{23}\tau \Bigl ({\bf t}+2[a_2^{-1}]_{\rm o}
+2[a_3^{-1}]_{\rm o}\Bigr )
\tau \Bigl ({\bf t}+2[a_1^{-1}]_{\rm o}\Bigr ) +
\\ \\
\phantom{aaaaaaaaaaaaaa}
+\, \tilde a_{31}\tau \Bigl ({\bf t}+2[a_3^{-1}]_{\rm o}
+2[a_1^{-1}]_{\rm o}\Bigr )
\tau \Bigl ({\bf t}+2[a_2^{-1}]_{\rm o}\Bigr ) 
\\ \\
\phantom{aaaaaaaaaaaaaaaaaaaaa}
+\, \tilde a_{12}\tilde a_{23}\tilde a_{31}
\tau \Bigl ({\bf t}+2[a_1^{-1}]_{\rm o}
+2[a_2^{-1}]_{\rm o}+2[a_3^{-1}]_{\rm o}\Bigr )
\tau ({\bf t})=0,
\end{array}
\eeq
where
\beq\label{bb4}
\tilde a_{ij}=\frac{a_i-a_j}{a_i+a_j}.
\eeq
Note that the simplest solution is $\tau ({\bf t})=1$.
This follows from the easily verified identity
$$
\tilde a_{12}+\tilde a_{23}+\tilde a_{31}+
\tilde a_{12}\tilde a_{23}\tilde a_{31}=0.
$$
Equation (\ref{bb3}) first appeared in the work
\cite{Miwa82}.
As is proved in \cite{Shigyo13}, it is in fact
equivalent to the whole BKP hierarchy. The proof is based on the same
ideas as that for the KP hierarchy but one should deal with
Pfaffians instead of determinants.

Equation (\ref{bb3}) can be also understood as a fully
discrete BKP equation. For
any function $f({\bf t})$ set
\beq\label{g4b}
f(s_1, s_2, s_3)\equiv f \Bigl ({\bf t}+2s_1[a_1^{-1}]_{\rm o}
+2s_2[a_2^{-1}]_{\rm o}
+2s_3[a_3^{-1}]_{\rm o}\Bigr ), \qquad s_i\in \ZZ ,
\eeq
similarly to (\ref{g4}), and employ the
same short-hand notation as in (\ref{g5}). Then equation (\ref{bb3})
acquires the form
\beq\label{bb5}
\tilde a_{12}\tau_{12}\tau_3 +
\tilde a_{23}\tau_{23}\tau_1 +
\tilde a_{31}\tau_{31}\tau_2 +
\tilde a_{12}\tilde a_{23}\tilde a_{31}\tau_{123}\tau =0.
\eeq
As in the KP case, if one considers this equation as a discrete
one, without any concern about analytic properties of solutions,
values of the coefficients are not that important because they
can be made arbitrary by means of the simple transformation 
given in section \ref{section:KP}.

The higher functional relations of the BKP hierarchy (BKP analogs
of equations (\ref{mkp13}) and (\ref{mkp13a})) were derived in
the work \cite{Shigyo13} as corollaries of 
equations (\ref{bb2}). In the most general and suggestive form, 
they can be written in terms of Pfaffians
of the skew-symmetric matrices $\tilde a_{ij}$ (\ref{bb4}) and
\beq\label{M}
M_{ij}^{\pm}=\tilde a_{ij}\, \frac{\tau \Bigl (
{\bf t}\pm 2[a_i^{-1}]_{\rm o}
\pm 2[a_j^{-1}]_{\rm o}\Bigr )}{\tau ({\bf t})}, \quad
i,j =1, \ldots , 2m
\eeq
in the following way:
\beq\label{M1}
\begin{array}{l}
\displaystyle{
\frac{\tau \Bigl ({\bf t}\pm 
2\sum\limits_{i=1}^{2m}[a_i^{-1}]_{\rm o}\Bigr )}{\tau ({\bf t})}=
\frac{\mbox{Pf}_{1\leq i,j\leq 2m}
(M_{ji}^{\pm})}{\mbox{Pf}_{1\leq i,j\leq 2m}
(\tilde a_{ji})}.}
\end{array}
\eeq
Some information about Pfaffians is collected in Appendix A.
The Pfaffian downstairs can be explicitly evaluated:
\beq\label{M3}
\mbox{Pf}
(\tilde a_{ji})=\prod_{i>j}^{2m}\tilde a_{ij}=
\prod_{i>j}^{2m}\frac{a_i-a_j}{a_i+a_j}=:
\Delta^*_{2m}(a_1, \ldots , a_{2m}),
\eeq
and regarded as a Pfaffian 
analog of the Vandermonde determinant. Equation (\ref{M1}) is
written for an even number of the points $a_i$. To obtain a similar
equation for an odd number of them, equal to $2m-1$, one should
simply tend $a_{2m}$ in (\ref{M1}) to $\infty$. 
In this limit, the skew-symmetric matrices still remain to be
of size $2m \times 2m$ but
the matrix elements $\tilde a_{2m, i}$ and
$\tilde a_{j, 2m}$ become equal to $\pm 1$. 
In particular,
equation (\ref{bb3}) is the simplest particular case of (\ref{M1})
corresponding to the choice $m=2$, $a_{4}=\infty$.

The BA function is defined by the formula
\beq\label{bb6}
\psi ({\bf t}, z)=e^{\xi_{\rm o}({\bf t}, z)}
\frac{\tau ({\bf t}-2[z^{-1}]_{\rm o})}{\tau ({\bf t})}.
\eeq
Note that the dual BA function, if defined as in (\ref{mkp1}),
is just $\psi ({\bf t}, -z)$ and thus is not an independent 
function. In fact another definition of the dual BA function
is possible
(with the help of the dressing operator; 
see, for example, \cite{Z21}) but it also gives a function
that is not independent from the function $\psi ({\bf t}, z)$
defined by equation (\ref{bb6}). As in the KP case, the BA function
satisfies linear equations. The simplest one follows from
equation (\ref{bb3}) rewritten in a special way. Namely, 
set $a_3=z$, then, using the definition (\ref{bb6}),
one is able, after some simple transformations, to 
represent equation (\ref{bb3}) in the form of a
linear difference equation for the BA function. In the short-hand
notation it reads:
\beq\label{bb7}
\psi_2 -\psi_1=\tilde a_{12}\, \frac{\tau \tau_{12}}{\tau_1\tau_2}
\, (\psi_{12}-\psi ).
\eeq
Note that comparing to the analogous equations (\ref{mkp4})
for the discrete KP hierarchy it has four terms, rather that three,
associated with four vertices of a  
plaquette of the square lattice.

Like in the KP case, the BA function depends on the spectral
parameter $z$, and enters the $z$-independent 
equation (\ref{bb7}) linearly.
This allows one to introduce a general solution $\varphi ({\bf t})$ 
to the linear
equation (\ref{bb7}) in the same way as in (\ref{mkp2a}), i.e.,
by integrating the BA function with an arbitrary function 
(or a distribution) of $z$:
\beq\label{bb8}
\varphi ({\bf t})=\int_{\CCC}\psi ({\bf t}, z)\rho (z)d^2 z.
\eeq
(Again, we assume that $\rho (z)$ has a compact support.)
So, the function $\varphi ({\bf t})$ satisfies the linear
equation (\ref{bb7}):
\beq\label{bb7a}
\varphi_2 -\varphi_1=\tilde a_{12}\, 
\frac{\tau \tau_{12}}{\tau_1\tau_2}
\, (\varphi_{12}-\varphi ).
\eeq
The BA function is just one special solution of this equation.

Equations of the form (\ref{bb7a}) written for each pair
of the discrete variables $s_1, s_2, s_3$ form an overdetermined 
linear system whose compatibility condition is just the bilinear
equation (\ref{bb5}). To see this, we rewrite (\ref{bb7a}) 
in the form
\beq\label{bb7b}
\varphi_{\alpha \beta}=\varphi +\tilde a_{\alpha \beta}^{-1}\,
\frac{\tau_{\alpha}\tau_{\beta}}{\tau \tau_{\alpha \beta}}
(\varphi_{\beta} -\varphi_{\alpha})
\eeq
for $\alpha, \beta =1,2,3$ ($\alpha \neq \beta$) 
and shift the third 
variable $s_{\gamma}$ (which is the same in 
all terms in (\ref{bb7b}))
by 1: $s_{\gamma}\to s_{\gamma}+1$. (As before,
$(\alpha \beta \gamma )$ is any 
cyclic permutation of $(123)$.) Then (\ref{bb7b}) reads
\beq\label{bb7c}
(\varphi_{\alpha \beta})_{\gamma}
=\varphi_{\gamma} +\tilde a_{\alpha \beta}^{-1}\,
\frac{\tau_{\alpha \gamma}\tau_{\beta \gamma}}{\tau_{\gamma}
\tau_{\alpha \beta \gamma}}
(\varphi_{\beta \gamma} -\varphi_{\alpha \gamma}).
\eeq
Compatibility of the three equations (\ref{bb7b})
means that 
\beq\label{bb7d}
(\varphi_{12})_3 =
(\varphi_{23})_1=(\varphi_{31})_2.
\eeq
Expressing $\varphi_{\beta \gamma}$
and $\varphi_{\alpha \gamma}$ in the right-hand side of (\ref{bb7c})
through $\varphi , 
\varphi_{\alpha}, \varphi_{\beta}, \varphi_{\gamma}$ 
with the help of equations (\ref{bb7b}), and substituting the results 
into (\ref{bb7d}),
one can see that $\varphi$ cancels from these equalities while
the other terms give the compatibility conditions
\beq\label{bb9}
R(\varphi_1 -\varphi_2)=R(\varphi_2 -\varphi_3)=
R(\varphi_3 -\varphi_1)=0,
\eeq
where
$$
R=\tau \tau_{123}+(\tilde a_{12}\tilde a_{31})^{-1}
\tau_1 \tau_{23}+(\tilde a_{23}\tilde a_{12})^{-1}
\tau_2 \tau_{31}+(\tilde a_{31}\tilde a_{23})^{-1}
\tau_3 \tau_{12}.
$$
Therefore, the compatibility requires that $R=0$.
This is precisely equation (\ref{bb5}).

\subsection{The discrete modified BKP hierarchy}

Our next goal is to derive an equation that would contain 
the $\varphi$-functions at neighboring points of 
a 3D lattice only,
such that the tau-function does not participate in it explicitly.
It is thus going to be a BKP analogue of the discrete
mKP equation (\ref{mkp5}), and by analogy, might be called 
the discrete modified BKP (mBKP) equation. However, we will see
that the form of this equation suggests 
to simultaneously regard it as a BKP analogue of the 
discrete Schwarzian KP equation (\ref{s5}).
To derive it, we fix three parameters $a_1, a_2, a_3$, with
the associated discrete variables being $s_1, s_2, s_3$,
and write equation (\ref{bb7a}) for each pair of
indices $\alpha , \beta =1,2,3$ in the form
\beq\label{bb9a}
\frac{\varphi_{\beta}-\varphi_{\alpha}}{\varphi_{\alpha \beta}-
\varphi}=\tilde a_{\alpha \beta}\, 
\frac{\tau \tau_{\alpha \beta}}{\tau_{\alpha}\tau_{\beta}}.
\eeq
Let $\{\alpha , \beta , \gamma \}$ be any cyclic permutation
of the indices $\{1,2,3\}$. Shifting $s_{\gamma}\to s_{\gamma}+1$
in (\ref{bb9a}) (with $\gamma \neq \alpha , \beta$),
it is easy to observe that the combination
$$
\left. \Bigl (
\frac{\varphi_{\beta \gamma}-
\varphi_{\alpha \gamma}}{\varphi_{\alpha \beta \gamma}-\varphi_{\gamma}}
\Bigr )\right /
\Bigl (\frac{\varphi_{\beta}-
\varphi_{\alpha}}{\varphi_{\alpha \beta }-\varphi}
\Bigr )
=\frac{\tau_{\alpha}\, 
\tau_{\beta}\, \tau_{\gamma}\, \tau}{\tau_{\alpha \beta}
\tau_{\beta \gamma}\tau_{\gamma \alpha}\tau_{\alpha \beta \gamma}}
$$
is symmetric with respect to all permutations of the
indices. Therefore, we obtain the discrete equations for the 
$\varphi$-function of the form
\beq\label{bb10}
\frac{(\varphi_{23}-\varphi_{13})
(\varphi_{12}-\varphi )}{(\varphi_{123}-\varphi_{3})
(\varphi_{2}-\varphi_{1})}=
\frac{(\varphi_{31}-\varphi_{21})
(\varphi_{23}-\varphi )}{(\varphi_{123}-\varphi_{1})
(\varphi_{3}-\varphi_{2})}=
\frac{(\varphi_{12}-\varphi_{23})
(\varphi_{13}-\varphi )}{(\varphi_{123}-\varphi_{2})
(\varphi_{1}-\varphi_{3})}
\eeq
or
\beq\label{bb11a}
\begin{array}{l}
\displaystyle{
\frac{(\varphi_{12}-\varphi )(\varphi_{1}-\varphi_{123})
(\varphi_{23}-\varphi_{13} )
(\varphi_{3}-\varphi_{2})}{(\varphi_{23}-\varphi )
(\varphi_{3}-\varphi_{123})
(\varphi_{31}-\varphi_{21} )
(\varphi_{2}-\varphi_{1})}=1,}
\\ \\
\displaystyle{
\frac{(\varphi_{12}-\varphi )(\varphi_{3}-\varphi_{2})
(\varphi_{1}-\varphi_{123} )
(\varphi_{23}-\varphi_{13})}{(\varphi_{2}-\varphi_{1} )
(\varphi_{32}-\varphi )
(\varphi_{3}-\varphi_{123} )
(\varphi_{13}-\varphi_{12})}=1.}
\end{array}
\eeq
Discrete equations of this form first appeared in \cite{Schief03},
where their geometric meaning was also clarified.
It is easy to see that they are invariant under 
fractional-linear (M\"obius)
transformations of the form (\ref{s6}). Therefore, they can be
simultaneously regarded as BKP analogues of the discrete mKP equation
(because they are written for the wave functions, like in the mKP case)
and of the discrete Schwarzian KP equation (because they are
invariant under the M\"obius transformations). 
It should be also noted that these equations do not contain 
the ''lattice spacing'' parameters $a_i$ explicitly, and hence
should be invariant under any permutation of the variables
$s_1, s_2, s_3$. Indeed, under the permutations, the two
equations in (\ref{bb11a}) are transformed into each other.
So, for the BKP hierarchy we have two equations for one and the same
function rather than one. However, as it follows
from the construction, these equations are compatible and have a lot of
common solutions. 

\subsection{Continuum limit of the discrete modified BKP 
equation}
\label{section:continuum-limit-BKP}

As before, the continuum limit of equations (\ref{bb11a}) should be
taken in steps. At the first step we tend $a_1 \to \infty$
keeping the other two parameters fixed. This gives:
\beq\label{cb1}
\begin{array}{l}
\displaystyle{
\p_{t_1}\log \frac{\varphi_{23}-\varphi}{\varphi_3 -\varphi_2}
=\Delta_3 \Bigl (\frac{\p_{t_1}
(\varphi_2+\varphi )}{\varphi_2-\varphi}\Bigr ),}
\\ \\
\displaystyle{
\Delta_2 \Bigl (\frac{\p_{t_1}
(\varphi_3+\varphi )}{\varphi_3-\varphi}\Bigr )
=\Delta_3 \Bigl (\frac{\p_{t_1}
(\varphi_2+\varphi )}{\varphi_2-\varphi}\Bigr ),}
\end{array}
\eeq
where $\Delta_3 f(s_1, s_2 , s_3)=f(s_1, s_2 , s_3+1)$.
The further limit as $a_2\to \infty$ is obtained by expansion 
of these equations up to the first non-vanishing order in
$a_2^{-1}\to 0$. The two equations in (\ref{cb1}) give the 
same result:
\beq\label{cb2}
6\, \p_{t_1}\Bigl (\frac{\p_{t_1}
(\varphi_3+\varphi )}{\varphi_3-\varphi}\Bigr )=
\Delta_3 \Bigl 
(\frac{\p_{t_1}^{3}\varphi -\p_{t_3}\varphi }{\p_{t_1}\varphi }
\Bigr ).
\eeq
Finally, we tend $a_3\to \infty$ in (\ref{cb2}). To obtain a
non-trivial equation from (\ref{cb2}) in this limit, one should
expand the both sides up to the order $a_3^{-3}$. The necessary 
calculations are direct but rather long. 
In the variables $x=t_1, y=t_3, t=t_5$ the equation reads:
\beq\label{cb3}
\begin{array}{l}
\displaystyle{
-\frac{9}{5}\, \frac{\varphi_t}{\varphi_x}=\frac{1}{5}\,
\Bigl ((S[\varphi ])_{xx}+4 S[\varphi ]\Bigr )}
\\ \\
\phantom{aaaaaaaa}\displaystyle{
+\, \frac{\varphi_{xy}\varphi_{xx}}{\varphi_x^2}
-2\frac{\varphi_{yxx}}{\varphi_x} -\Bigl (
\frac{\varphi_y}{\varphi_x}\Bigr )^2 +
\frac{\varphi_{y}\varphi_{xxx}}{\varphi_x^2}-
\frac{\varphi_{y}\varphi^2_{xx}}{\varphi_x^3}
+\p_x^{-1}\Bigl (\frac{\varphi_{xxx}-\varphi_y}{\varphi_x}
\Bigr ),}
\end{array}
\eeq
where $S[\varphi ]$ is the Schwarzian derivative given by
(\ref{s10a}). 
Let us mention that
$$
(S[\varphi ])_{xx}+4 S[\varphi ]=
\frac{\varphi^{\rm V}}{\varphi '}-5\,
\frac{\varphi ^{''}\varphi^{\rm IV}}{(\varphi ')^2}+
5\Bigl (\frac{\varphi^{''}}{\varphi '}\Bigr )^2 
\frac{\varphi^{'''}}{\varphi'},
$$
where prime means the $x$-derivative.
The M\"obius-invariance of equation (\ref{cb3}) can be directly
verified. As was already mentioned, equation (\ref{cb3}) is
simultaneously the Schwarzian analog of the BKP equation.

\subsection{B\"acklund-Darboux transformations}
\label{section:BDBKP}

Equation
(\ref{mkp9}) suggests to define BD transformations
for the tau-function of the BKP hierarchy following the analogy
with the KP case. That is to say, let us represent a solution to
the linear equation (\ref{bb7a}) in the form
\beq\label{B1}
\varphi ({\bf t})=\frac{\hat \tau ({\bf t})}{\tau ({\bf t})},
\eeq
where $\hat \tau$ is the BD-transformed tau-function
(below we shall see that it is a BKP tau-function indeed).
From (\ref{bb8}) we see that
\beq\label{B2}
\hat \tau ({\bf t})=\int_{\CCC} e^{\xi_{\rm o}({\bf t}, z)}
\tau \Bigl ({\bf t}-2[z^{-1}]_{\rm o}\Bigr )\rho (z)d^2 z.
\eeq

To prove that $\hat \tau$ is a BKP tau-function (i.e. that it
obeys equation (\ref{bb5}) if $\tau$ does so), we substitute
(\ref{B1}) into equation (\ref{bb7a}). After cancellations,
we obtain the following equation
connecting $\tau$ and $\hat \tau$:
\beq\label{B3}
\hat \tau_2 \tau_1 -\hat \tau_1 \tau_2 =
\tilde a_{12}(\hat \tau_{12} \tau -\hat \tau \tau_{12}).
\eeq
As before, let $\{\alpha , \beta , \gamma \}$ be any cyclic permutation
of the indices $\{1,2,3\}$. For any cyclic permutation,
write the equation of the form (\ref{B3}):
\beq\label{B4}
\hat \tau_\beta \tau_\alpha -\hat \tau_\alpha \tau_\beta =
\tilde a_{\alpha \beta}(\hat \tau_{\alpha \beta} 
\tau -\hat \tau \tau_{\alpha \beta})
\eeq
and multiply its both sides by $\hat \tau_\gamma$. Summing the 
three equations obtained in this way, we arrive at the relation
\beq\label{B5}
\tau \Bigl (\tilde a_{12}\hat \tau_{12}\hat \tau_3 +
\tilde a_{23}\hat \tau_{23}\hat \tau_1 +
\tilde a_{31}\hat \tau_{31}\hat \tau_2 \Bigr )=
\hat 
\tau \Bigl (\tilde a_{12}\hat \tau_{3} \tau_{12} +
\tilde a_{23}\hat \tau_{1} \tau_{23} +
\tilde a_{31}\hat \tau_{2} \tau_{31} \Bigr ).
\eeq
In a similar way, multiplying the 
both sides of (\ref{B4}) 
by $\tau_\gamma$ and summing the resulting equations, we arrive at
\beq\label{B6}
\tau \Bigl (\tilde a_{12}\hat \tau_{12}\tau_3 +
\tilde a_{23}\hat \tau_{23}\tau_1 +
\tilde a_{31}\hat \tau_{31}\tau_2 \Bigr )=
\hat 
\tau \Bigl (\tilde a_{12}\tau_{3} \tau_{12} +
\tilde a_{23}\tau_{1} \tau_{23} +
\tilde a_{31}\tau_{2} \tau_{31} \Bigr ).
\eeq
Next, we proceed by shifting $s_{\gamma}\to s_{\gamma}+1$ in 
(\ref{B4}) which results in
\beq\label{B7}
\hat \tau_{\beta \gamma} \tau_{\alpha \gamma} -
\hat \tau_{\alpha \gamma} \tau_{\beta \gamma}=
\tilde a_{\alpha \beta}(\hat \tau_{\alpha \beta \gamma} 
\tau_{\gamma} -\hat \tau_{\gamma} \tau_{\alpha \beta \gamma}).
\eeq
Multiplying both sides of this relation by 
$\hat \tau_{\alpha \beta}$ and summing the resulting equations,
we get
\beq\label{B8}
\tau_{123} \Bigl (\tilde a_{12}\hat \tau_{12}\hat \tau_3 +
\tilde a_{23}\hat \tau_{23}\hat \tau_1 +
\tilde a_{31}\hat \tau_{31}\hat \tau_2 \Bigr )= 
\hat \tau_{123} \Bigl (\tilde a_{12}\tau_{3} \hat \tau_{12} +
\tilde a_{23}\tau_{1} \hat \tau_{23} +
\tilde a_{31}\tau_{2} \hat \tau_{31} \Bigr ).
\eeq
At last, multiplying the both sides of (\ref{B7}) by 
$\tau_{\alpha \beta}$ and summing the three equations,
we obtain the relation
\beq\label{B9}
\hat \tau_{123} \Bigl (\tilde a_{12}\tau_{12}\tau_3 +
\tilde a_{23}\tau_{23}\tau_1 +
\tilde a_{31}\tau_{31}\tau_2 \Bigr )=
\hat 
\tau_{123} \Bigl (\tilde a_{12}\hat \tau_{3} \tau_{12} +
\tilde a_{23}\hat \tau_{1} \tau_{23} +
\tilde a_{31}\hat \tau_{2} \tau_{31} \Bigr ).
\eeq
Now, equation (\ref{bb5}) states that the right-hand side of
(\ref{B6}) (and, therefore, its left-hand side as well) 
is equal to
$$
-\tilde a_{12}\tilde a_{23}\tilde a_{31} \tau \tau_{123} \hat \tau .
$$
Comparing with (\ref{B8}), we see that it holds
$$
\tilde a_{12}\hat \tau_{12}\hat \tau_3 +
\tilde a_{23}\hat \tau_{23}\hat \tau_1 +
\tilde a_{31}\hat \tau_{31}\hat \tau_2 = 
-\tilde a_{12}\tilde a_{23}\tilde a_{31} \hat \tau \hat \tau_{123},
$$
which is the four-term bilinear equation (\ref{bb5})
for the BKP tau-function. We have proved that the transformation
$\tau \to \hat \tau$ of the form (\ref{B2}) sends 
any solution to the BKP hierarchy to another
solution, i.e., it is indeed a BD
transformation.

Similarly to the KP case, given a sequence
of functions $\{\rho_1 (z), \rho_2(z), \ldots \}$,
we can repeat the transformation of the form (\ref{B2})
many times, defining the chain of BD transformations
$\tau (0, {\bf t})=\tau ({\bf t}) \to \tau (1, {\bf t})
\to \tau (2, {\bf t})\to \ldots$ by the recursion relation
\beq\label{B10}
\tau (n+1, {\bf t})=\int_{\CCC} e^{\xi_{\rm o}({\bf t}, z)}
\rho_n (z)\tau \Bigl (n, {\bf t}-2[z^{-1}]_{\rm o}
\Bigr )d^2 z, \quad n\geq 0.
\eeq
Equation (\ref{B3}) for the chain can be written in the form
\beq\label{B12}
\tau_2(n+1) \tau_1(n) -\tau_1(n+1) \tau_2(n) =
\tilde a_{12}\Bigl (\tau_{12}(n+1) \tau (n) -
\tau (n+1)\tau_{12}(n)\Bigr ),
\eeq
where $n=0,1,2, \ldots \, $.
Remarkably, equation (\ref{B12}) has essentially the same 
structure as the four-term bilinear equation (\ref{bb5}) 
for the BKP tau-function\footnote{Values of
the numerical constants in front of each term are different
in these equations but this discrepancy is
not that very important since they can be changed by a 
simple redefinition
of the tau-function that consists in multiplying it by exponential
function of a quadratic form in the independent variables.}.
Namely, the discrete ``variable'' $n$ that numbers 
sequential steps of the BD chain in equation (\ref{B12})
enters in the same way as the third discrete variable $s_3$
in (\ref{bb5}).

The BD transformations, as defined above, are of the forward type.
A crucial difference, comparing to the KP case, is that 
any analogs of backward transformations do not exist in the
BKP setup. There exist different possibilities to extend
the chain to the backward direction so that equation (\ref{B12})
remains valid. The simplest one is to
put $\tau (-n)=\tau (0)$ for all $n\geq 1$. 
Note, however, that equation (\ref{B12}) is invariant
under the change $n \to -n$. Therefore, we can extend the chain
to the backward direction in the symmetric way 
by putting $\tau (-n)=\tau (n)$ for
$n\geq 0$, or $\tau (-n)=\tau (n-1)$
for $n\geq 0$. 

Explicitly solving recurrence relation (\ref{B10}), we have:
\beq\label{B11}
\tau (n, {\bf t})=\int_{\CCC}\! \ldots \! \int_{\CCC}
\Bigl (\prod_{i=1}^{n} d^2 z_i e^{\xi_{\rm o}({\bf t}, z_i)}
\rho_i (z_i) \Bigr ) \Bigl (
\prod_{i>j}^{n}\frac{z_i -z_j}{z_i +z_j}\Bigr ) \,
\tau \Bigl (0, {\bf t}-2\sum_{l=1}^n [z_i^{-1}]_{\rm o}\Bigr )
\eeq
which is valid for both even and odd $n$.
With the help of equation (\ref{M1}), the right-hand side 
can be expressed in terms of the Pfaffian:
\beq\label{B11a}
\frac{\tau (2m, {\bf t})}{\tau ({\bf t})}=
\mbox{Pf}_{ij}\Bigl [\int_{\CCC} \! \int_{\CCC}
\rho_i({\bf t}, z_i)\rho_j({\bf t}, 
z_j)\, M^-_{ji}\, d^2 z_i d^2z_j \Bigr ],
\quad 1\leq i,j\leq 2m,
\eeq
where the skew-symmetric 
matrix $M_{ij}^-$ is defined in (\ref{M1}), i.e.,
\beq\label{Ma}
M_{ij}^{\pm}=\tilde z_{ij}\, \frac{\tau \Bigl (
{\bf t}\pm 2[z_i^{-1}]_{\rm o}
\pm 2[z_j^{-1}]_{\rm o}\Bigr )}{\tau ({\bf t})}
\eeq
and for brevity we have denoted
$
\rho_i({\bf t}, z)=:
e^{\xi_{\rm o}({\bf t}, z)}\rho_i (z).
$
This is the BKP analogue of equation (\ref{b6}). Equation 
(\ref{B11a}) is written for $\tau (n)$ with an even non-negative 
$n=2m$ but can be easily adopted for the case of an odd $n=2m-1$.
For this, one should omit the terms
$\xi_{\rm o}({\bf t}, z_{2m})$ in the exponential
factors and formally put $\rho_{2m}(z)$ equal to the
(formal) delta-function with support at infinity after that. 
Informally, this means that
one should put $z_{2m}=\infty$ in the matrix $M_{ij}^-$.
The skew-symmetric matrix under Pfaffian is still of size
$2m\times 2m$ but the elements $(i, 2m)$ and $(2m, j)$ 
are represented as single integrals with the density
functions $\rho_i$, $\rho_j$ respectively rather than double ones.

Let us mention some important particular cases of equations
(\ref{B11}) and (\ref{B11a}). They 
simplify considerably if the initial 
tau-function is taken to be the trivial solution, i.e.
if one puts $\tau (0, {\bf t})=1$. In this case $M^-_{ij}=
\tilde z_{ij}$, and (\ref{B11a}) reads
\beq\label{B11b}
\frac{\tau (2m, {\bf t})}{\tau ({\bf t})}=
\mbox{Pf}_{ij}\left [\int_{\CCC} \! \int_{\CCC}
\rho_j({\bf t}, z)\, 
\Bigl (\frac{z-w}{z+w}\Bigr )\rho_i({\bf t}, w)
\, d^2 w d^2z \right ],
\quad 1\leq i,j\leq 2m,
\eeq
The special choice $\rho_j(z)=\delta ({\rm Im}\, z)
\theta ({\rm Re}\, z)  z^j
\omega (z)$, where the distribution
$\delta ({\rm Im}\, z)
\theta ({\rm Re}\, z)$ is the delta-function concentrated 
on the ray $\RR_{+}$, and $\omega (z)$ is some continuous function,
converts the multiple integral in (\ref{B11}) to
\beq\label{B11c}
\begin{array}{rcl}
\tau (n, {\bf t})&=&
\displaystyle{ \int_{\RRR_{+}}\! \ldots \! \int_{\RRR_{+}}
\prod_{i>j}^{n}\frac{x_i -x_j}{x_i +x_j}\,
\prod_{l=1}^{n} x_l^l e^{\xi_{\rm o}({\bf t}, x_l)}
\omega (x_l)  d x_l }
\\ && \\
&=& \displaystyle{\frac{1}{n!}
\int_{\RRR_{+}}\! \ldots \! \int_{\RRR_{+}}
\prod_{i>j}^{n}\frac{(x_i -x_j)^2}{x_i +x_j} \,
\prod_{l=1}^{n}  e^{\xi_{\rm o}({\bf t}, x_l)}
\omega (x_l)  d x_l }.
\end{array}
\eeq
This is the partition function of the 
Bures statistical ensemble whose connection with 
the BKP hierarchy was addressed in \cite{HL17} in some details.
Equation (\ref{B11b}) specialized to this case reads:
\beq\label{B11d}
\frac{\tau (2m, {\bf t})}{\tau ({\bf t})}=
\mbox{Pf}_{ij}\left [\int_{\RRR_{+}} \! \int_{\RRR_{+}}
x^i y^j \, 
\Bigl (\frac{x-y}{x+y}\Bigr )
\omega ({\bf t}, x)\omega ({\bf t}, y)
\, dx dy \right ],
\eeq
where
$\omega ({\bf t}, x)=:
e^{\xi_{\rm o}({\bf t}, x)}\omega (x)$. For odd $n=2m-1$ this
formula needs some modifications. The details can be found in
\cite{HL17}. 
Applications to Pfaffian point processes were discussed in
\cite{WL19}.

Choosing the density functions in (\ref{B11}) (with
$\tau (0, {\bf t})=1$) to be distributions
concentrated at a finite number of distinct 
isolated points in $\CC$, one obtains multi-soliton solutions
to the BKP hierarchy and their generalizations. For example,
one may take
\beq\label{B13}
\rho_i(z)=\delta^{(2)}(z-p_i)+\alpha_i \delta^{(2)}(z-q_i).
\eeq
Here we will not go into further details. For the 
construction of the soliton solutions to the BKP hierarchy
in terms of Pfaffians see \cite{Hirota,HT96}.

\subsection{Two and three chains of the 
B\"acklund-Darboux transformations}

Let us consider two BD transformations ${\bf B}_1$,
${\bf B}_2$ with the density
functions $\rho_1(z), \, \rho_2(z)$ with supports 
on some compact domains ${\sf D}_1$, ${\sf D}_2$
respectively. Applying first 
${\bf B}_1$, then ${\bf B}_2$, we have:
\beq\label{bc5}
\tau^{(12)}({\bf t})=
\int_{{\sf D}_2}\! dz_2^2 \, e^{\xi_{\rm o}({\bf t}, z_2)}
\rho_2(z_2)
\! \int_{{\sf D}_1} \! dz_1^2 \, e^{\xi_{\rm o}({\bf t}, z_1)}
\rho_2(z_1)\,
\frac{z_2-z_1}{z_2+z_1}\, \tau \Bigl ({\bf t}-
2[z_1^{-1}]_{\rm o}-2[z_2^{-1}]_{\rm o}\Bigr ).
\eeq
Because of the singularity at $z_1=-z_2$ the order of integrals
in general can not be interchanged. However, if we require that
\beq\label{bc6}
{\sf D}_1 \cap (-{\sf D}_2)=\emptyset ,
\eeq
then the singularity is absent, the integrals over $z_1$ and $z_2$ 
can be
interchanged, and we see that the opposite order of the transformations
(first ${\bf B}_2$, then ${\bf B}_1$) gives the same result as in
(\ref{bc5}) but with sign minus. So, in this case the 
transformations anticommute: 
${\bf B}_1 {\bf B}_2 =-{\bf B}_2 {\bf B}_1$. This (anti)commutation
law receives its natural interpretation in the fermionic 
construction based on neutral Fermi-operators (see section 
\ref{section:neutral} below).

Now, given $N$ sequences of density functions 
$\{\rho_1^{(\alpha )}(z), \rho_2^{(\alpha )}(z), \ldots \, \}$,
$\alpha =1, \ldots , N$ such that their supports satisfy
the conditions
${\sf D}_i^{(\alpha )}\cap (-{\sf D}_i^{(\beta )})=\emptyset$ 
for all $\alpha \neq \beta \,$\footnote{To satisfy these conditions,
it is enough to assume that the supports belong to the
right upper quadrant of the complex plane which we denote
as $\CC_{+}^{+}$ ($x\geq 0, \, y>0$).}, 
we can consider $N$ chains of the BD transformations in the way
explained in section
\ref{section:severalBacklundKP}. In the similar way, we can
introduce discrete variables $n_{\alpha}\in \ZZ$ that count 
the number of BD transformations made in the ``$\alpha$th direction''.
Below we use the short-hand 
notation introduced in that section: $\tau^{(\alpha \beta )}$
etc. for the tau-functions obtained by sequential application
of different BD transformations.
Using the integral representation of the BD transformations
and the basic bilinear equation (\ref{bb5}), it is not 
difficult to obtain the following equations:
\beq\label{bc7}
\tau_1^{(\alpha )}\tau^{(\beta )}-
\tau_1^{(\beta )}\tau^{(\alpha )}+\epsilon_{\alpha \beta}
\Bigl (\tau_1^{(\alpha \beta )}\tau^{}-
\tau_1 \tau^{(\alpha \beta )}\Bigr )=0,
\eeq

\beq\label{bc8}
\epsilon_{\alpha \beta}\tau^{(\alpha \beta )}\tau^{(\gamma )}
+ 
\epsilon_{\beta \gamma}\tau^{(\beta \gamma )}\tau^{(\alpha )}
+
\epsilon_{\gamma \alpha}\tau^{(\gamma \alpha)}\tau^{(\beta )}+
\epsilon_{\alpha \beta}\epsilon_{\beta \gamma}
\epsilon_{\gamma \alpha}
\tau^{(\alpha \beta \gamma )}\tau =0.
\eeq

\noindent
The last equation has the same structure as the basic
bilinear equation (\ref{bb5}), but with different coefficients
equal to $\pm 1$. This equation should be compared with the 
equation for the tau-function of the multi-component BKP
hierarchy\footnote{Its construction was very briefly 
outlined in \cite{DJKM82} and requires more attention.}.

\section{The approach based on free fermions}
\label{section:fermions}

A very instructive and suggestive approach to 
integrable hierarchies such as KP and BKP is the operator technique 
based on the quantum field
theory of free massless fermions. It was suggested in the works by 
Kyoto school in 1983 \cite{DJKM83,JM83}. In particular,
the approach based on free fermions allows one to represent
the BD transformations in a very simple and clear form, making
all statements related to them almost obvious.

\subsection{Charged fermions (KP)}
\label{section:charged}

Let us begin with a brief remainder of how the KP hierarchy looks
like in terms of free fermions. For more details see the 
original papers \cite{DJKM83,JM83} and the detailed 
review \cite{AZ12}.

\subsubsection{The algebra of Fermi-operators and vacuum states}

Let $\boldpsi_n , \boldpsistar_{n}$, 
$n\in \ZZ$, be free Fermi-operators 
with the standard anti-commutation relations
\beq\label{fermop1}
[\boldpsi_n , \boldpsi_m ]_+ = [\boldpsistar_n, 
\boldpsistar_m]_+=0, \quad
[\boldpsi_n , \boldpsistar_m]_+=\delta_{mn}.
\eeq
The Fermi-operators carry a charge:
the charge of $\boldpsi$ is $+1$ and the charge of $\boldpsistar$
is $-1$.
They generate
an infinite dimensional Clifford algebra ${\cal A}$. 
Elements of the algebra are linear combinations
of all possible products of the charged
Fermi-operators $\boldpsi_n$, $\boldpsistar_n$ 
with different indices (monomials). 
They are multiplied in accordance with the anticommutation
relations (\ref{fermop1}).

We will also use the
generating series of the Fermi-operators defined as
\beq\label{ferm0}
\boldpsi (z)=\sum_{k\in \z}\boldpsi_k z^k, \quad \quad
\boldpsistar (z)=\sum_{k\in \z}\boldpsistar_k z^{-k}.
\eeq
They are 
fermionic operator fields in the complex plane $\CC$ of the
variable $z$. 

Next, we introduce the right vacuum state $\left |0\rbr$ as
a ``Dirac sea'', where all negative mode states are empty
and all non-negative ones are occupied:
$$
\boldpsi_n \rvac =0, \quad n< 0; \quad \quad \quad
\boldpsistar_n \rvac =0, \quad n\geq 0.
$$
Similarly, the dual (left ) vacuum state has the properties
$$
\lvac \boldpsistar_n  =0, \quad n< 0; \quad \quad \quad
\lvac \boldpsi_n  =0, \quad n\geq 0.
$$
With respect to the vacuum $\rvac$, the operators $\boldpsi_n$ with
$n<0$ and $\boldpsistar_n$ with $n\geq 0$ are annihilation operators
while the operators $\boldpsistar_n$ with $n<0$ and
$\boldpsi_n$ with $n\geq 0$ are creation operators (the latter
ones create holes).
We also need ``shifted'' Dirac vacua $\rvacn$ and $\lvacn$
defined as
\beq\label{vacdefr}
\rvacn = \left \{
\begin{array}{l}
\boldpsi_{n-1}\ldots \boldpsi_1 \boldpsi_0 \rvac , \,\,\,\,\, n> 0
\\ \\
\boldpsistar_n \ldots \boldpsistar_{-2}
\boldpsistar_{-1}\rvac , \,\,\,\,\, n<0,
\end{array} \right.
\eeq
\beq\label{vacdefl}
\lvacn = \left \{
\begin{array}{l}
\lvac \boldpsistar_{0}\boldpsistar_{1}\ldots 
\boldpsistar_{n-1} , \,\,\,\,\, n> 0
\\ \\
\lvac \boldpsi_{-1}\boldpsi_{-2}\ldots \boldpsi_{n} , \,\,\,\,\, n<0.
\end{array} \right.
\eeq
This definition implies that the charge of the right vacuum
$\rvacn$ is $n$, and the charge of the left vacuum 
$\lvacn$ is $-n$.

The vacuum expectation value $\lvac (\ldots )\rvac$ is a
Hermitian linear form on the Clifford algebra ${\cal A}$ fixed
by the normalization
$
\left. \lvac \! 0 \right > =1.
$
Then, from the commutation relations (\ref{fermop1}) 
and definitions of the ``shifted'' Dirac vacua (\ref{vacdefr}), (\ref{vacdefl}) it follows that 
$\left. \lvacn \! n\right > =1$ for any $n$.
Bilinear combinations of fermions
satisfy the properties 
$\lvacn \boldpsi_i \boldpsi_j\rvacn = \lvacn 
\boldpsistar_i \boldpsistar_j \rvacn =0$
for all $i,j$ and
$$
\lvacn \boldpsi_i \boldpsistar_j\rvacn =\delta_{ij}\quad
\mbox{for $j<n$}, \quad \quad
\lvacn \boldpsi_i \boldpsistar_j\rvacn =0 \quad \mbox{for $j\geq n$}.
$$
The expectation value of any operator with non-zero charge 
is zero. As is well known, the expectation values of products of
Fermi-operators with zero charge 
can be determined by various forms of the Wick theorem (see
\cite{AZ12} for details).
For example, we have:
\beq\label{ex2}
\lvacn \boldpsistar (w_1)\ldots \boldpsistar (w_m)
\boldpsi (z_m)\ldots \boldpsi (z_1)\rvacn =
\det_{1\leq i,j \leq m}
\lvacn \boldpsistar (w_i) \boldpsi (z_j)\rvacn ,
\eeq
where
\beq\label{ex1}
\lvacn \boldpsistar (w) \boldpsi (z)\rvacn =
\frac{z^nw^{1-n}}{w-z}
\eeq
(assuming that $|w|>|z|$).

Neutral bilinear combinations
$\displaystyle{\sum_{mn} a_{mn}\boldpsistar_m \boldpsi_n}$
of the Fermi-operators, with certain conditions
on the matrix $a_{mn}$, generate an
infinite-dimensional Lie algebra \cite{JM83}.
Exponentiating such expressions, one obtains
an infinite dimensional group (a version
of $GL(\infty )$).
Elements of this group can be represented
in the form
\begin{equation}\label{gg}
{\bf G}=\exp \Bigl (
\sum_{i, k \in {\z }}a_{ik}\boldpsistar_i \boldpsi_k\Bigr ).
\end{equation}
Following the paper \cite{AZ12}, we call them (invertible)
group-like elements. Their main property is the following
commutation relations with the $\boldpsi$- and 
$\boldpsistar$-operators:
\beq\label{GG}
\boldpsi_n {\bf G}=\sum_{l\in \z}R_{nl}{\bf G}\boldpsi_l,
\quad
{\bf G}\boldpsistar_n=\sum_{l\in \z}
R_{ln}\boldpsistar_l {\bf G},
\eeq
where the matrix $R$ is $R=e^{a}$. 

\subsubsection{The KP tau-function as an expectation value}

Consider the following bilinear operators with zero charge:
\beq\label{Jka}
{\bf J}_k =\sum_{j\in \z}\boldpsi_j 
\boldpsistar_{j+k}\,, \quad 
k\geq  1.
\eeq
They play especially important role in the applications
to the integrable hierarchies of the KP type. These 
operators are positive Fourier modes\footnote{The 
precise definition requires a normal ordering, 
which is not essential for the operators (\ref{Jka})
with $k\neq 0$, see \cite{AZ12}.} of the 
``current operator'' ${\bf J}(z)$ which is given by
the (normally ordered) product of the Fermi-fields:
${\bf J}(z)=\boldpsi (z)\boldpsistar (z)$.
It is easy to 
see that $[{\bf J}_k , \boldpsi _m]=\boldpsi_{m-k}$, 
$[{\bf J}_k , \boldpsistar _m]=-\boldpsistar_{m+k}$.
Let ${\bf t}=\{t_1, t_2, t_3, \ldots \}$ be the set of 
parameters which are going to be the KP times. Set
\beq\label{ferm2}
{\bf J}({\bf t})= \sum_{k\geq 1}t_k {\bf J}_k.
\eeq
It is not difficult to prove that
\beq\label{ferm3}
e^{{\bf J}({\bf t})}\boldpsi (z)
e^{-{\bf J}({\bf t})}=e^{\xi ({\bf t} ,\, z)}\boldpsi (z),
\quad \quad
e^{{\bf J}({\bf t})}\boldpsistar (z)
e^{-{\bf J}({\bf t})}=e^{-\xi ({\bf t} , \,z)}\boldpsistar (z),
\eeq
where $\xi ({\bf t}, z)$ is given by (\ref{xi}).

As it has been established in the works of the Kyoto school,
the expectation values of 
group-like elements are $\tau$-functions
of integrable hierarchies of nonlinear differential equations, i.e.,
they obey an infinite set of bilinear equations 
of the Hirota-Miwa type 
\cite{Hirota81,Miwa82} which eventually 
follow from the Wick theorem
for expectation values of products of the Fermi-operators.

The KP tau-function is defined as the vacuum expectation value
\beq\label{tau1}
\tau ({\bf t})=\lvac e^{{\bf J}({\bf t})}{\bf G}\rvac ,
\eeq  
where ${\bf G}$ is any group-like element of the form (\ref{gg})
(or of a more general form given below).
The mKP tau-function is defined as
\beq\label{tau2}
\tau (n,{\bf t})=\lvacn e^{{\bf J}({\bf t})}{\bf G}\rvacn .
\eeq  
For each $n$ it is also a KP tau-function as a function
of the times ${\bf t}$.
The proof 
that the functions having this form satisfy the bilinear relation 
of the KP hierarchy in the generating integral form
(\ref{g1}) is based on the operator bilinear identity
\beq\label{tau2a}
\sum_{n\in \z}  \boldpsi_n {\bf G}\otimes
\boldpsistar_{n} {\bf G}=\sum_{n\in \z} {\bf G}\,
\boldpsi_n \otimes {\bf G} \boldpsistar_{n} ,
\eeq
or
\beq\label{tau2b}
\sum_{n\in \z}  \boldpsistar_n {\bf G}\otimes
\boldpsi_{n} {\bf G}=\sum_{n\in \z} {\bf G}\,
\boldpsistar_n \otimes {\bf G} \boldpsi_{n} ,
\eeq
which is valid for the group-like elements ${\bf G}$ 
with zero charge 
of the form (\ref{gg}) (and also for certain operators
with a nonzero charge of a more general form given below).
The identities (\ref{tau2a}), (\ref{tau2b}) immediately
follow from the commutation relations (\ref{GG}).
Equation (\ref{tau2a}) is to be understood as the identity
\beq\label{tau2c}
\sum_{n\in \z}\left <U\right | \boldpsi_n {\bf G}\left |V \right >
\left <U'\right | \boldpsistar_n {\bf G}\left |V' \right >=
\sum_{n\in \z}\left <U\right | {\bf G}\boldpsi_n \left |V \right >
\left <U'\right | {\bf G}\boldpsistar_n \left |V' \right >
\eeq
valid for any states $\left <U\right |$, $\left <U'\right |$,
$\left |V\right >$, $\left |V'\right >$ from the fermionic Fock
space. Choosing $\left |V\right >=\left |V'\right >=\rvac$,
$\left <U\right |=\left <1\right | e^{{\bf J}({\bf t})}$, 
$\left <U'\right |=\left <-1\right | e^{{\bf J}({\bf t'})}$, 
we see that
the right-hand side of (\ref{tau2c}) vanishes, and the rest 
can be written as
\beq\label{tau2d}
\oint_{C_{\infty}}\left <1\right |e^{{\bf J}({\bf t})}
\boldpsi (z){\bf G}\rvac 
\left <-1\right |e^{{\bf J}({\bf t'})}
\boldpsistar (z){\bf G}\rvac \frac{dz}{z}=0.
\eeq
To bring it to the form (\ref{g1}),
one should use the commutation relations (\ref{ferm3}) and 
the bosonization rules
\beq\label{bos}
\begin{array}{l}
\lvacn \boldpsi (z)e^{{\bf J}({\bf t})}
=z^{n-1} \left < n-1\right | e^{{\bf J}({\bf t}-[z^{-1}])},
\\ \\
\lvacn \boldpsistar (z)e^{{\bf J}({\bf t})}
=z^{-n} \left < n+1\right | e^{{\bf J}({\bf t}+[z^{-1}])}
\end{array}
\eeq
(equations (2.93) in \cite{AZ12}). 
Note that the charge of both sides are $-(n-1)$ and $-(n+1)$
in the first and the second equation respectively.
The repeated use of the rules (\ref{bos})
gives more general bosonization formulas:
\beq\label{bos1}
\begin{array}{l}
\displaystyle{
\lvacn \boldpsi (z_m)\ldots \boldpsi (z_1)e^{{\bf J}({\bf t})}
\! =\! \Bigl (\prod_{l=1}^{n-m}z_l^{n-m}\Bigr ) 
\Delta_m (z_1, \ldots , z_m)
\left <n\! -\! m\right | \exp \Bigl ({\bf J}\Bigl ({\bf t}\! -\! 
\sum_{i=1}^{m}
[z_{m}^{-1}]\Bigr )\Bigr ),}
\\ \\
\displaystyle{
\lvacn \boldpsistar (z_m)\ldots \boldpsistar (z_1)e^{{\bf J}({\bf t})}
\!\! =\!\! \Bigl (\prod_{l=1}^{n-m}z_l^{-\! n\! - \!m +\! 1}\Bigr ) 
\Delta_m (z_1, \ldots , z_m)
\left <n\! +\! m\right | \exp \Bigl ({\bf J}\Bigl ({\bf t}\! +\! 
\sum_{i=1}^{m}
[z_{m}^{-1}]\Bigr )\Bigr ),}
\end{array}
\eeq
where $\Delta_m$ is the Vandermonde determinant given by 
(\ref{vandermonde}).

The BA functions (\ref{mkp1}) are expressed through the vacuum
expectation values by the following formulas:
\beq\label{tau3}
\begin{array}{l}
\displaystyle{
\psi ({\bf t}, z)=\frac{\left <1\right | e^{{\bf J}({\bf t})}
\boldpsi (z){\bf G}\rvac}{\lvac e^{{\bf J}({\bf t})}{\bf G}\rvac},}
\\ \\
\displaystyle{
\psi^* ({\bf t}, z)=\frac{\left <-1\right | e^{{\bf J}({\bf t})}
\boldpsistar (z){\bf G}\rvac}{z 
\lvac e^{{\bf J}({\bf t})}{\bf G}\rvac}.}
\end{array}
\eeq
The BA kernel (\ref{s1}) is given by
\beq\label{BAkernel}
\Psi ({\bf t}; z,w)=\frac{\left <0\right | e^{{\bf J}({\bf t})}
\boldpsi (z)
\boldpsistar (w)\, 
{\bf G}\rvac}{w\lvac e^{{\bf J}({\bf t})}{\bf G}\rvac}.
\eeq
It is assumed in (\ref{tau3}), (\ref{BAkernel})
that the element ${\bf G}$ has zero charge.
To see that these representations agree 
with formulas (\ref{mkp1}) and (\ref{s1}), one should use the
bosonization rules.

Along with KP tau-functions given by (\ref{tau1}) with
${\bf G}$ of the form (\ref{gg}) more general KP tau-functions
can be introduced as expectation values of certain 
fermionic group-like operators with a definite nonzero charge. 
In \cite{AZ12} such operators are called non-invertible group-like
elements. 
Namely, let 
${\bf w}_i$ be a linear combination of the operators
$\boldpsi_k$ and ${\bf w}^*_i$ be a linear combination
of the operators $\boldpsistar_k$. The element
$$
{\bf W}^{M,N}=
{\bf w}^*_M\ldots {\bf w}^*_1 \, 
{\bf w}_N\ldots {\bf w}_1
$$
has charge $N-M$. The operator ${\bf W}^{M,N}$ as well as 
the product
${\bf W}^{M,N}{\bf G}$ with ${\bf G}$ of the form (\ref{gg})
are examples of non-invertible group-like elements of charge
$N-M$. Although neither ${\bf w}_i$ nor ${\bf w}^*_i$ 
satisfy (\ref{GG}), it is easy to check that 
they nevertheless obey the basic
bilinear operator identity (\ref{tau2a}), which is enough
for derivation of the bilinear equation for the tau-function.
With the help of it, it is not difficult to show (see, e.g. 
\cite{AZ12})
that the function
\beq\label{tau4}
\tau ({\bf t})=\left <N\! -\! M\right | e^{{\bf J}({\bf t})}
{\bf W}^{M,N}{\bf G} \left |\vphantom{N-M}0\right >
\eeq  
is a KP tau-function, too. In the corresponding expressions 
for the BA functions the left vacuum should be shifted 
accordingly, to make the whole expression (including the
vacua and the Fermi-operators) neutral.

\subsubsection{The BD transformations} 
\label{section:BDchargedfermions}

Using the representation (\ref{tau4}) of KP tau-functions,
it is easy to define the BD transformations. Introduce the
operators
\beq\label{tau5}
\boldPsi =\int_{\CCC}\boldpsi (z)\rho (z)d^2z,\quad
\boldPsistar =\int_{\CCC}\boldpsistar (z)z^{-1}\rho^* (z)d^2z,
\eeq
where $\rho$ and $\rho^*$ are density functions, as in (\ref{mkp2a}).
The forward BD transformation that sends the tau-function
(\ref{tau1}) to $\hat \tau$ given by
(\ref{mkp9b}) is defined as
\beq\label{tau6}
\hat \tau ({\bf t})=\left <1\right | e^{{\bf J}({\bf t})}
\boldPsi \, {\bf G} \left |0\right >.
\eeq
Hereafter we assume that the element ${\bf G}$ has zero charge;
otherwise the left vacuum should be shifted accordingly.
Similarly, the backward BD transformation from $\tau$ to $\check \tau$
is given by
\beq\label{tau6a}
\check \tau ({\bf t})=\left <-1\right | e^{{\bf J}({\bf t})}
\boldPsistar \, {\bf G} \left |0\right >.
\eeq

A chain of the BD transformations is defined by sequential action
of the operators of the form (\ref{tau5}).
Namely, introduce the operators
\beq\label{tau5a}
\boldPsi_j =\int_{\CCC}\boldpsi (z)\rho_j (z)d^2z,\quad
\boldPsistar_j =\int_{\CCC}\boldpsistar (z)z^{-1}\rho^*_j (z)d^2z,
\eeq
then the tau-function (\ref{b4}) and (\ref{b4a}) can be 
represented as
\beq\label{tau7}
\tau (n, {\bf t})=\left <n\right | e^{{\bf J}({\bf t})}
\boldPsi_n \ldots \boldPsi_1 \, {\bf G} \left |0\right >, 
\quad n\geq 0,
\eeq
and
\beq\label{tau7a}
\hspace{1.0cm}
\tau (-m, {\bf t})=\left <-m\right | e^{{\bf J}({\bf t})}
\boldPsistar_m \ldots \boldPsistar_1 \, {\bf G} \left |0\right >,
\quad m\geq 0.
\eeq
This is easy to see with the help of bosonization formulas
(\ref{bos1}).
One can also consider a composition of $n$ forward and $m$ 
backward BD transformations:
\beq\label{tau7b}
\tau (m, n, {\bf t})=\left <n-m\right | e^{{\bf J}({\bf t})}
\boldPsistar_m \ldots \boldPsistar_1 \,
\boldPsi_n \ldots \boldPsi_1 
\, {\bf G} \left |0\right >.
\eeq
The representation of BD transformations in terms of Fermi-operators 
makes it obvious that 
any two forward and any two backward 
transformations anticommute while a forward transformation
anticommutes with a backward one only 
if supports of the corresponding
density functions do not intersect (because in this case 
$\boldPsi_j$ anticommutes with $\boldPsistar_k$).
The case $m=n$ in (\ref{tau7b}) 
corresponds to a composition of $n$ 
sequential binary BD transformations, and the Wick theorem
implies that the right-hand side of (\ref{tau7b}) in this case
is the determinant of the $n\times n$ matrix whose matrix
elements are the BA kernels at different points, as in equation
(\ref{s1a}).

The determinant formulas like (\ref{b6}) and (\ref{b6a})
as well as bilinear relations of the form (\ref{mkp10}), 
(\ref{mkp12}) follow, eventually, from one or another form
of the Wick theorem for free fermions\footnote{A detailed 
exposition of different forms of the Wick theorem with proofs
can be found in the review \cite{AZ12}.}.

\subsection{Neutral fermions (BKP)}
\label{section:neutral}

The BKP hierarchy, like the KP one, admits a description
in terms of free massless fermions. However, in this case we need 
neutral fermions $\boldphi_k$, $k\in \ZZ$ and the corresponding
neutral fields
$$
\boldphi (z)=\sum_{k\in \z}\boldphi_k z^k.
$$
This formalism was developed in \cite{JM83,DJKM82}. 
In our presentation, we also follow the work \cite{LO15}.

\subsubsection{The algebra of neutral Fermi-operators 
and vacuum states}
 
The neutral Fermi-operators satisfy the anticommutation 
relation
\beq\label{p1}
[\boldphi_m , \boldphi_n]_{+}=(-1)^m \delta_{m, -n}.
\eeq
We thus see that $\boldphi_m^2=0$ for all $m\neq 0$.
The operator $\boldphi_0$ is special: its square is nonzero.
At $n=m=0$ we have from (\ref{p1}): 
$$\boldphi_0^2 =\frac{1}{2}.$$
The algebra ${\cal B}$ of neutral fermions consists of linear
combinations of all possible products of the neutral 
Fermi-operators $\boldphi_n$ (monomials). Elements of the
algebra are multiplied in accordance with the anticommutation
relations (\ref{p1}). One can define parity ${\sf p}\in \ZZ_2$ 
of some elements of the algebra ${\cal B}$ by the following rules:
$$
{\sf p}(1)=0, \quad
{\sf p}(\boldphi_k)=1, \quad {\sf p}(\boldphi_k \boldphi_l)=0
\quad \mbox{for all $k,l\in \ZZ$}
$$
and
$$
{\sf p}({\bf a}{\bf b})={\sf p}({\bf a})+{\sf p}({\bf b}) 
\;\; (\mbox{mod $2$})
$$
for any two elements ${\bf a}, {\bf b}\in {\cal B}$ with a definite
parity. For the neutral Fermi-operators, parity is a characteristic 
analogous to charge for the charged ones. Elements 
${\bf a}\in {\cal B}$ with ${\sf p}({\bf a})=1$ will be called
${\it odd}$, and those with ${\sf p}({\bf a})=0$ {\it even}.
Any element of 
the algebra ${\cal B}$ can be represented as a sum of 
even and odd elements.

The right and left vacua can be defined by the requirements
\beq\label{p2}
\boldphi_n\rvac =0 \;\; \mbox{for} \;\; n<0,
\qquad
\lvac \boldphi_n =0 \;\; \mbox{for} \;\; n>0.
\eeq
However, these properties do not yet fix the vacuum states uniquely,
because given vacuum states $\rvac , \lvac$, the states
$\left |0'\right >=\boldphi_0 \rvac$ and
$\left <0'\right |=\lvac \boldphi_0 $ also satisfy conditions
(\ref{p2}). One of possibilities to choose the vacuum states is
using the embedding of the algebra ${\cal B}$ into the 
Clifford algebra ${\cal A}$ of charged fermions 
and identify the
vacuum state and its dual with the right and left 
Dirac vacua for the 
latter\footnote{This is done in detail, e.g., in \cite{LO15}.}.
Here we adopt this ``physical'' point of view.
Some details are given in Appendix B of the present paper.
For a more detailed (and more rigorous) treatment of the algebra
${\cal B}$ and the vacuum states see \cite{DJKM82}.

Assuming that the right and left vacuum states are fixed,
we define the vacuum expectation value $\lvac (\ldots )\rvac$ 
by the properties
that a) $\lvac {\bf a} \rvac =0$ for all odd ${\bf a}\in {\cal B}$
and b) for even elements it holds
\beq\label{p4}
\begin{array}{rcl}
\lvac \left. 0\right > &=& 1, 
\\ \\
\lvac \boldphi_i \boldphi_j \rvac 
 & =& \left \{
\begin{array}{l}
(-1)^j \delta_{i, -j}, \;\;\; j>0,
\\ \\
\frac{1}{2}\, \delta_{i0}, \;\; j=0,
\\ \\
0 \;\;\,\, , \quad j<0.
\end{array}\right.
\end{array}
\eeq
We note that
$$
\lvac \boldphi_0 {\bf b}\rvac =\lvac  {\bf b}\boldphi_0\rvac
\quad \mbox{or} \quad 
2\lvac \boldphi_0 {\bf b}\boldphi_0 \rvac =
\lvac  {\bf b}\rvac
$$
for any ${\bf b}\in {\cal B}$.
In fact the explicit form of the right-hand sides of (\ref{p4})
follows from the anticommutation relations (\ref{p1}) and 
conditions (\ref{p2}). For monomials of a higher even degree,
the expectation values can be found using the Wick theorem.
Let ${\bf w}_1, \ldots , {\bf w}_{2m}$ be linear combinations
of the Fermi-operators $\boldphi_k$, then it holds:
\beq\label{p5}
\lvac {\bf w}_1, \ldots , {\bf w}_{2m}\rvac
=\sum_{\sigma} (-1)^{\sigma} \lvac {\bf w}_{\sigma (1)}
 {\bf w}_{\sigma (2)}\rvac \, \ldots \, 
 \lvac {\bf w}_{\sigma (2m-1)}
 {\bf w}_{\sigma (2m)}\rvac ,
\eeq
where the sum goes over all permutations $\sigma$ of the set
$\{1, \ldots , 2m\}$ such that 
$$
\sigma (1)<\sigma (2), \, \sigma (3)<\sigma (4), \, \ldots , \,
\sigma (2m-1)<\sigma (2m)
$$ 
and
$
\sigma (1)<\sigma (3) < \ldots < \sigma (2m-1),
$
$(-1)^{\sigma}$ is the signature of the permutation $\sigma$.

In particular, we have:
\beq\label{p6}
\lvac \boldphi (z)\boldphi (w)\rvac =
\frac{1}{2}\, \Bigl (\frac{z-w}{z+w}\Bigr ),
\eeq
and multi-point correlation functions of the neutral fields  
are given by
the Pfaffian of the skew-symmetric matrix
\beq\label{Bij}
\tilde z_{ij}=\frac{z_i -z_j}{z_i+z_j}, \quad
i,j=1, \ldots , 2m.
\eeq
We have from (\ref{p6}):
$$
\lvac \boldphi (z_i)\boldphi (z_j)\rvac =
\frac{1}{2} \, \tilde z_{ij},
$$
and it follows from the Wick theorem (\ref{p5}) that
\beq\label{p7}
\begin{array}{l}
\lvac \boldphi (z_1), \ldots , \boldphi (z_{2m})\rvac =
\mbox{Pf}\, (\frac{1}{2}\tilde z_{ij}).
\end{array}
\eeq
The Pfaffian can be explicitly found (see (\ref{M3})), 
and the result is
\beq\label{p8}
\lvac \boldphi (z_1), \ldots , \boldphi (z_{2m})\rvac =
\frac{1}{2^m}\prod_{i<j}^{2m}
\frac{z_i-z_j}{z_i+z_j}.
\eeq

Bilinear combinations
$\displaystyle{\sum_{mn} b_{mn}\boldphi_m \boldphi_n}$
of the neutral Fermi-operators, with certain conditions
on the matrix $b_{mn}$, which we will not discuss here,
generate an
infinite-dimensional Lie algebra \cite{JM83}.
Exponentiating these expressions, one obtains
an infinite dimensional group. 
Elements of this group can be represented
in the form
\begin{equation}\label{bgg}
{\bf g}=\exp \Bigl (\frac{1}{2}
\sum_{i, k \in {\z }}b_{ik}\boldphi_i \boldphi_k\Bigr )
\end{equation}
with a skew-symmetric matrix $b_{ik}$. (Without loss of 
generality, its diagonal elements can be put equal to zero.)
By analogy with (\ref{gg}) we call them group-like elements.

\subsubsection{The BKP tau-function as an expectation value}

Consider the following even bilinear operators:
\beq\label{Jkb}
{\bf J}_k^{\rm B} =\frac{1}{2}\sum_{j\in \z}(-1)^{j-1}
\boldphi_j \boldphi_{-j-k}\,, \quad 
k\geq  1.
\eeq
It is easy to see that ${\bf J}_k^{\rm B} =0$ for all even nonzero 
$k$.
Let ${\bf t}=\{t_1, t_3, t_5, \ldots \}$ be the set of 
parameters which are going to be the BKP times. Set
\beq\label{ferm2b}
{\bf J}^{\rm B}
({\bf t})= \sum_{k\geq 1, \, {\rm odd}}t_k {\bf J}^{\rm B}_k.
\eeq
It is not difficult to prove that $[{\bf J}_k^{\rm B}, \boldphi (z)]
=z^k \boldphi (z)$, hence
\beq\label{ferm3b}
e^{{\bf J}^{\rm B}({\bf t})}\boldphi (z)
e^{-{\bf J}^{\rm B}({\bf t})}=e^{\xi_{\rm o} ({\bf t} ,\, z)}
\boldphi (z),
\eeq
where $\xi_{\rm o} ({\bf t}, z)$ is the same as in (\ref{xib}).
We also note that ${\bf J}^{\rm B}({\bf t})\rvac =0$,
and hence 
$e^{{\bf J}^{\rm B}({\bf t})}\rvac =\rvac $.

Along with elements of the form (\ref{bgg}) it is necessary 
to consider elements of the following more general form:
\beq\label{p9}
{\bf g}'={\bf v}_1\ldots {\bf v}_{2m}{\bf g},
\eeq
where ${\bf v}_i$ are linear combinations of the Fermi-operators
$\boldphi_k$. General BKP tau-functions are given by
\beq\label{p10}
\tau ({\bf t})=\lvac e^{ {\bf J}^{\rm B} ({\bf t})}
\, {\bf g}' \rvac 
\eeq
with ${\bf g}'$ of the form (\ref{p9}). 
The proof 
that such functions satisfy the bilinear relation 
of the BKP hierarchy in the generating integral form
(\ref{bb1}) is based on the operator bilinear identity
\beq\label{p10a}
\sum_{n\in \z} (-1)^n \boldphi_n {\bf g}\otimes
\boldphi_{-n} {\bf g}=\sum_{n\in \z} (-1)^n {\bf g}\,
\boldphi_n \otimes {\bf g} \boldphi_{-n} ,
\eeq
which is valid for the group-like elements ${\bf g}$ 
of the general form (\ref{p9}) and also for elements
of the form (\ref{p9}). Equation (\ref{p10a}) is to be
understood as the identity
\beq\label{p10b}
\sum_{n\in \z}(-1)^n
\left <U\right | \boldphi_n {\bf g}\left |V \right >
\left <U'\right | \boldphi_{-n} {\bf g}\left |V' \right >=
\sum_{n\in \z}(-1)^n 
\left <U\right | {\bf g}\boldphi_n \left |V \right >
\left <U'\right | {\bf g}\boldphi_{-n} \left |V' \right >
\eeq
valid for any states $\left <U\right |$, $\left <U'\right |$,
$\left |V\right >$, $\left |V'\right >$ from the fermionic Fock
space for neutral fermions. 
Choosing $\left |V\right >=\left |V'\right >=\rvac$,
$\left <U\right |=\left <0\right | \boldphi_0
e^{{\bf J}^{\rm B}({\bf t})}$, 
$\left <U'\right |=\left <0\right | \boldphi_0 
e^{{\bf J}^{\rm B}({\bf t'})}$, we have from (\ref{p10b}):
$$
\begin{array}{l}
\displaystyle{
\sum_{n\in \z} (-1)^n \lvac \boldphi_0 
e^{{\bf J}^{\rm B}({\bf t})}\boldphi_n {\bf g}\rvac
\lvac \boldphi_0 
e^{{\bf J}^{\rm B}({\bf t'})}\boldphi_{-n} {\bf g}\rvac }
\\ \\
\displaystyle{\phantom{aaaaaaaaaaaaaaaaaaaa}
=\lvac \boldphi_0 
e^{{\bf J}^{\rm B}({\bf t})} {\bf g}\boldphi_0\rvac
\lvac \boldphi_0 
e^{{\bf J}^{\rm B}({\bf t'})} {\bf g}\boldphi_0\rvac }
\end{array}
$$
or
$$
\begin{array}{l}
\displaystyle{
\oint_{C_{\infty}}e^{\xi_{\rm o}({\bf t}-{\bf t '}, z)}
\lvac \boldphi_0 \boldphi (z) 
e^{{\bf J}^{\rm B}({\bf t})}{\bf g}\rvac
\lvac \boldphi_0 \boldphi (-z) 
e^{{\bf J}^{\rm B}({\bf t'})}{\bf g}\rvac }
\\ \\
\displaystyle{\phantom{aaaaaaaaaaaaaaaaaaaaaa}
=
\lvac \boldphi_0 
e^{{\bf J}^{\rm B}({\bf t})} {\bf g}\boldphi_0\rvac
\lvac \boldphi_0 
e^{{\bf J}^{\rm B}({\bf t'})} {\bf g}\boldphi_0\rvac .}
\end{array}
$$
To bring this equation to the form (\ref{bb1}), 
one should use the ``neutral'' analogs
of bosonization rules (\ref{bos}):
\beq\label{p12}
\begin{array}{l}
\lvac \boldphi (z)e^{ {\bf J}^{\rm B} ({\bf t})}=
\lvac \boldphi_0 e^{ {\bf J}^{\rm B} ({\bf t}-2[z^{-1}]_{\rm o})},
\\ \\
\lvac \boldphi_0
\boldphi (z)e^{ {\bf J}^{\rm B} ({\bf t})}=\frac{1}{2}
\lvac e^{ {\bf J}^{\rm B} ({\bf t}-2[z^{-1}]_{\rm o})}
\end{array}
\eeq
(see equations (6.5) in \cite{JM83}). Note that the second
equation here can be rewritten in a form 
which does not contain the pre-factor $\frac{1}{2}$, and thus looks 
closer to the first equation:
\beq\label{p12a}
\left <0'\right |\boldphi (z)e^{ {\bf J}^{\rm B} ({\bf t})}=
\left <0'\right | 
\boldphi_0 e^{ {\bf J}^{\rm B} ({\bf t}-2[z^{-1}]_{\rm o})},
\eeq
where $\left <0'\right |=\lvac \boldphi_0$.
By induction, one can prove the following more general
bosonization formulas:
\beq\label{p12b}
\begin{array}{l}
\displaystyle{
\lvac \boldphi (z_1) \, \ldots \, \boldphi (z_{2m})
e^{ {\bf J}^{\rm B} ({\bf t})}=\frac{1}{2^{m}}
\prod_{i<j}^{2m}\frac{z_i-z_j}{z_i+z_j}\,
\lvac \exp \left (
{\bf J}^{\rm B} \Bigl ({\bf t}-2\sum_{l=1}^{2m}[z_l^{-1}]_{\rm o}
\Bigr ) \right ),}
\\ \\
\displaystyle{
\lvac \boldphi (z_1) \, \ldots \, \boldphi (z_{2m-1})
e^{ {\bf J}^{\rm B} ({\bf t})}=\frac{1}{2^{m-1}}
\prod_{i<j}^{2m-1}\frac{z_i-z_j}{z_i+z_j}\,
\left < 0'\right | 
\exp \left (
{\bf J}^{\rm B} \Bigl ({\bf t}\! -\! 
2\sum_{l=1}^{2m-1}[z_l^{-1}]_{\rm o}
\Bigr ) \right ),}
\end{array}
\eeq
where $m\geq 1$. Acting by both sides of the first
formula to the right vacuum $\rvac$, we reproduce 
(\ref{p8}).

The BA function is expressed as follows:
\beq\label{p11}
\psi ({\bf t}, z)=2\, 
\frac{\lvac \boldphi_0
e^{ {\bf J}^{\rm B} ({\bf t})}\, \boldphi (z){\bf g}' 
\rvac}{\lvac e^{ {\bf J}^{\rm B} ({\bf t})}\, {\bf g}' \rvac}.
\eeq
To see that it is indeed connected with the tau-function 
in the same way as in (\ref{bb6}), one should use the 
bosonization rules.

\subsubsection{The BD transformations} 

Consider the operator
\beq\label{P1}
\boldPhi =\int_{\CCC} \boldphi (z) \rho (z) d^2 z,
\eeq
where $\rho$ is the same function (or distribution) as in 
(\ref{bb8}). Using this operator, the BD transformation (\ref{B2})
for the BKP tau-function can be represented as
\beq\label{P2}
\hat \tau ({\bf t})=2\lvac \boldphi_0 
e^{ {\bf J}^{\rm B} ({\bf t})}
\boldPhi \,
{\bf g}' \rvac ,
\eeq
that is easy to check. The fact that $\hat \tau$ is a BKP tau-function,
too, is clear because the right-hand side of (\ref{P2}) has
the form (\ref{p10}).

To define a chain of sequential transformations of this type,
$$\tau ({\bf t})=\tau (0,{\bf t}) \to  \tau (1,{\bf t}) \to
\tau (2,{\bf t})\to \ldots ,$$
we fix some half-infinite sequence of density functions
$\{ \rho_1 , \rho_2 , \rho_3, \ldots \}$ and introduce the operators
\beq\label{P3}
\boldPhi_i =\int_{\CCC} \boldphi (z) \rho_i (z) d^2 z, \quad
i=1,2,3, \ldots .
\eeq
Equation (\ref{P2}) is then rewritten as
\beq\label{P2a}
\tau (1, {\bf t})=2\lvac \boldphi_0 
e^{ {\bf J}^{\rm B} ({\bf t})}
\boldPhi_1 \,
{\bf g}' \rvac ,
\eeq
with $\tau (0, {\bf t})=\lvac e^{ {\bf J}^{\rm B} ({\bf t})}
\, {\bf g}' \rvac $. The next tau-function then is
\beq\label{P4}
\tau (2, {\bf t})=2\lvac  
e^{ {\bf J}^{\rm B} ({\bf t})}
\boldPhi_2 \boldPhi_1 \,
{\bf g}' \rvac .
\eeq
For general positive $n$, we have:
\beq\label{P5}
\tau (n , {\bf t}) = \, \lvac 
(2\boldphi_0)^{n}e^{ {\bf J}^{\rm B} ({\bf t})}
\boldPhi_{n} \ldots \boldPhi_1 
{\bf g}' \rvac .
\eeq
Note that $(2\boldphi_0)^n =2^{n/2}$ for even $n>0$ and
$(2\boldphi_0)^n =2^{(n+1)/2}\boldphi_0$ for odd $n>0$.
One can easily check that the right-hand side of (\ref{P5})
coincides with (\ref{B11}).
From the representation (\ref{P5}) it is clear that, in contrast
to the case of charged fermions, 
any two BD transformations in general do not anticommute (a
sufficient condition for anticommutativity of 
two transformations ${\bf B}_1$, ${\bf B}_2$ is empty
overlap between supports of $\rho_1 (z)$ and $\rho_2(-z)$).

\subsubsection{BD transformations for BKP versus those for KP}

It is known that the BKP tau-functions are square roots of 
certain KP tau-functions in which all ``even'' times are
put equal to 0. This fact allows us to connect the BD transformations
for these two hierarchies. The connection can be most easily
established using the technique of free fermions. Specifically,
we need the embedding of the algebra ${\cal B}$ of neutral fermions
into the Clifford algebra ${\cal A}$ of the charged ones. 
This realization of the algebra ${\cal B}$ is described in
Appendix B. In this section we use the notation and definitions
from Appendix B.

The important fact that allows one to connect tau-functions
for the KP and BKP hierarchies is 
that the operators
${\bf g}$, ${\bf J}^{\rm B}({\bf t})$ commute with
${\bf \hat g}$, ${\bf \hat J}^{\rm B}({\bf t})$.
Let $\tau^{\rm BKP}({\bf t})$ be a BKP tau-function:
\beq\label{apb10}
\tau^{\rm BKP} ({\bf t})=\lvac (2\boldphi_0)^m
e^{ {\bf J}^{\rm B} ({\bf t})} \boldPhi_1 \ldots \boldPhi_m
\, {\bf g} \rvac ,
\eeq
where 
\beq\label{P1a}
\boldPhi_i =\int_{\CCC_{+}^{+}} 
\boldphi (z) \rho^{\rm B}_i (z) d^2 z,
\eeq
with some density functions or distributions 
$\rho^{\rm B}_i (z)$ with supports in the right upper quadrant.
This tau-functions can be thought of as the result of $m$ 
sequential BD
transformations for the BKP hierarchy.
The same representation holds with the 
algebra of the $\boldphihat$-operators in the right-hand side.
Then we can write:
\beq\label{apb11}
\begin{array}{c}
\Bigl (\tau^{\rm BKP} ({\bf t})\Bigr )^2
=\lvac (2\boldphi_0)^m e^{ {\bf J}^{\rm B} ({\bf t})}
\boldPhi_1 \ldots \boldPhi_m
\, {\bf g} \rvac  \lvac (2\boldphihat_0)^m
e^{ {\bf \hat J}^{\rm B} ({\bf t})}
\boldPhihat_1 \ldots \boldPhihat_m
\, {\bf \hat g} \rvac
\\ \\
\displaystyle{
=i^{m(m+1)}\lvac (2\boldphi_0)^m (2\boldphihat_0)^m
e^{ {\bf J}^{\rm B} ({\bf t}) + 
{\bf \hat J}^{\rm B} ({\bf t})}
\prod_{j=1}^m (\boldPhi_j \boldPhihat_j)
\, {\bf g} {\bf \hat g}\rvac }
\\ \\
\displaystyle{
=(-2i)^m\lvac 
e^{ {\bf J} ({\bf t})}
\prod_{j=1}^m (\boldPhi_j \boldPhihat_j)
\, {\bf G}\rvac },
\end{array}
\eeq
where ${\bf G}={\bf g}{\bf \hat g}$.
The product $\boldPhi_j \boldPhihat_j$ can be transformed
using the formulas from Appendix B:
\beq\label{trans}
\begin{array}{c}
\displaystyle{
\boldPhi_j \boldPhihat_j=\int_{\CCC_{+}^{+}}\! \int_{\CCC_{+}^{+}}
\rho^{\rm B}_j(z)\rho^{\rm B}_j(w)\boldphi (z)\boldphihat (w)
d^2 z d^2 w}
\\ \\
\displaystyle{
=\frac{i}{2}\int_{\CCC_{+}^{+}}\! \int_{\CCC_{+}^{+}}
\rho^{\rm B}_j(z)\rho^{\rm B}_j(w)
\Bigl (\boldpsi (z)+\boldpsistar (-z)\Bigr )
\Bigl (\boldpsi (w)-\boldpsistar (-w)\Bigr )
d^2 z d^2 w}
\\ \\
\displaystyle{
=i\Bigl (\int_{\CCC_{+}^{+}}
\rho^{\rm B}(z)\boldpsistar (-z)d^2 z\Bigr )
\Bigl (\int_{\CCC_{+}^{+}}
\rho^{\rm B}(w)\boldpsi (w)d^2 w\Bigr ).
}
\end{array}
\eeq
Therefore, we conclude that
\beq\label{trans1}
\boldPhi_j \boldPhihat_j=i\boldPsi_j \boldPsistar_j,
\eeq
where
\beq\label{trans2}
\boldPsi_j = \int_{\CCC}\rho_j (z)\boldpsi (z)d^2 z ,
\quad
\boldPsistar_j = \int_{\CCC}\rho_j^* (z)\boldpsistar (z)d^2 z,
\eeq
and
\beq\label{trans3}
\rho_j (z)=\rho_j^{\rm B}(z) , \qquad 
\rho_j^* (z)=\rho_j^{\rm B}(-z).
\eeq
Plugging this into (\ref{apb11}), we conclude that
\beq\label{trans4}
\Bigl (\tau^{\rm BKP} ({\bf t})\Bigr )^2=2^m
\lvac  e^{ {\bf J} ({\bf t})}
\prod_{j=1}^m (\boldPsi_j \boldPsistar_j)
\, {\bf G}\rvac 
=\tau^{\rm KP}(t_1, 0, t_3, 0, t_5, \ldots ).
\eeq
This means that the BKP tau-function is the square root 
of the KP tau-function corresponding to the group-like element 
$\displaystyle{\prod_{j=1}^m (\boldPsi_j 
\boldPsistar_j){\bf g}{\bf \hat g}}$, with all even times being put
equal to 0. The insertion $\boldPsi_j 
\boldPsistar_j$ implements the binary BD transformation 
of the KP tau-function, with special density functions given by
(\ref{trans3}). Therefore, any BD transformation of the BKP type
is equivalent to a binary BD transformation of the KP type 
with specially chosen density functions, as in (\ref{trans3}).

\section{Conclusion}

In this paper, we have revisited the world of B\"acklund-Darboux (BD)
transformations for integrable hierarchies of nonlinear 
partial differential equations such as KP, mKP, BKP and their 
difference analogs. We have tried to develop a unified approach
based on the bilinear formalism, where the main hero is the 
tau-function. We have seen that when iterated, the BD transformations
form infinite or half-infinite chains, with $j$th step of the chain
being defined with the help of a density function 
$\rho_j(z)$ (which can be a distribution). Remarkably, neighboring
members of such a chain are connected by bilinear 
equations belonging to 
the same hierarchy (more precisely, to its discrete analog), whose
form does not depend on a particular choice of the density
functions at each step. In fact the idea that BD transformations
yield an integrable discretization of the hierarchy in question
is very old and it is hard to point out any ``primordial''
reference. For the BKP hierarchy, this idea was first realized
in \cite{NS98}\footnote{I thank the referee who informed me
about this work.}.

In conclusion, let us note that
for some important applications, the chains of sequential BD
transformations (finite or infinite) are of a special interest.
The most important and interesting
cases of the whole construction are the cases when the density
functions have supports concentrated either 
on 1D lines (such as, for example, $\RR$ or $\SSSS$) 
or at some distinct points
in $\CC$, i.e., when they are distributions rather than 
continuous functions in the plane. 

The former possibility, when the density corresponding to 
each step of the BD chain is concentrated on 1D curves
in the 2D plane, is relevant to the integrable
properties of the theory of random 
matrices\footnote{The literature on random matrices is enormous.
Here we only mention the classical book \cite{Mehta},
where the connections with integrable equations were not
discussed, and the papers 
\cite{Morozov,Mironov1,Mironov2,KMMOZ91} 
devoted to integrability of models of random matrices.}. 
In known examples,
the curves are either the real line $\RR$ for matrices with
real eigenvalues
(or $\RR_{+}$ in some cases) or the unit circle $\SSSS$
for unitary matrices. In these applications, the discrete
variable $n\in \ZZ_{+}$ numbering sequential steps of the BD chain
is just size of the random matrices. In this case 
the chain of BD transformations
is half-infinite, and for applications to quantum
gravity and string theory of particular interest is the limit
$n \to \infty$ taken in one or another sophisticated way
(for example, the double-scaling limit).

Meanwhile, 
integrable properties of random matrices with complex 
eigenvalues (for example, normal matrices) require an access
to a similar theory of BD transformations for more general integrable
hierarchies such as 2D Toda lattice. We hope to address this
issue in a separate publication. 

The applications to quantum integrable models
arise in the cases when the density functions 
corresponding to the BD transformations are
concentrated at some isolated points
in $\CC$ chosen in a special way such that the tau-functions
at each level of the BD chain 
are polynomials or ``trigonometric polynomials'' of the 
first variable $x$ identified with the quantum spectral\
parameter (for quantum models with rational and 
trigonometric $R$-matrices
respectively). Unlike the applications to matrix models, 
of the main interest here are {\it finite} chains 
of BD transformations, with the discrete variable $n$ varying from
$0$ to $N$, where the $N$ comes from 
the underling symmetry algebra of the quantum model 
($U(gl_N)$ or $U_q(gl_N)$). Another difference is that 
the BD transformations of the chain should be performed
in the backward direction, from $\tau (N)$ to $\tau (0)=1$,
thus ``undressing'' the original problem to the trivial
one. The sequential
steps of the chain correspond to the levels of the nested
Bether ansatz procedure. Given a fixed $\tau (N)$ as a polynomial
(or trigonometric polynomial) whose roots are inhomogeneity
parameters of the spin chain, the procedure 
of undressing it to the trivial one compatible with the mKP
hierarchy, has a finite number of solutions, 
each of which corresponding 
to different eigenstates of the quantum spin chain (on a finite
1D lattice). Details on this procedure can be found in
\cite{Z25}.

As we have seen in section \ref{section:BacklundKP}, there are
two types of BD transformations for the mKP hierarchy: forward
and backward ones. 
The scheme outlined in the previous paragraph is related 
to the forward transformations which appear as avatars of the
nested Bethe ansatz for quantum models based on classical
symmetry algebras and their $q$-deformations. The results
of \cite{KSZ08,Z08,TZZ15} suggest that
for graded quantum magnets (based on superalgebras, for example,
$gl(N|M))$ one should include into play also the backward BD
transformations defined by equation (\ref{b11}), with the 
BD chain consisting of $N$ forward and $M$ backward 
transformations. We anticipate that for $M>0$ 
the operator realization 
of the BD transformations given in section
\ref{section:BDchargedfermions} might be useful and instructive.
We plan to address the details elsewhere. 

Last but not least, there is a tempting problem to investigate 
whether there are some 
applications of the BD transformations 
and their chains for the BKP hierarchy
(see section \ref{section:BDBKP}) to any quantum
integrable models, in the analogy with the mKP case outlined above.
Perhaps, to uncover the connection (if it does exist in one or
another form), one should consider generalized 
quantum spin chains based on 
symmetry algebras of type B rather than A.

\section*{Acknowledgments}

\addcontentsline{toc}{section}{Acknowledgments}

I thank J. Harnad whose questions and comments on 
my recent paper \cite{Z25} gave me a strong motivation 
to revisit the world of B\"acklund-Darboux transformations and
try to understand them better. Also, it was J. Harnad who
drew my attention to the seminal papers \cite{KS02,Schief03}.
Thanks are also due to A. Orlov and V. Prokofev 
for valuable discussions and to R. Willox 
for informing me about the articles \cite{WTS97,WTLS98}. 
An illuminating discussion with
A. Veselov, after which the idea that B\"acklund-Darboux 
transformations need a better understanding first came to my mind,
is also gratefully acknowledged. 

This work is an output of the research project 
``Symmetry. Information. Chaos''
implemented as a part of the Basic Research Program at 
National Research University Higher School 
of Economics (HSE University).

\section*{Appendix A: Pfaffians}
\def\theequation{A\arabic{equation}}
\setcounter{equation}{0}

\addcontentsline{toc}{section}{Appendix A: Pfaffians}

Here we briefly recall some facts about Pfaffians which are
necessary for reading sections \ref{section:BKP} and 
\ref{section:fermions}. More details can be found, e.g.,
in the book \cite{Hirota-book}.

Odd-dimensional skew-symmetric matrices are degenerate, their
determinants are equal to 0. Determinant 
of an even-dimensional skew-symmetric matrix is always a full square.
The Pfaffian of an even-dimensional 
$2m\! \times \! 2m$ skew-symmetric matrix
$A$ is square root of its determinant.
It is given by the explicit formula
$$
\mbox{Pf}\, (A) =\sum_{\sigma} (-1)^\sigma A_{i_1i_2}
A_{i_3i_4}\ldots A_{i_{2m-1}i_{2m}},
$$
where the sum is taken over all permutations $\sigma$ of the indices 
$1,2, \ldots , 2m$ such that $$i_1<i_2, \, 
i_3<i_4, \ldots , i_{2m-1}<i_{2m}, \qquad 
i_1<i_3<\ldots < i_{2m-1}$$ 
and $(-1)^{\sigma}$ is signature of the permutation. 
For example, at $m=1$ we have
$\mbox{Pf}\,( A) =A_{12}$ and at $m=2$
$\mbox{Pf}\,( A) =A_{12}A_{34}-A_{13}A_{24}+A_{14}A_{23}$. 
The recurrence relation for the Pfaffians is
$$
\mbox{Pf}(A)=\sum_{k=2}^{N}\,
A_{1k}\mbox{Pf}(A_{\hat 1, \hat k}),
$$
where $A$ is a skew-symmetric matrix $A$ of an even size
$N\times N$ and $A_{\hat 1, \hat k}$ is the 
matrix $A$ with the first and $k$th columns and rows removed.

Let $A$ be an $N \times N$ skew-symmetric matrix with $N=2m$.
The following properties of its Pfaffian are useful in
applications:
$$
\begin{array}{l}
\mbox{Pf} (A)=(-1)^m \, \mbox{Pf} (A^{\rm t}), 
\quad \mbox{where} \;\; A^{\rm t}_{ij}=A_{ji},
\\ \\
\mbox{Pf} (\lambda A)=\lambda^{m} \, 
\mbox{Pf} (A), \quad \lambda \in \CC ,
\\ \\
\mbox{Pf} (B A B^{\rm t})=\det (B)\, \mbox{Pf} (A) \quad
\mbox{for arbitrary $N\times N$ matrix $B$}.
\end{array}
$$

It is known that tau-functions of integrable hierarchies 
of type A (such as KP  and 2D Toda) are determinants (in general
case, of infinite-dimensional matrices or operators) while 
those for hierarchies of types B and D are Pfaffians of some
(possibly, infinite dimensional) skew symmetric matrices. 

\section*{Appendix B: Neutral fermions from the charged ones}
\def\theequation{B\arabic{equation}}
\setcounter{equation}{0}

\addcontentsline{toc}{section}{Appendix B: 
Neutral fermions from the charged ones}

It is instructive to realize the algebra 
${\cal B}$ of neutral fermions dealt with in
section \ref{section:neutral} as a subalgebra of the Clifford 
algebra ${\cal A}$ of the charged fermions discussed in section
\ref{section:charged}. This embedding 
was first suggested in 1983 in the seminal paper 
\cite{JM83} and reviewed many times 
after that. Below we follow the paper \cite{LO15}.

In fact the neutral Fermi operators can be realized in terms
of the charged ones in the following way:
\beq\label{apb0}
\begin{array}{l}
\displaystyle{
\boldphi_n =\frac{1}{\sqrt{2}}\, \Bigl (\boldpsi_n +(-1)^n
\boldpsistar_{-n}\Bigr ), \; \quad n\in \ZZ , }
\\ \\
\displaystyle{
\boldphihat_n =\frac{i}{\sqrt{2}}\, \Bigl (\boldpsi_n -(-1)^n
\boldpsistar_{-n}\Bigr ), \quad n\in \ZZ .}
\end{array} 
\eeq
Then it is easy to check that the operators $\boldphi_n$
and $\boldphihat_n$
defined in this way obey the anticommutation relations
(\ref{p1}) and the $\boldphi$-operators anti-commute
with the $\boldphihat$-operators:
\beq\label{apb1}
[\boldphi_m , \boldphi_n]_{+}=
[\boldphihat_m , \boldphihat_n]_{+}=(-1)^m \delta_{m, -n},
\quad
[\boldphi_m , \boldphihat_n]_{+}=0.
\eeq
In particular, $\boldphi_0^2=\boldphihat_0^2=\frac{1}{2}$.
For the fermionic fields we have:
\beq\label{apb2}
\begin{array}{l}
\displaystyle{
\boldphi (z) =\frac{1}{\sqrt{2}}\, \Bigl (\boldpsi (z) +
\boldpsistar (-z)\Bigr ), }
\\ \\
\displaystyle{\boldphihat (z) =\frac{i}{\sqrt{2}}\, 
\Bigl (\boldpsi (z) -
\boldpsistar (-z)\Bigr ), }
\end{array} 
\eeq
Note that
\beq\label{apb5}
\boldphi_m \boldphi_n +
\boldphihat_m \boldphihat_n =
(-1)^n \boldpsi_m \boldpsistar_{-n} +
(-1)^m \boldpsistar_{-m} \boldpsi_{n}.
\eeq

Let $\rvac$, $\lvac$ be the dual Dirac vacua for the charged
fermions introduced in section \ref{section:charged}. Then it
is easy to check that the operators $\boldphi_n$, $\boldphihat_n$
defined as above act to the vacuum states as follows:
\beq\label{apb3}
\begin{array}{l}
\boldphi_n \rvac =\boldphihat_n \rvac =0, \quad n<0,
\\ \\
\lvac \boldphi_n  =\lvac \boldphihat_n =0  , \quad n>0,
\end{array}
\qquad
\begin{array}{l}
\boldphi_0 \rvac =-i\boldphihat_0 \rvac =
\frac{1}{\sqrt{2}}\boldpsi_0 \rvac ,
\\ \\
\lvac \boldphi_0  =i\lvac \boldphihat_0 =
\frac{1}{\sqrt{2}}\lvac \boldpsistar_0 .
\end{array}
\eeq
These formulas allow one to identify the vacuum states 
$\rvac$, $\lvac$ for the charged fermions with the vacuum
states $\rvac$, $\lvac$ introduced in section \ref{section:neutral}
for the neutral fermions\footnote{We note in passing that
the other vacuum state $\left |0'\right >$ introduced 
in section \ref{section:neutral} is 
$\left |0'\right >=\frac{1}{\sqrt{2}}\left |1\right >$.}.
In our opinion, this identification makes some 
constructions from section 
\ref{section:neutral} less artificial. 

We can introduce the group-like elements
\begin{equation}\label{apb6}
{\bf g}=\exp \Bigl (\frac{1}{2}
\sum_{i, k \in {\z }}b_{ik}\boldphi_i \boldphi_k\Bigr ),
\quad
{\bf \hat  G}=\exp \Bigl (\frac{1}{2}
\sum_{i, k \in {\z }}b_{ik}\boldphihat_i \boldphihat_k\Bigr )
\end{equation}
with a skew-symmetric matrix $b_{ik}$.
They are simultaneously serve as group-like elements for the 
algebra ${\cal A}$, as well as their product 
${\bf G}={\bf g} {\bf \hat g}$. Moreover, the elements
${\bf g}$, $ {\bf \hat g}$ obviously commute with each other. 

An important special case of group-like elements are the 
``evolution operators''
$e^{{\bf J}^{\rm B}({\bf t})}$, $e^{{\bf \hat J}^{\rm B}({\bf t})}$,
where
\beq\label{apb7}
{\bf J}^{\rm B}({\bf t})=\sum_{k\geq 1}t_k {\bf J}_k^{\rm B},
\quad
{\bf \hat J}^{\rm B}({\bf t})=\sum_{k\geq 1}t_k {\bf \hat J}_k^{\rm B}
\eeq
and
\beq\label{apb8}
{\bf J}_k^{\rm B} =\frac{1}{2}\sum_{j\in \z}(-1)^{j-1}
\boldphi_j \boldphi_{-j-k}\,, \quad 
{\bf \hat J}_k^{\rm B} =\frac{1}{2}\sum_{j\in \z}(-1)^{j-1}
\boldphihat_j \boldphihat_{-j-k}, \quad
k\geq  1.
\eeq
Note that ${\bf J}_k^{\rm B}={\bf \hat J}_k^{\rm B}=0$ for all even
positive $k$, so the sums in (\ref{apb7}) can be taken over 
all $k\geq 1$, with only times with odd indices entering the game.
Moreover, we have
\beq\label{apb9}
{\bf J}^{\rm B}({\bf t})+{\bf \hat J}^{\rm B}({\bf t})=
{\bf J}({\bf t}), 
\eeq
where ${\bf J}({\bf t})$ in the right-hand side is the 
evolution operator (\ref{ferm2}) for charged fermions, and it is
implied that all times with even indices in the set ${\bf t}$ 
are put equal to 0.

At last, let us show that the operator bilinear identity
(\ref{p10a}) for neutral fermions follows from the identities
(\ref{tau2a}), (\ref{tau2b}) for charged fermions.
Using the embedding ${\cal B}\subset {\cal A}$, we can rewrite
(\ref{p10a}) as
\beq\label{apb12}
\begin{array}{ll}
&\displaystyle{
\sum_{n\in \z}(-1)^n \Bigl (\boldpsi_n +(-1)^n \boldpsistar_{-n}
\Bigr ){\bf G} \otimes 
\Bigl (\boldpsi_{-n} +(-1)^n \boldpsistar_{n}
\Bigr ){\bf G}}
\\ &\\
=&\displaystyle{\sum_{n\in \z}(-1)^n {\bf G}
\Bigl (\boldpsi_n +(-1)^n \boldpsistar_{-n}
\Bigr ) \otimes 
{\bf G} \Bigl (\boldpsi_{-n} +(-1)^n \boldpsistar_{n}
\Bigr )},
\end{array}
\eeq
where ${\bf G}={\bf g}{\bf \hat g}$ is the group-like element
from the algebra ${\cal A}$. After opening the brackets we see that
half of the terms cancel due to identities 
(\ref{tau2a}), (\ref{tau2b}) and what is left is to prove that
\beq\label{apb13}
\sum_{n\in \z}(-1)^n \boldpsi_{n}{\bf G}\otimes 
\boldpsi_{-n}{\bf G}=
\sum_{n\in \z}(-1)^n {\bf G} \boldpsi_{n}\otimes 
{\bf G} \, \boldpsi_{-n},
\eeq
\beq\label{apb13a}
\sum_{n\in \z}(-1)^n \boldpsistar_{n}{\bf G}\otimes 
\boldpsistar_{-n}{\bf G}=
\sum_{n\in \z}(-1)^n {\bf G} \boldpsistar_{n}\otimes 
{\bf G} \, \boldpsistar_{-n}.
\eeq
Of course for arbitrary group-like ${\bf G}$ this does not take place.
However, for special ${\bf G}$ of the form ${\bf G}={\bf g}
{\bf \hat g}$ which corresponds to solutions of the BKP hierarchy
one can prove both (\ref{apb13}), (\ref{apb13a}). 

Let us present
some details for (\ref{apb13}). We need to prove that 
$(-1)^n \boldpsi_{-n}$ commutes with ${\bf G}$ in the same
way as $\boldpsistar_n$, i.e.,
$$
(-1)^n {\bf G}\boldpsi_{-n}=\sum_{l\in \z}(-1)^{l}R_{ln}
\boldpsi_{-l}{\bf G},
$$
or
\beq\label{apb14}
{\bf G} \boldpsi_n =\sum_{l\in \z}(-1)^{n+l}
R_{-l, -n}\boldpsi_l {\bf G}.
\eeq
Comparing with (\ref{GG}), we see that we should prove
the following symmetry of the $\ZZ \! \times \! \ZZ$ matrix $R_{nl}$:
\beq\label{apb15}
R_{nl}^{-1}=(-1)^{n+l}R_{-l, -n}.
\eeq
Writing 
$$
\sum_{m,n}b_{mn}\boldphi_m \boldphi_n +
\sum_{m,n}b_{mn}\boldphihat_m \boldphihat_n =
\sum_{m,n}(-1)^m b_{-m, n}\boldpsistar_m \boldpsi_n +
\; \mbox{a $c$-number}
$$
(where the skew-symmetry of the matrix $b_{mn}$ is used),
we conclude that the matrix $a_{mn}$ for the element ${\bf G}$
is of the form
\beq\label{apb16}
\begin{array}{l}
a_{mn}=\frac{1}{2}\, (-1)^m b_{-m,n},
\end{array}
\eeq
from which it follows that
\beq\label{apb17}
a_{nm}=(-1)^{n+m+1} a_{-m, -n}.
\eeq
The symmetry (\ref{apb15}) of the matrix $R=e^a$ is a consequence
of this. The proof of (\ref{apb13a}) is similar.






\end{document}